\documentclass[alpha-refs]{wiley-article}

\usepackage{graphicx}
\usepackage[space]{grffile}
\usepackage{latexsym}
\usepackage{textcomp}
\usepackage{longtable}
\usepackage{tabulary}
\usepackage{booktabs,array,multirow}
\usepackage{amsfonts,amsmath,amssymb}
\usepackage{natbib}
\usepackage{url}
\usepackage{hyperref}
\hypersetup{colorlinks=false,pdfborder={0 0 0}}
\usepackage{etoolbox}
\newif\iflatexml\latexmlfalse

\AtBeginDocument{\DeclareGraphicsExtensions{.pdf,.PDF,.eps,.EPS,.png,.PNG,.tif,.TIF,.jpg,.JPG,.jpeg,.JPEG}}

\usepackage[utf8]{inputenc}
\usepackage[english]{babel}

\usepackage{siunitx}
\usepackage{caption}
\usepackage{subcaption}
\usepackage{epstopdf}

\iflatexml

\else

\paperfield{Data Assimilation}
  \abbrevs{EnKF, Ensemble Kalman Filter; BGM, Burger's Model; KSM, Kuromoto-Shivashinsky Model; RMSE, Root Mean Squared Error; HR, High Resolution; HRA, High Resolution Augmented; LR, Low Resolution}
\corraddress{Christian Sampson PhD, Department of Mathematics, University of North Carolina at Chapel Hill, Chapel Hill, NC , 27599, USA}
\corremail{Christian.Sampson@gmail.com}
\presentadd{Mathematics Department, University of North Carolina at Chapel Hill, Chapel Hill, NC, 27599 , USA}
\fundinginfo{Office of Naval Research,  Grant/Award Number: A18-0960 and N00014-18-1-2493; UK Natural Environment Research Council ,  Grant/Award Number: NCEO02004}

\fi

\papertype{Original Article}

\title{Ensemble Kalman Filter for non-conservative moving mesh solvers with a joint physics and mesh location update}

\author[1]{Christian Sampson}
\author[2,3]{Alberto Carrassi}
\author[4]{Ali Aydo\u{g}du }
\author[1]{Chris K.R.T Jones}

\affil[1]{University of North Carolina at Chapel Hill}
\affil[2]{Dept. of Meteorology and National Centre for Earth Observations, University of Reading, United Kingdom}
\affil[3]{Mathematical Institute, University of Utrecht, Netherlands}
\affil[4]{Centro Euro-Mediterraneo sui Cambiamenti Climatici, Bologna, Italy}
\runningauthor{Sampson, Carrassi, Aydogdu, Jones }

\begin{document}

\maketitle
\begin{abstract}
 {\small Numerical solvers using adaptive meshes can focus computational power on important regions of a model domain capturing important or unresolved physics. The adaptation can be informed by the model state, external information, or made to depend on the model physics. 
 In this latter case, one can think of the mesh configuration {\it as part of the model state}. If observational data is to be assimilated into the model, the question of updating the mesh configuration with the physical values arises. Adaptive meshes present significant challenges when using popular ensemble Data Assimilation (DA) methods. We develop a novel strategy for ensemble-based DA for which the adaptive mesh is updated along with the physical values. This involves including the node locations as a part of the model state itself allowing them to be updated automatically at the analysis step. This poses a number of challenges which we resolve to produce an effective approach that promises to apply with some generality. We evaluate our strategy with two testbed models in 1-d comparing to a strategy that does not update the mesh configuration. We find updating the mesh improves the fidelity and convergence of the filter. 
 We also present an extensive analysis on the performance of our scheme beyond just the RMSE error.   }

\end{abstract}

\clearpage
\section{Introduction}

Modern adaptive moving mesh schemes present significant advantages over traditional fixed mesh schemes in many geophysical applications. Adaptive meshes can focus resolution in places of interest in order to make better use of available computational power, \citet{huang2010adaptive}, or can be designed to optimise computational cost and accuracy based on external factors, an example being ship and acoustic receiver locations in the prediction of underwater noise pollution from oceanic shipping activity, see \citet{Trigg2018}. In some applications one may require the mesh to change as the system evolves to better represent the underlying physics, \citet{weller2010bams}. Adaptive meshes are typically governed by a set of rules suitable to the specific problem being solved.

There are many reasons why solving a geophysical problem in a Lagrangian frame may be appropriate, see {\it e.g.} \citet{asch2016data,jablonowski2004adaptive}. The associated numerical solver will inevitably be based on a moving mesh, and some of the advantages described above of a moving mesh are delivered a fortiori by the use of such a scheme that advects the nodes with the flow. For instance, this may have the effect of naturally concentrating nodes in locations of increased activity, or better resolving coherent structures. More specifically, if the nodes are advected with the flow, node clusters can provide information where gradients are large and sinks or eddies exist, likewise node deserts can indicate where gradients are small. When the nodes are advected by the flow in this way, it is almost inevitable that they will have to change in both number and location in order to maintain solution accuracy for the numerical solver.  When using an adaptive mesh that is governed by the model physics like this, the node locations and physical quantities are inexorably coupled. As a consequence the node locations can be considered part of the model state and of the model's solution history. 

They key point to note is that for computational models based on such Lagrangian solvers, the node locations encode underlying physics and therefore provide information about the overall model state. As such, they are all updated together under the model evolution. But the observational data reflect underlying physics also and we would therefore expect that the optimal incorporation of such data should update the node locations as well as the values of the physical state variables. We develop here a data assimilation scheme for achieving exactly this impact of data on the mesh itself.

The process of incorporating data into physical models is called Data Assimilation (DA). A survey of DA methods can be found in \citet{budhiraja2018}. DA has become an integral tool in the geosciences and meteorology improving numerical weather prediction and as a method for parameter estimation. A review of DA in the geosciences can be found in \citet{carrassi2018}. We will focus on ensemble methods \citet{evensen2009,houtekamer2016enkf} which make use of estimated statistics from an ensemble of model runs at an analysis time step. These methods are attractive when attempting to leverage the information that node locations carry through covariances estimated from the ensemble members. That information is specifically brought in through the cross covariances between the physical values and the node locations driven by those values. In the case where the nodes are advected with the flow, these cross covariances very closely match the spatial gradient of the fluid velocities across the model domain. This encodes extra and important physical information into the DA update step. We will use observations of the physical state to update both the physical values and node locations in our approach which, in this work, will come from a twin model experiment using the models outlined in Section~\ref{modelandmesh}. 

Adapting existing ensemble methods to adaptive moving meshes involves tackling some significant challenges. Ensemble DA methods rely on estimated statistics from the ensemble members and for success they must be statistically consistent. The main challenge is the fact that each ensemble member may have nodes in different locations, in different numbers, or both. Previous work along these lines has been carried out in \citet{bonan2017} for an adaptive mesh 1-d ice sheet model. In that work the adaptive mesh was conservative, in that each ensemble member has the same number of points. Observations of the ice sheet edge were also directly assimilated. In this work we consider updating node locations for a non-conservative adaptive mesh model using eulerian observations of the physical quantities of the "truth run", similar to how a satellite may take observations. A non-conservative mesh means that each ensemble member will have a {\it different number} of nodes in {\it different locations} requiring us to develop methodologies to obtain consistent and meaningful error covariance estimates. Any methodology we develop will necessarily have some disruptive effect on the individual ensemble members themselves in order to achieve a measure of statistical consistency between them. This may come through the addition or removal of nodes or the interpolation of values to specific locations. With this in mind, we define a successful method as one which improves the estimate of the truth over forecast with no DA and take special care to study the effects the method has on the ensemble members themselves. We discuss these effects in Section~\ref{results} and recommend considerations to minimise any negative effects depending on the application at hand and the desired prediction goal.     

Other ensemble approaches aimed at adaptive meshes have been developed. \citet{jain2018amr} study a tsunami model which uses an adaptively refined mesh taking the union of all meshes as reference mesh to which each ensemble member is interpolated to before the update step. In \citet{du2016ensemble} a model which uses 3D unstructured adaptive mesh model for geophysical flows \citet{maddison2011fluidity,davies2011fluidity} was considered and an EnKF developed which uses the idea of a reference mesh to carry out the analysis step. The reference mesh is chosen using the idea of super-meshing \citet{farrell2009conservative} and each ensemble member is interpolated to that fixed reference mesh before the analysis step. These previous studies all concern conservative adaptive meshes. 

However, in \citet{Aydodu2019npg} a fixed reference mesh is used in two 1-d models for which the mesh evolves with the flow and undergoes a ``remeshing'' step which injects new nodes should two be to far apart or removes nodes should two be too close together. This remeshing means that each ensemble member will likely have different numbers of points in different locations. In that work two reference meshes are used and are chosen based on the rules of the remeshing scheme. Ensemble members are mapped to the reference mesh before the update and mapped back to their previous meshes after. Our work goes further extending the update to the node locations themselves. We use the reference mesh only as a guide to match components of the state vector and augment our state vector with the node locations. A reasonable supposition is that avoiding the mapping scheme will help to lessen disruption of individual ensemble members providing for better estimates of the error covariances needed for the update step.

This paper is structured as follows, in Section~\ref{modelandmesh} we describe the model and adaptive mesh scheme we use in our twin model experiments. In Section~\ref{AMMENKFS} we outline the necessary ingredients for an EnKF on non conservative adaptive mesh models and describe the two implementations of such that we will compare. The results are presented Section~\ref{results} along with the optimised inflation parameters needed for the methods. We follow the results with a discussion in Section~\ref{results} on the cross covariances of the physical variables and node locations as well as the effect the adapted EnKF schemes have on the ensemble members themselves. We also present considerations on choosing the inflation parameters depending on the application and finally in Section \ref{conclusions} we present some concluding remarks and summary.

\section{Model and Mesh}\label{modelandmesh}

\subsection{Adaptive Mesh} \label{amm}
In this work we are interested in adaptive meshes that evolve with the flow of a physical system and which are non-conservative. We will make use of the same 1-d adaptive mesh scheme developed in \citet{Aydodu2019npg} as a prototype of 2-d, or 3-d, non-conservative adaptive mesh used in some modern numerical models, including the Lagrangian sea ice model neXtSIM \citet{rampal2016nextsim,rabatel2018impact}. 

The mesh itself is a 1-d mesh defined on the domain $D=[0,L)$ with nodes $\{z_1, z_2, \dots, z_N\}\in D$. It is assumed that $0\leq z_i<z_{i+1}<L$ and that the positions of the nodes satisfy criteria which define a valid mesh through two tolerance parameters $\delta_1, \delta_2$. A valid mesh is one for which, 
\begin{align}
    \delta_1 \leq |z_{i+1}-z_i| \leq \delta_2 \quad \forall i \in \mathbb{N}: 1\leq i < N-1 \label{valid1}\\
    \text{and} \quad \delta_1\leq |z_1+L-z_N| \leq \delta_2 \label{valid2}.
\end{align}
This criteria ensures that the mesh is periodic and that no two nodes are closer than $\delta_1$ or further apart than $\delta_2$. Moreover, $\delta_1$ and $\delta_2$ are chosen so that $\delta_1/\delta_2\geq2$ and are both divisors of $L$ (see \citet{Aydodu2019npg} for an extensive explanation and details on the assumptions).

The mesh points themselves evolve directly with the velocity ${\bf u}$ as,
\begin{align}
    \frac{{\rm d}z_i}{{\rm d}t}={\bf u}(t,z_i).
    \label{zevolove}
\end{align}
Equation~\eqref{zevolove} together with the physical model updating the velocity (along with any other model state variables) represents a coupled system of equations which can be solved alternately or simultaneously \citet{huang2010adaptive}. 

Given that the node locations are a function of time $z_i=z_i(t)$, it is clear that there will be instances when the criteria for a valid mesh given in Eqs.~\eqref{valid1} and \eqref{valid2} are violated. In such cases we need a suitable remeshing scheme to enforce our criteria which is given as follows. For each $i$ if $|z_{i+1}-z_{i}|<\delta_1$, $z_{i+1}$ is deleted. Alternately, if $|z_{i+1}-z_{i}|>\delta_2$ a new point $z^*$ is inserted at the mid point between $z_{i+1}$ and $z_i$ and the points are re-indexed according to their order from left to right. The most relevant consequence of this is that the number of nodes in the mesh is not constant. 

\subsection{Models and Observations} \label{models}
In this work we consider two models for use in our numerical experiments. The first is a diffusive form of Burgers' equation (BGM), \citet{burgers1948}:
\begin{equation}
   {\rm {\bf BGM}}:\quad \frac{{\rm \partial} u}{{\rm \partial} t}+u\frac{{\rm \partial} u}{{\rm \partial} z}=\nu \frac{{\rm \partial}^2u}{{\rm \partial} z^2}, \quad z\in [0,1), \label{BGM} 
\end{equation}   
with viscosity, $\nu=0.08$ and periodic boundary and initial conditions:
\begin{align}
   u(0,t)=u(1,t), \label{BGMBC} \\
   u(z,0)=\sin(2\pi z)+\frac{1}{2} \sin(\pi z). \label{BGMIC}
\end{align}
The Burgers equation has been used in several DA studies \citet{cohn1993dynamics, verlaan2001, pannekoucke2018parametric}. This model is of particular interest because of the steep gradients near the shock, a motivating reason to use an adaptive mesh.

The second model is a version of the Kuramoto-Sivashinsky (KSM) equation \citet{papageorgiou1991route} given by
\begin{equation}
  {\rm {\bf KSM}}: \quad  \frac{{\rm \partial} u}{{\rm \partial} t}+\nu\frac{{\rm \partial} u^4}{{\rm \partial} z^4}+\frac{{\rm \partial} u^2}{{\rm \partial} z^2}+u\frac{{\rm \partial} u}{{\rm \partial} z}=0 \quad z\in[0,2\pi). 
  \label{KSM} 
\end{equation}
The periodic boundary and initial condition are defined as:
\begin{align}
  BC: u(0,t)=u(2\pi,t),\label{KSMBC}\\
  IC: u(z,0)=-\sin(2\pi z).\label{KSMIC}
\end{align}
Here the viscosity $\nu=0.027$ is chosen so that we see chaotic behaviour in the model. Both models are solved using central differences and an Eulerian time stepping scheme with time steps of $10^{-3}$ for BGM and $10^{-5}$ for KSM. The tolerances used in the remeshing scheme outlined in section~\ref{amm} are $\delta_1=0.01$, $\delta=0.02$ for BGM and $\delta_1=0.02\pi$, $\delta_2=0.04 \pi$ for KSM. 

Observations of the physical values are generated from high resolution ``nature'' runs for both models. For the KSM model, there is an initial spin up to $T=20$ before observations are taken and the model state at that time is used to initialise the ensemble members in the DA experiments described in section~\ref{results}. Mean zero, Gaussian distributed, white noise is added to the observations for both models and experiments carried out with differing observation error standard deviation, $\sigma_{\rm o}$. The observations are Eulerian, {\it i.e.} they are taken on a fixed-in-time regularly spaced grid on the interval $[0,L)$ and at regular time intervals. The choice of regular spatial and temporal distributions for the data is done for the sake of simplicity and it can be relaxed without impact on the algorithm setup.


\section{EnKF for an Adaptive Moving Mesh model  - AMMEnKF }
\label{AMMENKFS}

The ensemble Kalman filter (EnKF) relies on estimates of error statistics using an ensemble of model runs assumed to be Gaussian distributed. The error estimates themselves are calculated using the state vector formed from each ensemble member. In the case of an Eulerian solver with a fixed mesh, this calculation is easily carried out as the number of nodes and their locations are the same for each ensemble member and thus, the dimension of the state vector is also the same for each ensemble member. In contrast for an adaptive moving mesh (AMM), the mesh node locations for each ensemble member will almost certainly be in different locations at an assimilation time. Further, due to the re-meshing outlined in Section~\ref{amm}, will have different numbers of nodes as well. This makes estimation of the error statistics less direct and lends to a need for the development of modified versions of the EnKF suited to models with solvers like those we consider here. 

For a non-conservative AMM solver we see two additional steps to be necessary each with their own important considerations. The first key step needed before applying the EnKF we refer to as {\it dimension matching}, this needed to provide consistent estimations of ensemble statistics. One would need to decide to add or remove points from ensemble members to achieve the same number of components among state vectors. In addition a sub-step is that of {\it component paring}, that is, how to assign which points, which may be in different locations, are to be compared in the state vectors. The second key step comes after applying the update and we refer to it as {\it dimension return}. This would involve deciding whether or not to remove points which were added, if they were, or if points were removed, whether or not to add points back into the ensemble members. Both of these steps have the potential to disrupt the ensemble statistics and should be tailored to the model and meshing schemes.   


One avenue toward an AMMEnKF involves the use of a reference mesh to which each ensemble member can be mapped and on which error statistics can be estimated. This has been explored originally in \citet{du2016ensemble} and \citet{Aydodu2019npg} in case of conservative and non-conservative meshes respectively. In \citet{Aydodu2019npg} the use of a reference mesh was explored in 1-d where the reference mesh itself is chosen based on the properties of the mesh adaptation scheme. In particular, two meshes were explored. The first is a high resolution (HR) mesh defined by the node proximity tolerance, $\delta_1$, which ensures {\it at most} one point from each ensemble member can be in any given interval of the partitioned domain. The second is a low resolution mesh (LR) defined by the node separation tolerance, $\delta_2$, which ensures each ensemble member has {\it at least} one point in any given interval of the partitioned domain. In both cases each ensemble member is mapped to the reference mesh before error statistics are calculated and then mapped back to their original meshes after the physical velocity values are updated in the analysis step. The mesh locations were not updated during analysis. 

However, the node positions themselves are driven by the physical flow and as such can be considered time dependent state variables. In this work we consider updating the node locations making use of the HR partitioning of the interval domain for the same models considered in \citet{Aydodu2019npg}. The key difference between the previous and the current works is that we now augment our state vector with the node locations and update them in the analysis step.  We are interested in exploring the use of the augmented state vector to leverage extra statistical information implied by the different meshes among the ensemble members. This is because, in this case, the mesh is connected to the physics and cross covariances between the physical values and the node locations says something about the system. Previous work for a conservative moving mesh was carried out in \citet{bonan2017}, there they also augment their state vector but avoid the issue of dimension matching.

We will, when needed, describe the methods in \citet{Aydodu2019npg} so that the reader may understand the relevant differences. In particular we focus on the HR method and refer to the augmented state vector as the HRA method. In both cases, HR and HRA, the analysis update is preceded and followed by two additional steps: (1) {\it dimension matching}, when the individual ensemble members (each on its own mesh) is projected onto the uniform, fixed-in-time, reference mesh, and, (2) {\it dimension return}, when the ensemble members are given each a mesh after their physical values (for HR) and their physical values and node locations (for HRA) have been updated. The full AMMEnKF procedure is detailed in the following subsection for both HR and HRA.

\subsection{Dimension matching}
\paragraph{{\bf HR Scheme}}\label{dimmatchHR}
In order to avoid the statistical consistency issues presented by having ensemble members with differing numbers of nodes at different locations, one can map each ensemble member to a reference mesh.  The reference mesh can be defined on the physical domain $[0,L)$ into $M$ intervals of equal length $\Delta \gamma$,
\begin{align}
    [0,L)=L_1\bigcup L_2 \bigcup \dots \bigcup L_m
\end{align}
where $L_i=[\gamma_i,\gamma_{i+1})$. In this case $\gamma_1=0$, $\gamma_i=(i-1)\Delta \gamma$ for each i. Further $\gamma_M=L-\Delta\gamma$ as $0$ and $L$ are identified on the periodic domain. The points $\gamma_i$ form the nodes of the reference grid.  

The reference grid is chosen in one of two ways, to ensure that each ensemble member has {\it at most} one point in each interval, $\Delta \gamma=\delta_1$, or that each ensemble member has {\it at least}, $\Delta \gamma=\delta_2$, one point in each interval. The former is referred to as the high resolution mesh (HR) and the latter the low resolution mesh (LR). 

\begin{figure*}
    \centering
        \includegraphics[width=\textwidth]{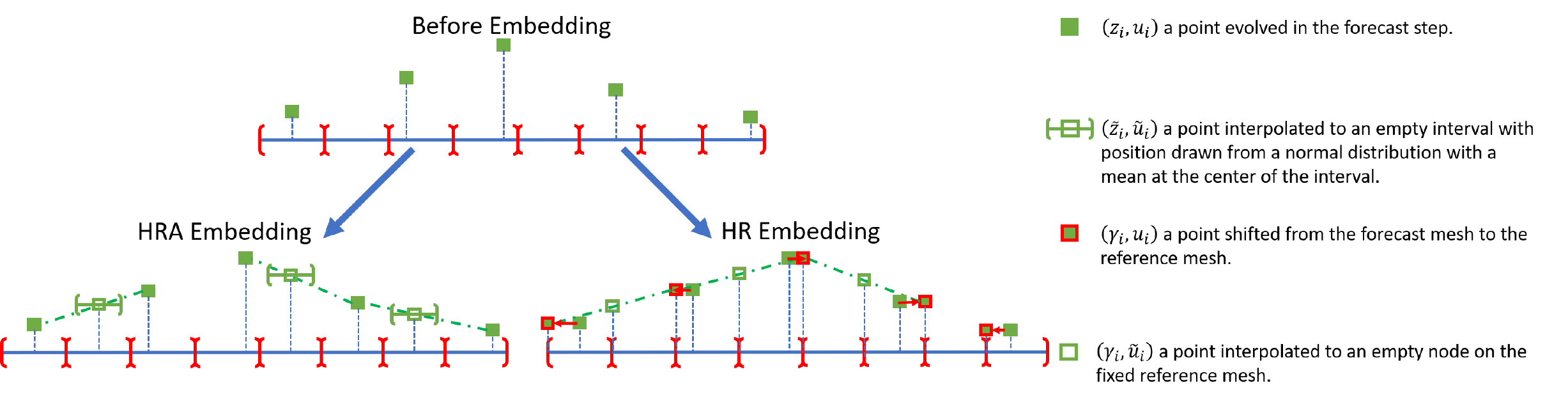}
    \caption{Illustration of the two dimension matching schemes. In the HR scheme, points are shifted to a fixed reference mesh with empty nodes interpolated to. In the HRA scheme, points are added to empty intervals and interpolated to.}\label{embeddings}
\end{figure*}

Here we focus on the HR mesh since we partition our physical domain in the same way. The mapping from an ensemble member to the HR mesh will take the $j^{th}$ ensemble member's state vector ${ \bf x}_j=(u_1 \dots u_N, z_1 \dots z_N)_j \in \mathbb{R}^{2N}$ to the vector,
\begin{align}
{\bf x}_j=\left(\widetilde{{\bf }u},{\bm\gamma}\right)^{\rm T}_j=(\widetilde{u}_1 \dots \widetilde{u}_M, \gamma_1 \dots \gamma_M)^{\rm T}_j \in \mathbb{R}^{2M} \quad \text{with} \quad M\geq N.
    \label{svHR}
\end{align}
Here $\widetilde{u}_i$ will be the physical value assigned to $\gamma_i$ through the introduction of a shifted mesh where $L_i\rightarrow \widetilde{L}_i=[\gamma_i-\delta_1/2,\gamma_i+\delta_1/2 )$ for $i=2,\dots M$. The first interval is taken to be $\widetilde{L}_1=[L-\delta_1/2,L]\bigcup[0,\delta_1 /2 )$ since we identify 0 and $L$. If there is a $z_k\in \widetilde{L_i}$, then set $\widetilde{u}_i=u_k$. If there is no such $z_k$ but $z_k\leq\gamma_i$ find $k$ such that $z_k\leq\gamma_i\leq z_{k+1}$ and set 
\begin{align}
    \widetilde{u}_i=\frac{u_k+u_{k+1}}{2}
    \label{setutildehr}
\end{align}
if there is no such $z_k$, then set 
\begin{align}
    \widetilde{u}_i=\frac{u_1+u_{N}}{2}.
    \label{setutildehrend}
\end{align}
This mapping is illustrated in the right branch of Fig.~\ref{embeddings}. Once each ensemble member has been mapped to the fixed reference grid, the standard EnKF can be applied. 


\paragraph{{\bf HRA Scheme}}
\label{dimmatchHRA}
In the HRA setting the reference mesh is also used to choose which nodes will be compared, but with out changing their locations. We partition the domain $D=\left[0,L\right)$ into $M$ subintervals ($L_i$) each of length $\delta_1$ so that $D=\bigcup_i L_i$. Since $\delta_1$ is the node proximity tolerance we are guaranteed that each subinterval will have {\it at most} one point in it. With this we can component match nodes which fall in the same subintervals. If an ensemble member does not have a point in a given subinterval we will insert one, a {\it ghost point}, based on the nearest neighbors.

In this approach we take the state vector of the $j^{th}$ ensemble member on the reference mesh to be of the form 
\begin{align}
{\bf x}_j=\left({\bf u},{\bf z}\right)^{\rm T}_j=\left(
u_1,u_2, \dots ,\widetilde{u_i},\dots,  u_M, z_1, z_2 ,\dots,\widetilde{z_i},\dots,z_M\right)^{\rm T}_j\in \mathbb{R}^{2M} \quad \text{with} \quad M\geq N.
\label{svHRA}
\end{align}
where $u_i$ or $\widetilde{u}_i$ would be the value of the velocity in the $i^{th}$ sub interval of the reference mesh. A value with no tilde would mean the ensemble member had a point in that interval while a tilde implies the member did not have a point in the $i^{th}$ interval and one was inserted and a physical value interpolated to that location. The location of an interpolated point is drawn from the Gaussian distribution, $\mathcal{N}\left(\frac{\gamma_i+\gamma_{i+1}}{2},\delta_1/2\right)$, with a check that the point drawn actually resides in the interval $L_i$, if not, we draw again until it does. This is illustrated in the left branch of Fig.~\ref{embeddings}. The choice of randomly sampling the node location is done to avoid biasing node locations in intervals that are empty amongst a large proportion of the ensemble members which can happen in areas of lower velocities and larger node spacing. However, it is possible that we end up having in invalid mesh in this process. Nevertheless, we do not enforce validity at this step as there will be many cases where no location in an empty interval can be chosen for which there is not a point with in $\delta_1$ near it. This is because the intervals themselves are of size $\delta_1$.

The physical value assigned to a ghost point $\widetilde{z}_i$ is calculated by linear interpolation as: 
\begin{align} 
\widetilde{u}_i=\frac{b}{a+b}u_l+\frac{a}{a+b}u_r && a=\widetilde{z}_i-z_l, b=z_r-\widetilde{z}_i
\label{physicalinterp}
\end{align}
where $(z_l,u_l)$ and $(z_r,u_r)$ are the closest nodes to $\widetilde{z}_i$ to the left and right respectively and $u_l,u_r$ the corresponding physical values at those nodes. This is done from left to right which does allow for the possibility that a nearest left neighbour may have been a ghost point. However, in the case that $\delta_2=2\delta_1$ we are guaranteed each empty interval will have a non-empty interval to its left and right. 



\subsection{Observation Operator}
\paragraph{{\bf HR scheme}}

For the HR method the observation operator applied to the $j^{th}$ ensemble member takes the form 
\begin{align}
h\left({{\bf x}_j}\right)=
\widetilde{u}_i+\frac{z_k^o-\gamma_i}{\gamma_{i+1}-\gamma_i}(\widetilde{u}_{i+1}-\widetilde{u}_i).
\label{obsopref}
\end{align}
Where $z_k^o$ is the observation location with $\gamma_i \leq z_k^o \leq \gamma_{i+1}$. 

\paragraph{{\bf HRA scheme}}

In a similar way we may define the observation operator for the HRA method as,
\begin{align}
h({\bf x}_j)=u_i+\frac{z_k^0-z_i}{z_{i+1}-z_i}(u_{i+1}-u_i).
\label{obsop}
\end{align}
Where either $z_i$, $z_{i+1}$, $u_i$ or $u_{i+1}$ could have a tilde if they were inserted due to the ensemble member having no value in the $i^{th}$ interval (see section~\ref{dimmatchHRA}).

This form of the observation operator means that we are not considering the location of the observation in the update, just the physical value. This is done since most geophysical measurements will not directly relate to a node position, since the nodes are not physical objects. Yet the physics does fundamentally drive node motion and the covariances between physical values and the node locations are non-zero in the error covariance matrix, described below and seen in Fig.~\ref{covuz}.


\subsection{Analysis using the EnKF}
Once the dimensions of the state vectors of each ensemble member have been matched the EnKF can be applied in the usual way. As in \citet{Aydodu2019npg} we will work using the stochastic version of the EnKF \citet{evensen2009}, and here as well the choice is not influential on the modification we propose for AMM models. Our method will apply equally if using deterministic EnKFs.   

Let us define the forecast ensemble matrix ${\bf E}^{\rm f}$ as
\begin{align}
    {\bf E}^{\rm f} = \left[ {\bf x}_1^{\rm f}, \dots , {\bf x}^{\rm f}_{ N^{\rm e}} \right] \in \mathbb{R}^{2M \times N^{\rm e}} . \label{errorcov}
\end{align}
Where the forecast state vectors ${\bf x}_j^{\rm f}$ takes the form as in Eq.~\eqref{svHR} for the fixed reference mesh case and Eq. ~\eqref{svHRA} for the augmented case where we also update node locations.  
In Eq.~\eqref{errorcov}, $M$ is the number of subintervals, $L_i$, which partition the domain $D$ into subintervals of size $\delta_1$ and $N^e$ is the number of ensemble members. The vectors ${\bf x}_j^{\rm f}$ are the dimension matched state vectors taken to be the columns of ${\bf E}^{\rm f}$. 

The forecast anomaly matrix ${\bf X}^{\rm f}$ takes the form
\begin{align}
 {\bf X}^{\rm f} = \frac{1}{\sqrt{N^{\rm e}-1}}\left[ {\bf x}_1^f-\bar{{\bf x}}^{\rm f}, \dots , {\bf x}^{\rm f}_{N^{\rm e}}-\bar{{\bf x}}^{\rm f} \right] ,
\end{align}
where $\bar{{\bf x}}^{\rm f}$ is the forecast ensemble mean defined as,
\begin{align}
    \bar{{\bf x}}^{\rm f}=\frac{1}{N^{\rm e}}\sum_{j=1}^{N^{\rm e}}  {\bf x}_j^{\rm f}.
\end{align}

In the stochastic EnKF the observations are treated as random variables so that each ensemble member is compared to a slightly different perturbation of the observation vector \citet{burgers1998analysis} . That is, given an observation vector ${\bf y}$ we generate $N^{\rm e}$ observations according to,
\begin{align}
    {\bf y}_j={ \bf y}+\epsilon_j \quad 1\leq j\leq N^{\rm e} \quad \epsilon_j \sim \mathcal{N}({\bf 0},{\bf R}),
\end{align}
where ${\bf R}$ is the covariance of the assumed zero mean, white-in-time observation noise $\epsilon$. We can then calculate the normalized anomaly ensemble of observations,
\begin{align}
    {\bf Y}_{\rm o} &= \frac{1}{\sqrt{N^{\rm e}-1}} \left[ {\bf y}_1-{\bf y}, \dots , {\bf y}_{N^{\rm e}}-{\bf y}\right] \\
     &=\frac{1}{\sqrt{N^{\rm e}-1}}\left[ \epsilon_1, \dots \epsilon_2\right],
\end{align}
which in turn defines the ensemble observation error covariance matrix,
\begin{align}
    {\bf R}^e={\bf Y}_{\rm o}\left({\bf Y}_{\rm o}\right)^{\rm T}.
\end{align}
We then define the observed ensemble-anomaly matrix using our observation operator $h$ as,
\begin{align}
    {\bf Y}=h({\bf E}^{\rm f})-h(\bar{{\bf E}}^{\rm f}),
\end{align}
where the operator $h$ is applied at each column of the matrix ${\bf E}^{\rm f}$.
This leads the Kalman Gain matrix, ${\bf K}$ to be,
\begin{align}
    {\bf K}={\bf X}^{\rm f}{\bf Y}^{\rm T}\left[\frac{1}{N^{\rm e}-1}{\bf Y}{\bf Y}^{\rm T}+{\bf R}^{\rm e}\right]^{-1}
    \label{kgain}
\end{align}
which is used, in the stochastic EnKF formulation, to individually update each ensemble member according to,
\begin{align}
    {\bf x}^{\rm a}_i={\bf x}^{\rm f}_i+{\bf K}\left[{\bf y}_i-h({\bf x}^{\rm f}_i)\right] \quad 1\leq i \leq N^e.
\end{align}
With the HRA method, however, there is the possibility that an ensemble member will have an invalid mesh after the update step. For this reason the re-meshing algorithm is applied to each ensemble member after updating. The remeshing is also tasked with handling points that have moved out of the domain; although not common, it can happen.

In this work we also make use of covariance multiplicative scalar inflation \citet{anderson1999monte} in which the ensemble forecast anomaly matrix is inflated as,
\begin{align}
    {\bf X}^{\rm f}\rightarrow \alpha {\bf X}^{\rm f},
\end{align}
with $\alpha \geq1$, before ${\bf X}^{\rm f}$ is used in the analysis update. This parameter is one that can be tuned through numerical experimentation, although approaches exist to make this task automatic and adaptive along the experiments (see {\it e.g.} \citet{raanes2019} and references therein). 
After updating each ensemble member the mean of each analysis can be used to obtain a best estimate of the physical state of the system.

\subsection{Dimension Return}\label{dimreturn}
After the update is complete each ensemble analysis vector has its dimension returned to its pre-analysis value. For the AMMEnKF-HR scheme this is a needed step as the structure of the adaptive mesh is removed during the update step and some kind of map back to the previous mesh state before the next forecast is necessary. In the AMMEnKF-HRA scheme the mesh itself is updated and and the remeshing scheme is applied to ensure a valid mesh.


\paragraph{{\bf HR scheme}}

\begin{figure*}
\centering
\includegraphics[width=\textwidth]{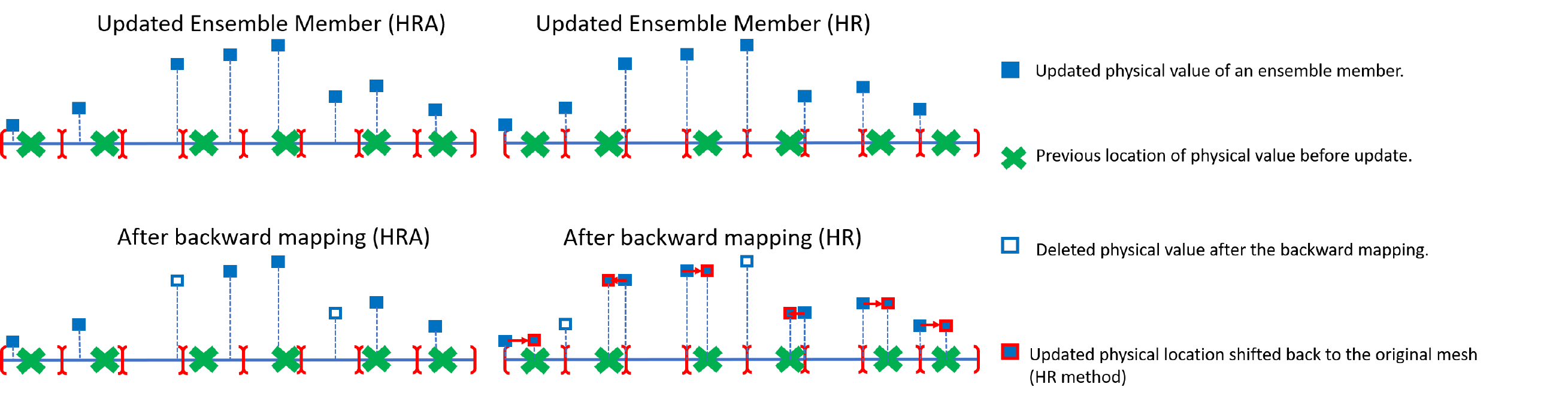}

\caption{The dimension return steps. In the HRA scheme points that occupy previously empty intervals are simply deleted. In the HR scheme points which came in from the forecast step are shifted back to their original mesh locations and points inserted at the dimension matching step are deleted.}
\label{backmap}
\end{figure*}

Following \citet{Aydodu2019npg}, in the HR case a backward map is used to return the updated ensemble members to their original meshes before forecasting again. In the forward mapping step, the mapping indices associating the nodes in the adaptive moving mesh with nodes in the reference mesh are stored in an array. These are the indices resulting from the projections on to the HR reference mesh. This allows us to map the updated physical values $\widetilde{{ \bf u}}^{\rm a}$ back to the mesh that the ensemble member came into the update step with, that is, the values updated at $\gamma_i$ are shifted back to their previous node locations. From there, the forecast is run until the next assimilation time step. It is notable that this can have the effect of introducing some amount of noise in each ensemble member as physical values determined at one location are moved to another. This is illustrated in the right branch of Fig.~\ref{backmap}. 

\paragraph{{\bf HRA scheme}}

In the HRA case, after the update and remeshing, nodes that are in intervals which were previously unoccupied by a point before the update step are deleted for each ensemble member using the stored indices as in the HR case. This last deletion is not specifically necessary to the scheme and performance with and without this step is essentially equivalent. However, we include this step in our analysis as there may be some applications where keeping the dimension of the ensemble members low is desirable during the forecast step. This process is diagrammed in Fig.~\ref{backmap}. 

A beneficial by-product of the mapping to and from the reference mesh in the HR scheme is that it induces additional variability among the physical values. This occurs when a value at one location is moved to another in the shift to and from the reference mesh.
The net effect is that the ensemble spread stays reasonably large, leading to the healthy functioning of the EnKF. 

\begin{figure*}
    \centering
    \begin{subfigure}[b]{0.49\textwidth}
        \includegraphics[width=\textwidth]{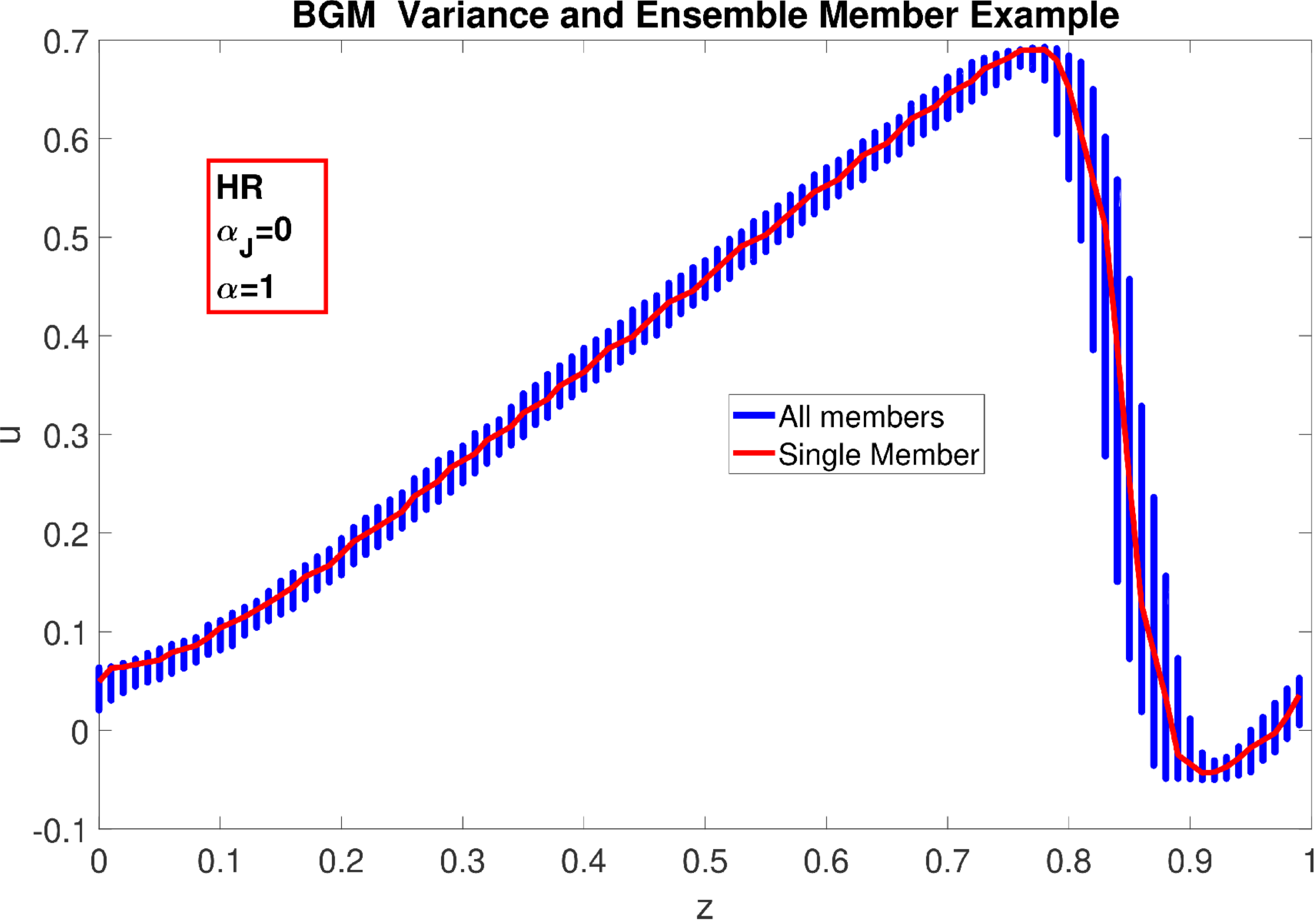}
         \label{INHERENT_0HR}
      \end{subfigure}
   \begin{subfigure}[b]{0.49\textwidth}
       \includegraphics[width=\textwidth]{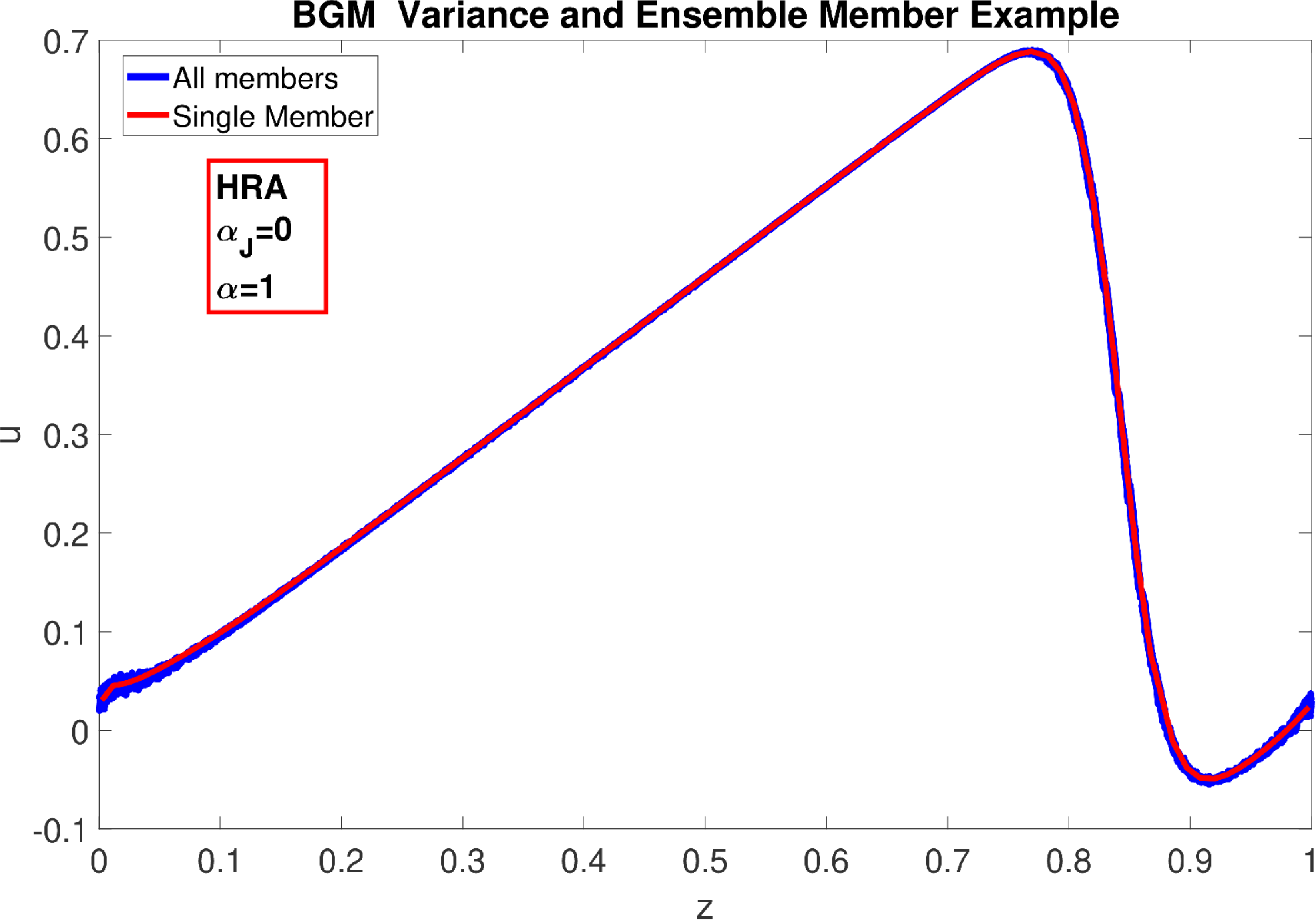}
        \label{INHERENT_0_11HR}
    \end{subfigure}
   
    \caption{Examples of the spread in the forecast ensembles and an example of an forecast ensemble member for the HR (left) and HRA (right) schemes. The forward and backward mapping of the HR scheme induces some inherent jitter increasing the spread. For the HRA case this does not happen and the ensemble members can collapse quickly. Also notable is the reduced smoothness in the ensemble member shown for the HR case caused by the mapping procedure. For these experiments $\alpha_{\rm j}=0,\alpha=1,\sigma_{\rm o}=0.01,N^{\rm e}=30$ and $I_{\rm m}=70$.}   \label{INHERENT}
\end{figure*}

This is not the case in the HRA method given that physical values and their locations are updated together. As a result, the spread of the ensemble when using the HRA scheme tends to be smaller than the HR case and in fact spread collapses quickly with the augmented HRA scheme. This behaviour is shown in Fig.~\ref{INHERENT}. 

While little spread could also be reflecting the desired analysis convergence to the truth, in practice it is a dangerous situation as it often induces the filter to underestimate the actual error, leading to filter divergence.  
We counteract this effect by adding white noise to the physical values, but leave the node locations unaltered. We shall refer to this process as jitter and it can be applied to each ensemble member after the update step. For a given ensemble member analysis vector ${\bf x}_j^{\rm a}$ the jitter is applied to its first $M$ components ({\it i.e.} to the physical values), according to,
\begin{align}
    {\bf u}_j^{\rm a}={\bf u}_j^{\rm a}+\left( \mathcal{N}(0,\bm{\sigma}_J), {\bf 0} \right)^{\rm T} \quad \text{with} \quad  \mathcal{N}(0,\bm{\sigma}_J), \ {\bf 0} \in \mathbb{R}^{M}\quad \text{and} \quad {\bm \sigma}_J=\alpha_J \max_{u_i,u_k \in {\bf u^{{\rm a}}}}|u_i-u_k|.
    \label{jitter}
\end{align}
The scalar parameter $\alpha_J$ regulates the amount of jitter. We take $0\leq\alpha_J\leq 1$ so that we add a percentage of the maximum difference between the physical values of the ensemble members. By having $\alpha_J$ dependent on the analysis field, the jitter is adaptive and is similar to an adaptive form of additive inflation \citet{anderson1999monte}. 

For comparison consistency, we also experimented by applying jitter in the HR method and found improvements in time averaged RMSE values for both schemes.

\begin{figure*}
    \centering
    \begin{subfigure}[b]{0.49\textwidth}
        \includegraphics[width=\textwidth]{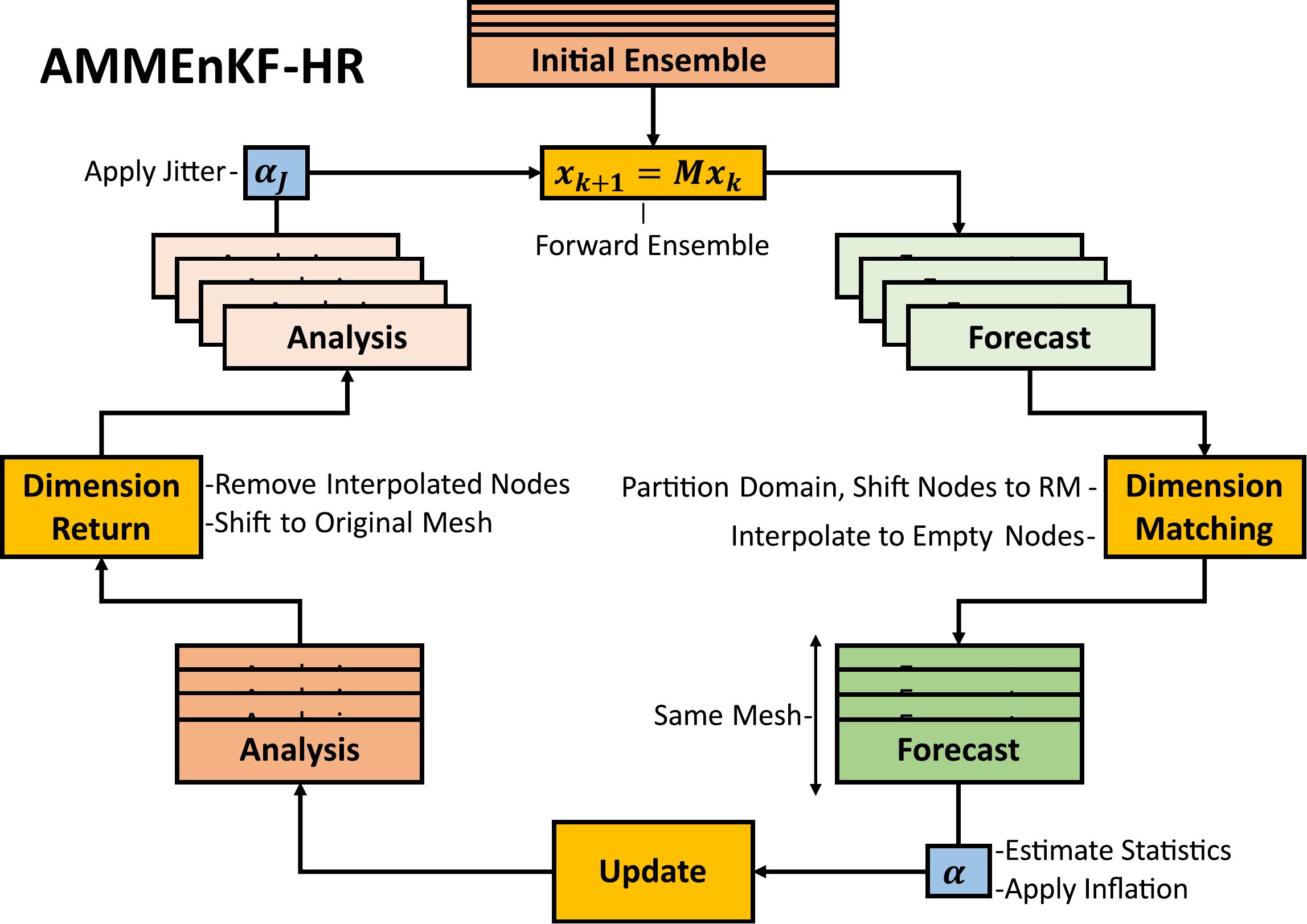}
         \label{HRFLOW}
      \end{subfigure}
   \begin{subfigure}[b]{0.49\textwidth}
       \includegraphics[width=\textwidth]{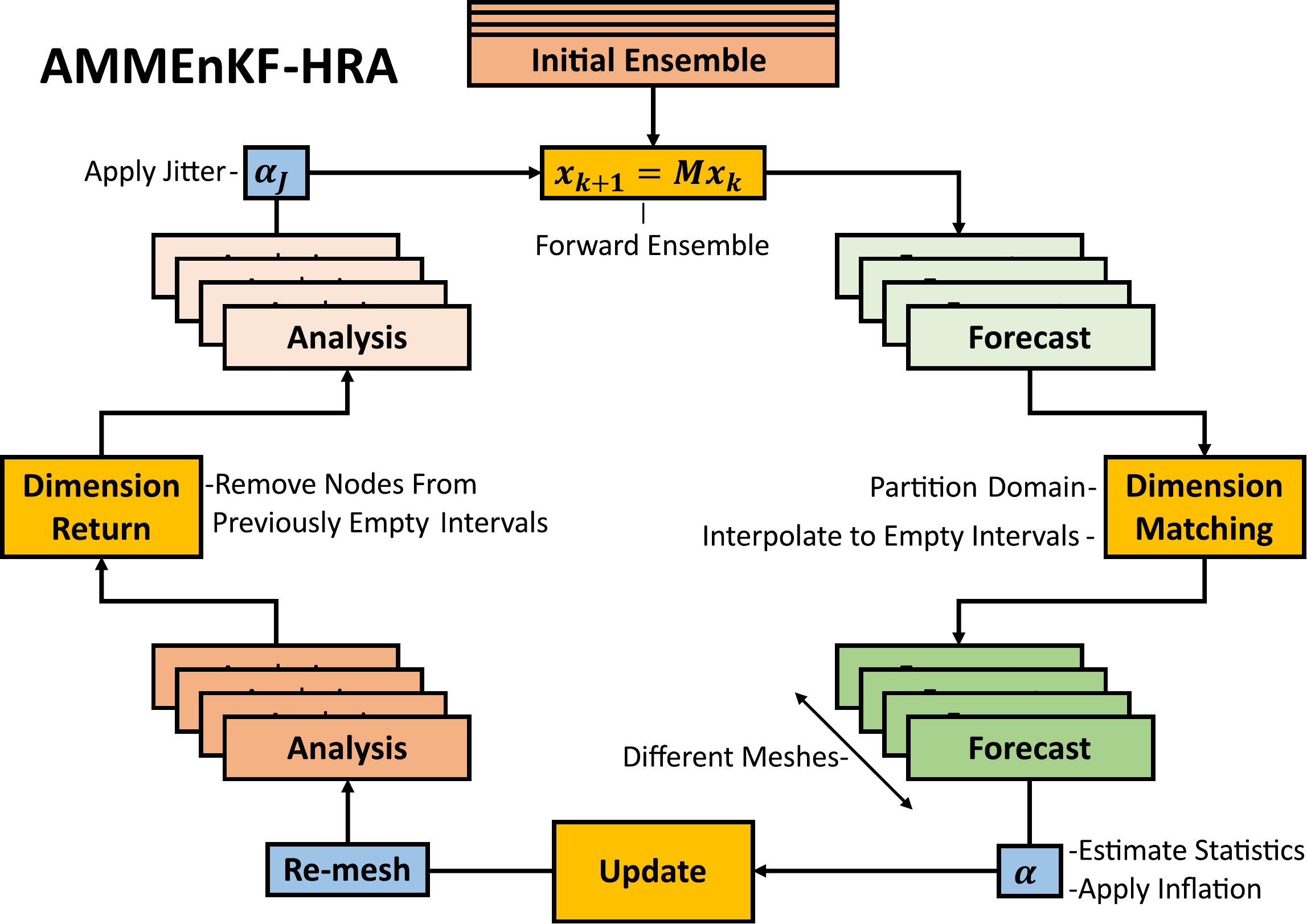}
        \label{HRAFLOW}
    \end{subfigure}
   
    \caption{Illustration of the AMMEnKF-HR (Left) and AMMEnKF-HRA (Right) schemes. The details of the dimension matching, update and dimension return as well as the addition of $\alpha$ and $\alpha_J$ can be found in section~\ref{AMMENKFS}}.   \label{Flows}
\end{figure*}

The HR and HRA algorithms are diagrammed in Fig.~\ref{Flows}.



\section{Results and Discussion}\label{results}

\begin{table}[h]
\centering
\begin{tabular}{l| l| l| l| l| l| l| l}
\hline
  &  $\nu$ & $\delta_1$, $\delta_2$ & $N$ & $T$ & $\Delta t$ & $N_{obs}$ \\
  \hline
  BGM & 0.008 & 0.01, 0.02 & 100 & 2 & 0.05 & 10 \\
  KSM & 0.027 & $0.02 \pi$, $0.04 \pi$ & 100 & 5 & 0.05 & 20 \\
  \end{tabular}
  \caption{The model parameter settings used in each of the DA experiments for the models described in section~\ref{models}. Here $\nu$ is the viscosity, $\delta_1$ and $\delta_2$ the node proximity and distance tolerances, $N$ the size of the reference mesh, $T$ the duration of the experiment, $\Delta t$ the integration time step, and $N_{obs}$ number of observations.}
  \label{expsetup}
\end{table}

In this section we present the results of numerical experiments designed to measure the performance of the two schemes with different parameter settings in terms of a time averaged RMSE. We make use of the BGM and KSM models described in section~\ref{models} for our experiments. For the BGM model, we run for a short time from $t=0$ to $t=2$ because of the rapid dissipation in fluid velocity with our chosen viscosity parameter. The time averaged RMSE in the DA experiments is taken after $t=1$. The ensemble members are initialised by perturbing the initial condition of the nature run. For the KSM model an initial spin up until $T=20$ is done which is used as the initial condition for the DA experiments that follow. With the initial condition provided by spin up, the model is then run for 5 more time steps and observations for the nature run are taken in those last 5 time steps. The ensemble members are initialised with a perturbed version of the initial condition provided after the 20 time step spin up and the time averaged RMSE is taken in the last 4 time steps of the 5 time step integration.  The dimension of the reference mesh for both schemes is $N=100$ however the dimension of the state-vector in the AMMEnKF-HRA scheme is twice that of the AMMEnKF-HR scheme since it has been augmented with the node locations.  The parameters used for the BGM and KSM models in these experiments are summarised in Table~\ref{expsetup}. It is also important to note that the observations are taken at fixed time and on fixed even intervals equally dividing the spatial domain for both model cases. In addition, random white noise with standard deviation $\sigma_{\rm o}$ is added to each observation and we vary this values in our experiments described below.

\subsection{Comparison between AMMEnKF-HR and AMMEnKF-HRA} 

We compare the performance of the AMMEnKF-HR introduced by \citet{Aydodu2019npg} (and recalled in section~\ref{AMMENKFS}), with that of the novel augmented formulation AMMEnKF-HRA also presented in section~\ref{AMMENKFS}. 

We will use two metrics to evaluate the performance. Together with the more standard RMSE of the analysis mean, we also consider the time average RMSE for the first spatial derivative of the analysis mean, ${\rm \partial}_x \widetilde{{ \bf u}}^{\rm a}$. The gradient of the analysis field allows us to assess how well each of the methodologies preserve derivative information. This is relevant for two reasons, the first is to evaluate if and how much applying jitter to the analysed ensemble members distorts their curve smoothness. The second is that the mapping scheme in the HR case can create artificially sharp changes in function values. This will happen when mapping to the reference mesh and when the analysis vector is mapped back to the original ensemble member mesh if the original node location is sufficiently far away from a reference mesh location. These sharp changes over the domain, due to the jitter, HR mapping, or both, can disrupt local rates of change with the risk of violating conservation rules, such as incompressibility ($\nabla \cdot { \bf u}=0$), for example. 
While we make no direct study of conservation laws in this work, we do evaluate the fidelity of the first derivative after the update step for each of these methods as a proxy for the potential violation of conservation laws in more realistic scenarios. In these results the time averaged RMSE's for the derivatives are obtained using the inflation parameters ($\alpha,\alpha_J$) that optimise the time averaged RMSE of the solution analysis mean. Depending on the situation, one may run similar experiments and choose a jitter and inflation that best preserve the first derivative if high fidelity of it is needed. In this work for these models though, there is not much difference in time averaged RMSE when using parameters that optimise the RMSE for the first derivative instead of solution itself. 

\begin{figure*}
    \centering
    \begin{subfigure}[b]{0.33\textwidth}
        \includegraphics[width=\textwidth]{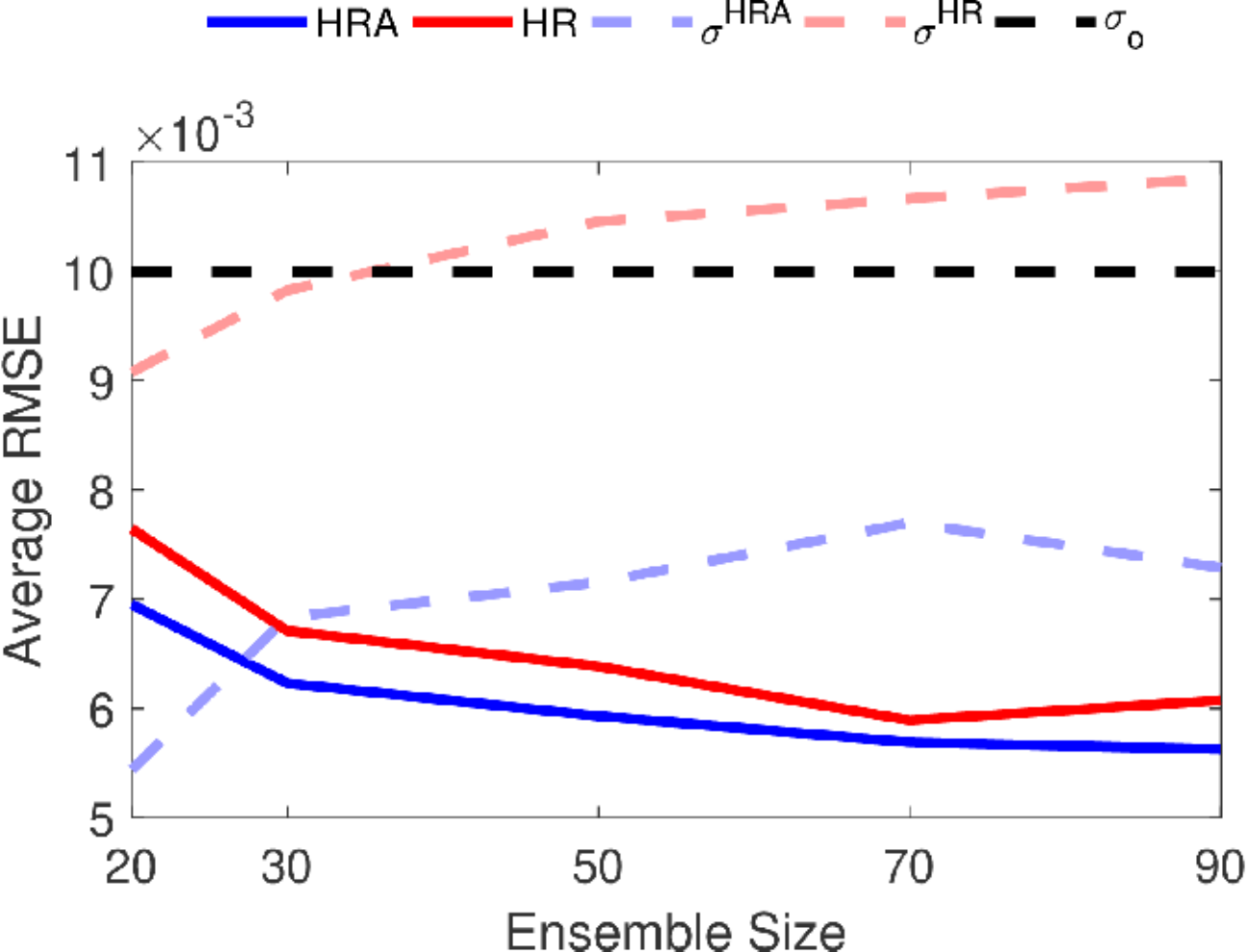}
        \label{ENSoptimalfunctionBGM}
      \end{subfigure}
   \begin{subfigure}[b]{0.33\textwidth}
       \includegraphics[width=\textwidth]{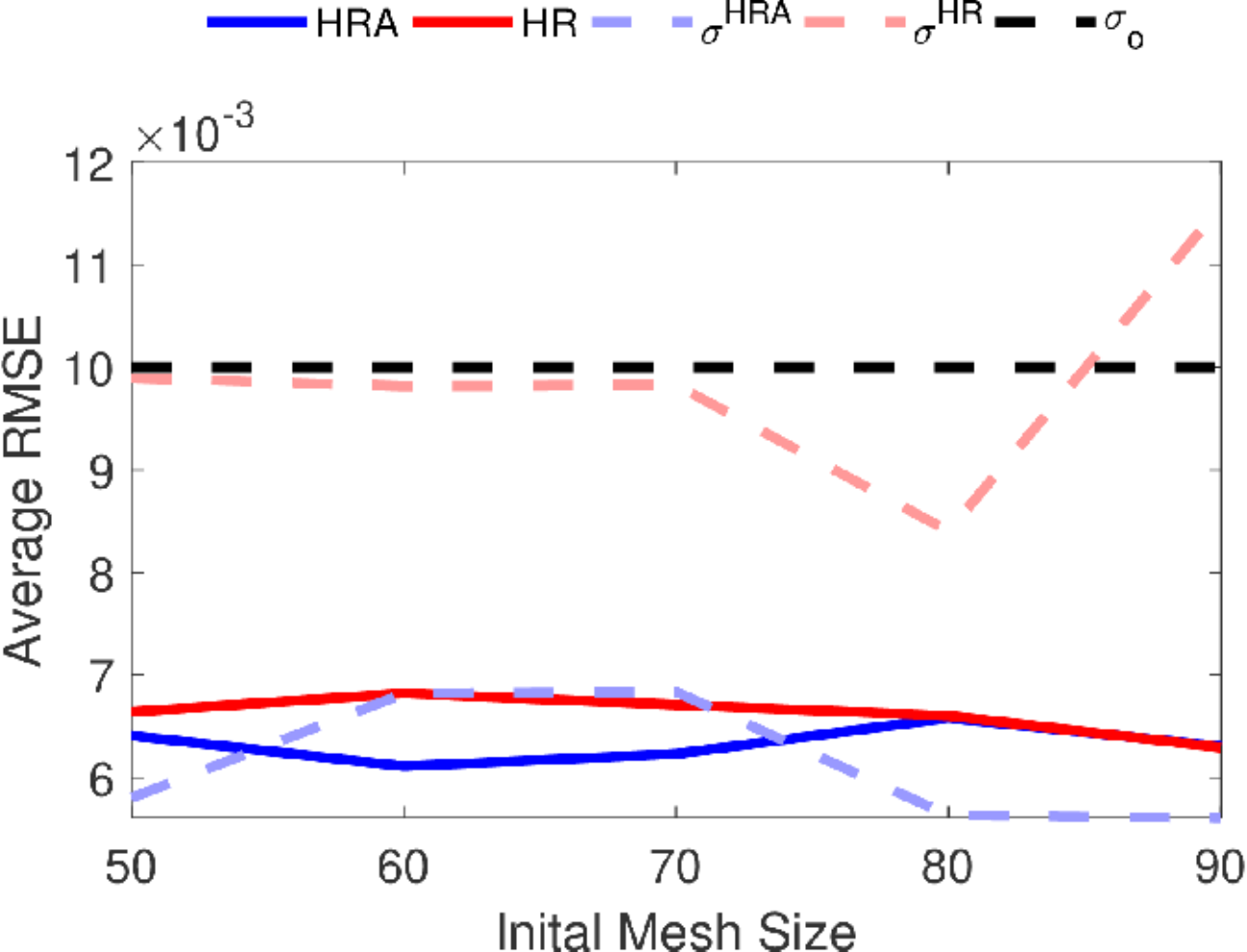}
        \label{INIoptimalfunctionBGM}
    \end{subfigure}
     \begin{subfigure}[b]{0.33\textwidth}
       \includegraphics[width=\textwidth]{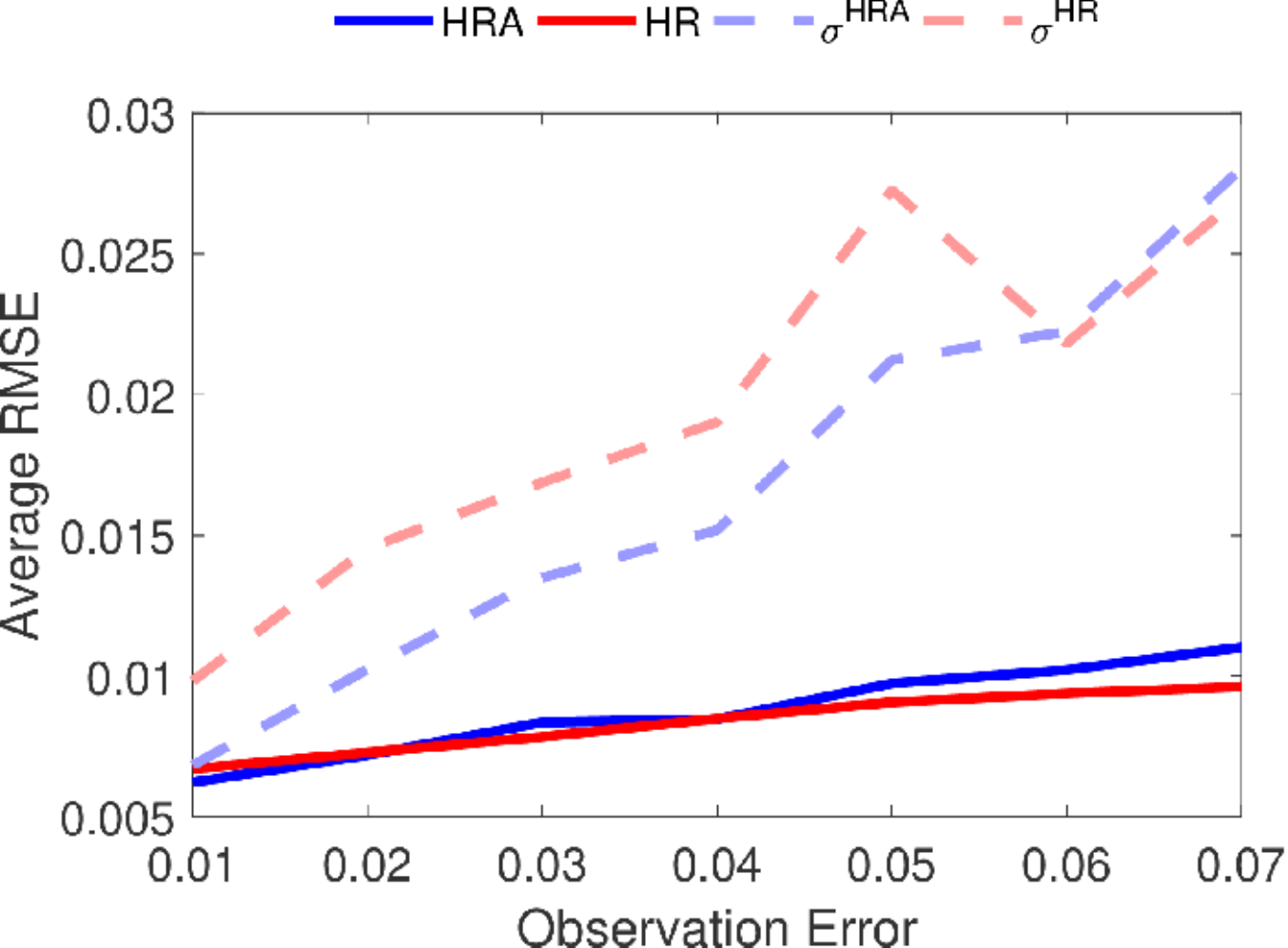}
        \label{OBSoptimalfunctionBGM}
    \end{subfigure}
    
       \begin{subfigure}[b]{0.33\textwidth}
        \includegraphics[width=\textwidth]{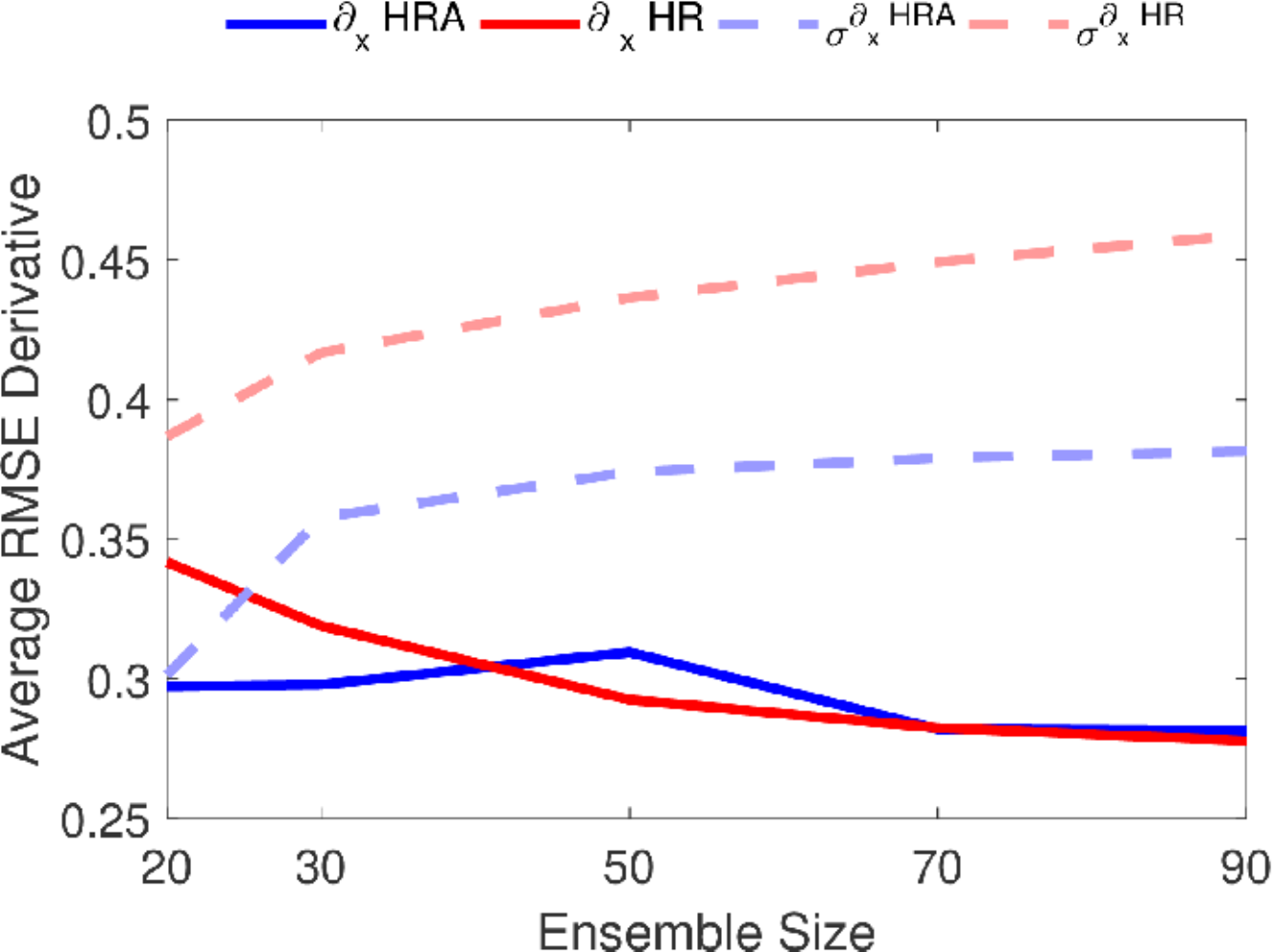}
           \label{ENSoptimalderivativeBGM}
        \end{subfigure}
   \begin{subfigure}[b]{0.33\textwidth}
       \includegraphics[width=\textwidth]{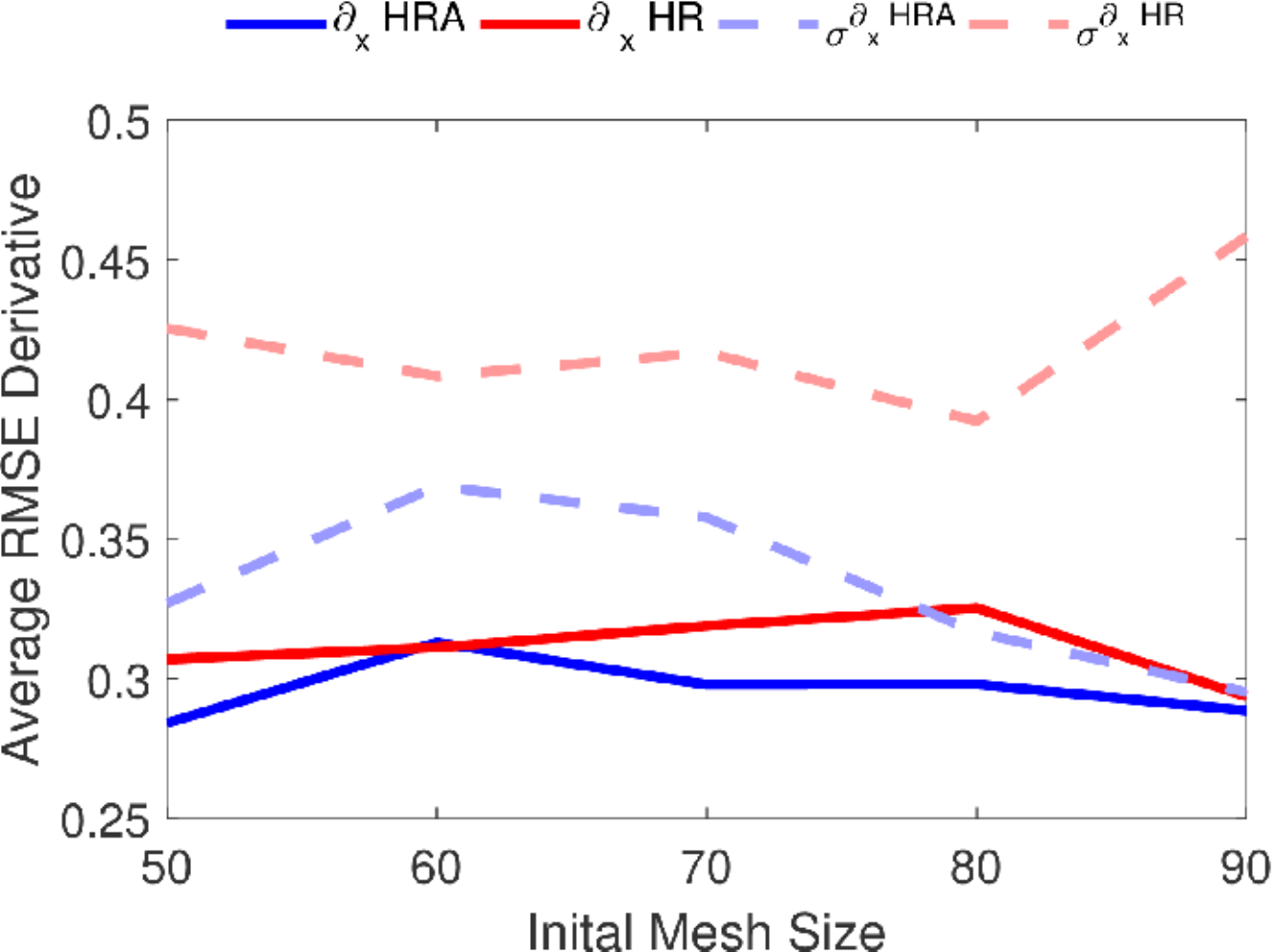}
        \label{INIoptimalderivativeBGM}
    \end{subfigure}
     \begin{subfigure}[b]{0.33\textwidth}
       \includegraphics[width=\textwidth]{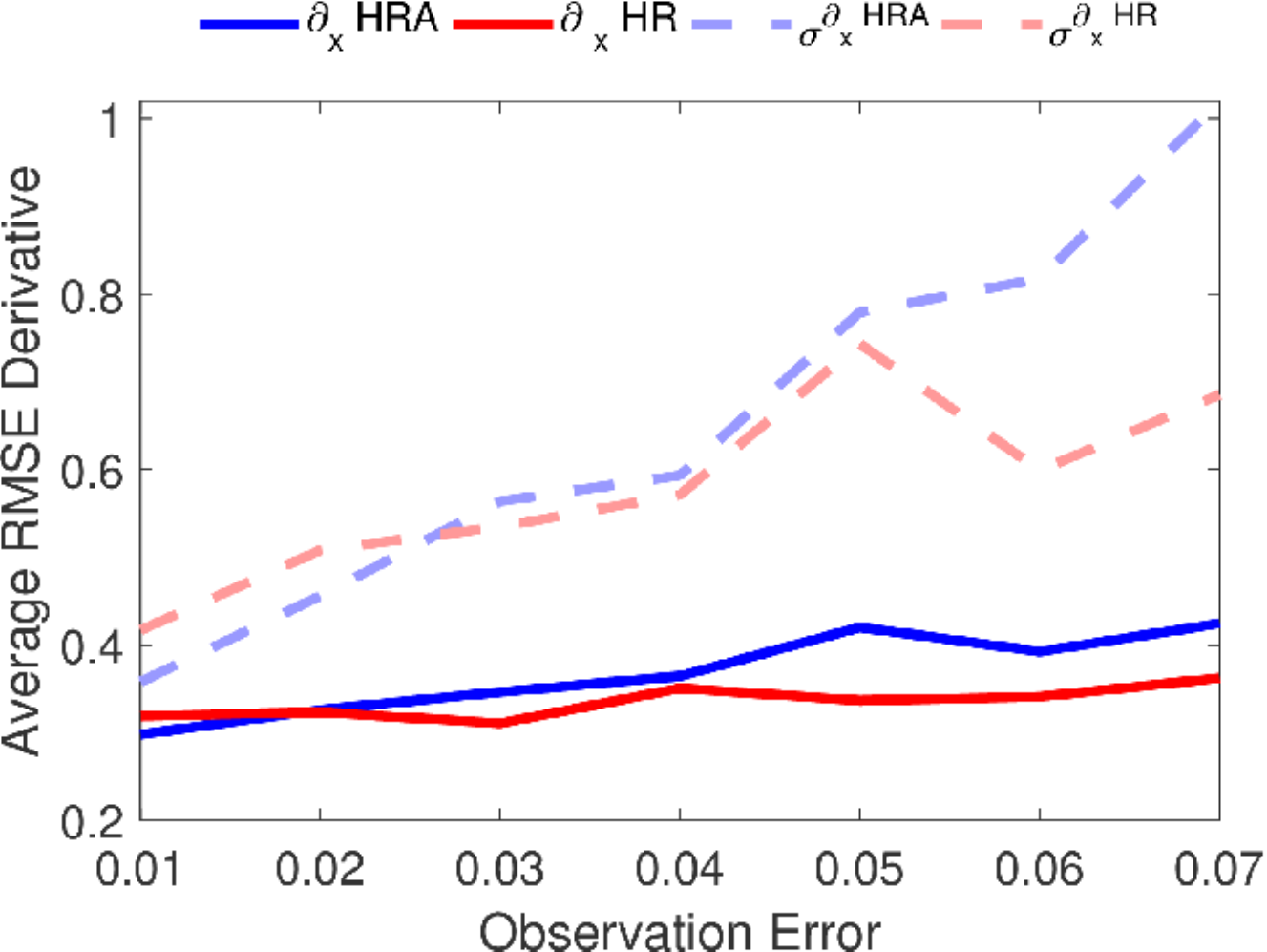}
        \label{OBSoptimalderivativeBGM}
    \end{subfigure}
 \caption{Results of the tuning experiments for the HR and HRA schemes with the BGM model. For each parameter shown along the x-axis, the optimal jitter and inflation are used to obtain a time averaged RMSE for the analysis (top panels). The time averaged RMSE values for the spatial derivatives (bottom panels) correspond to the parameter which optimise analysis RMSE (see text for details).}\label{optimalBGM}
\end{figure*}

\begin{figure*}
    \centering
    \begin{subfigure}[b]{0.33\textwidth}
        \includegraphics[width=\textwidth]{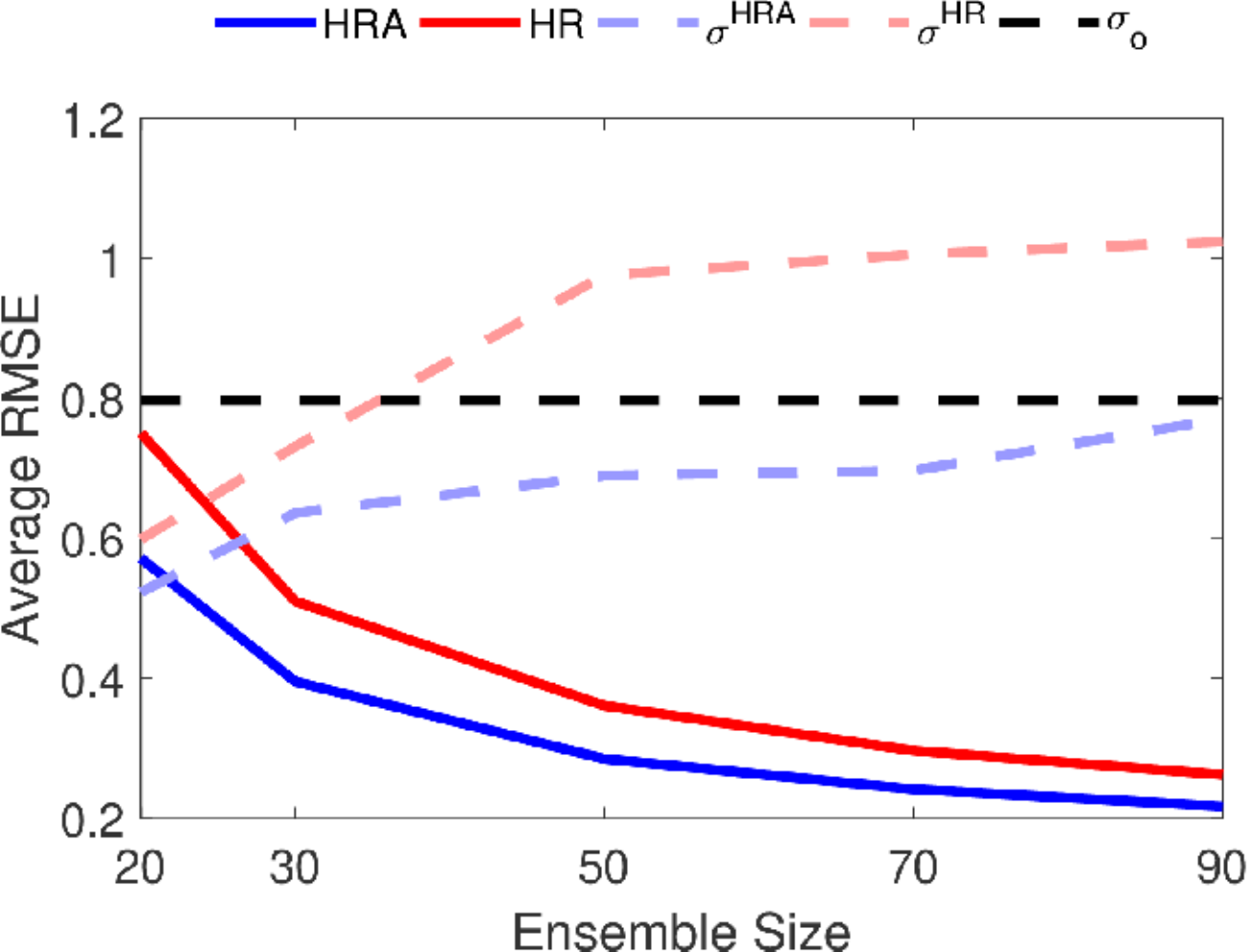}
         \label{ENSoptimalfunctionKSM}
      \end{subfigure}
   \begin{subfigure}[b]{0.33\textwidth}
       \includegraphics[width=\textwidth]{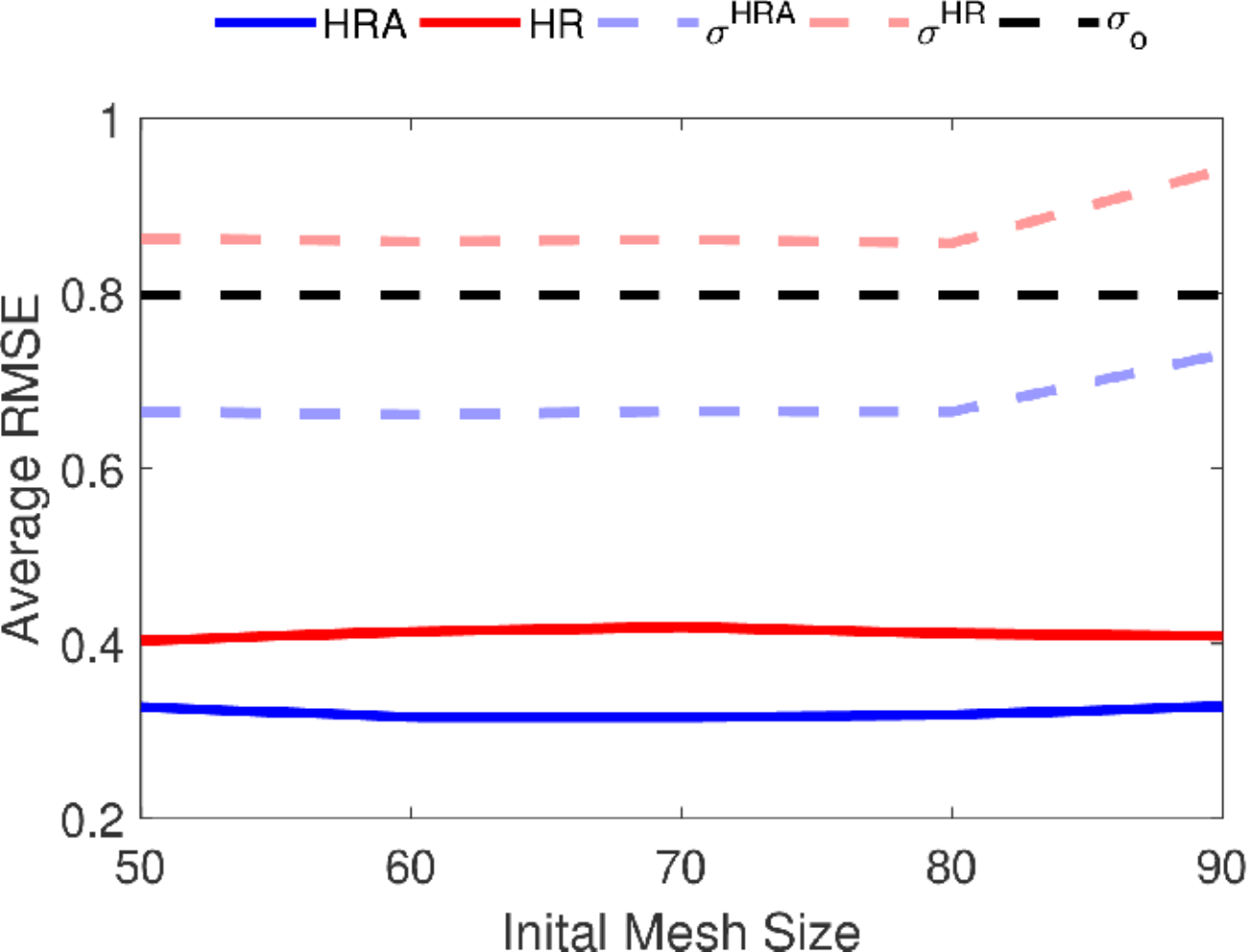}
        \label{INIoptimalfunctionKSM}
    \end{subfigure}
     \begin{subfigure}[b]{0.33\textwidth}
       \includegraphics[width=\textwidth]{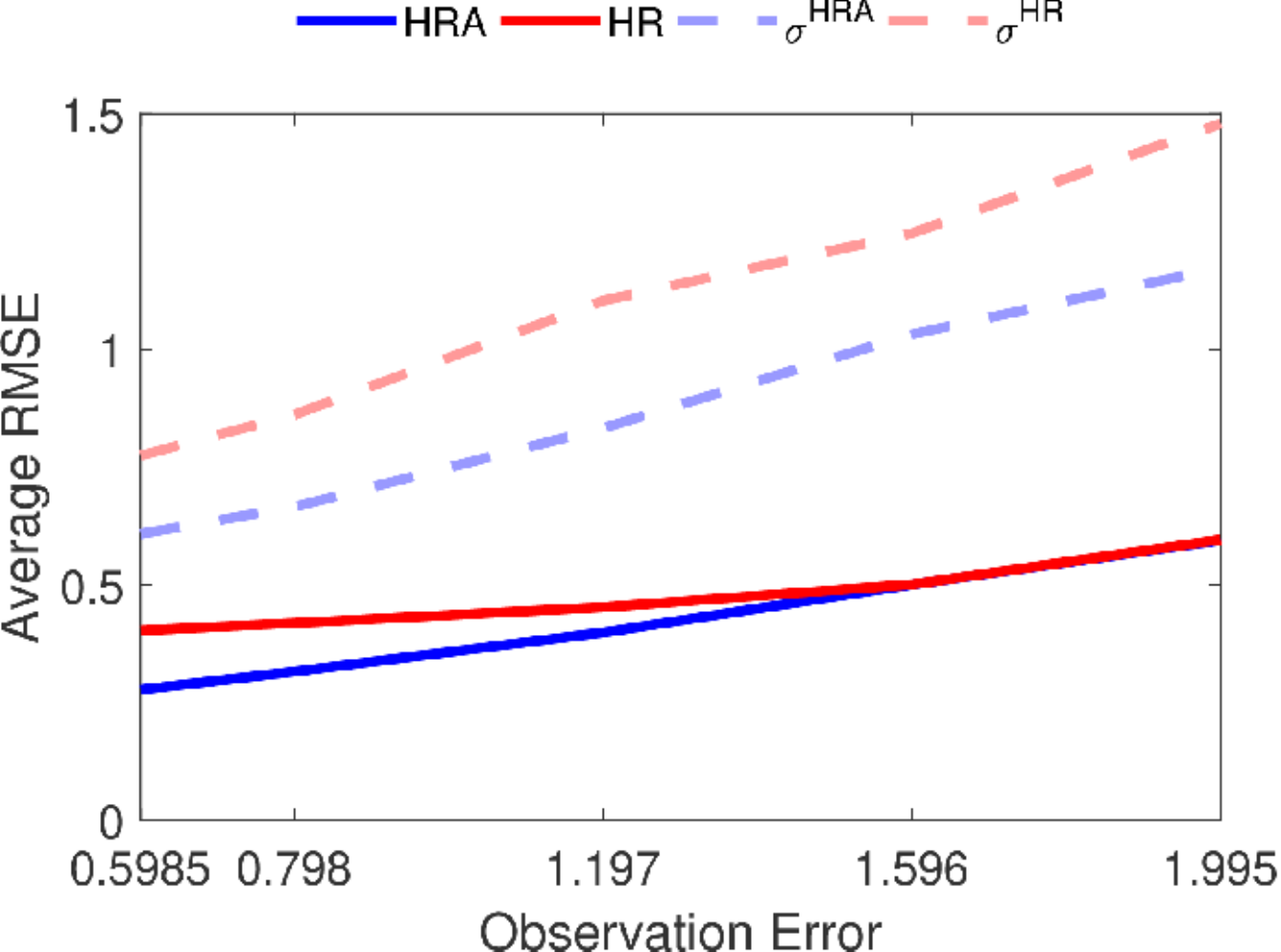}
        \label{OBSoptimalfunctionKSM}
    \end{subfigure}
    
       \begin{subfigure}[b]{0.33\textwidth}
        \includegraphics[width=\textwidth]{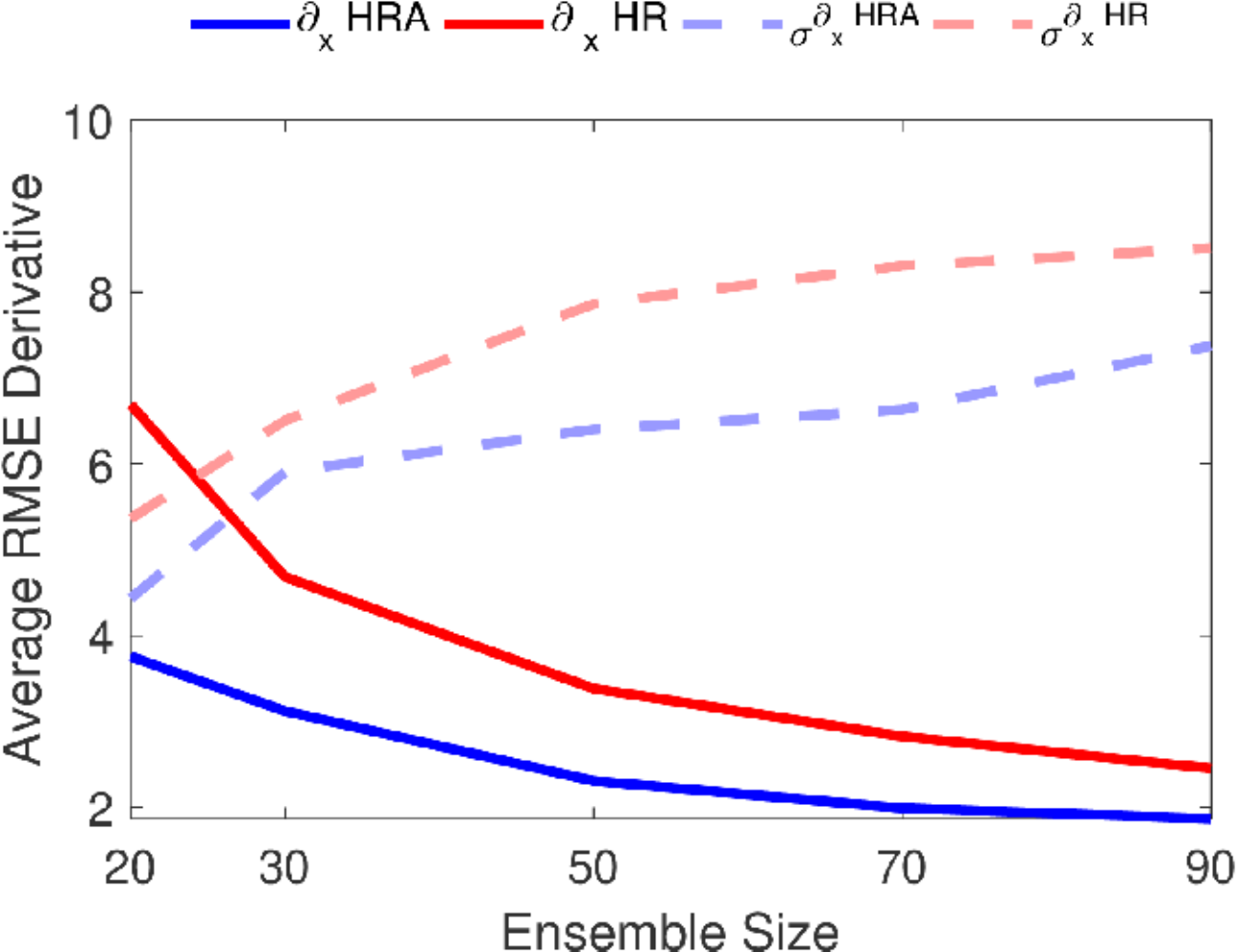}
           \label{ENSoptimalderivativeKSM}
        \end{subfigure}
   \begin{subfigure}[b]{0.33\textwidth}
       \includegraphics[width=\textwidth]{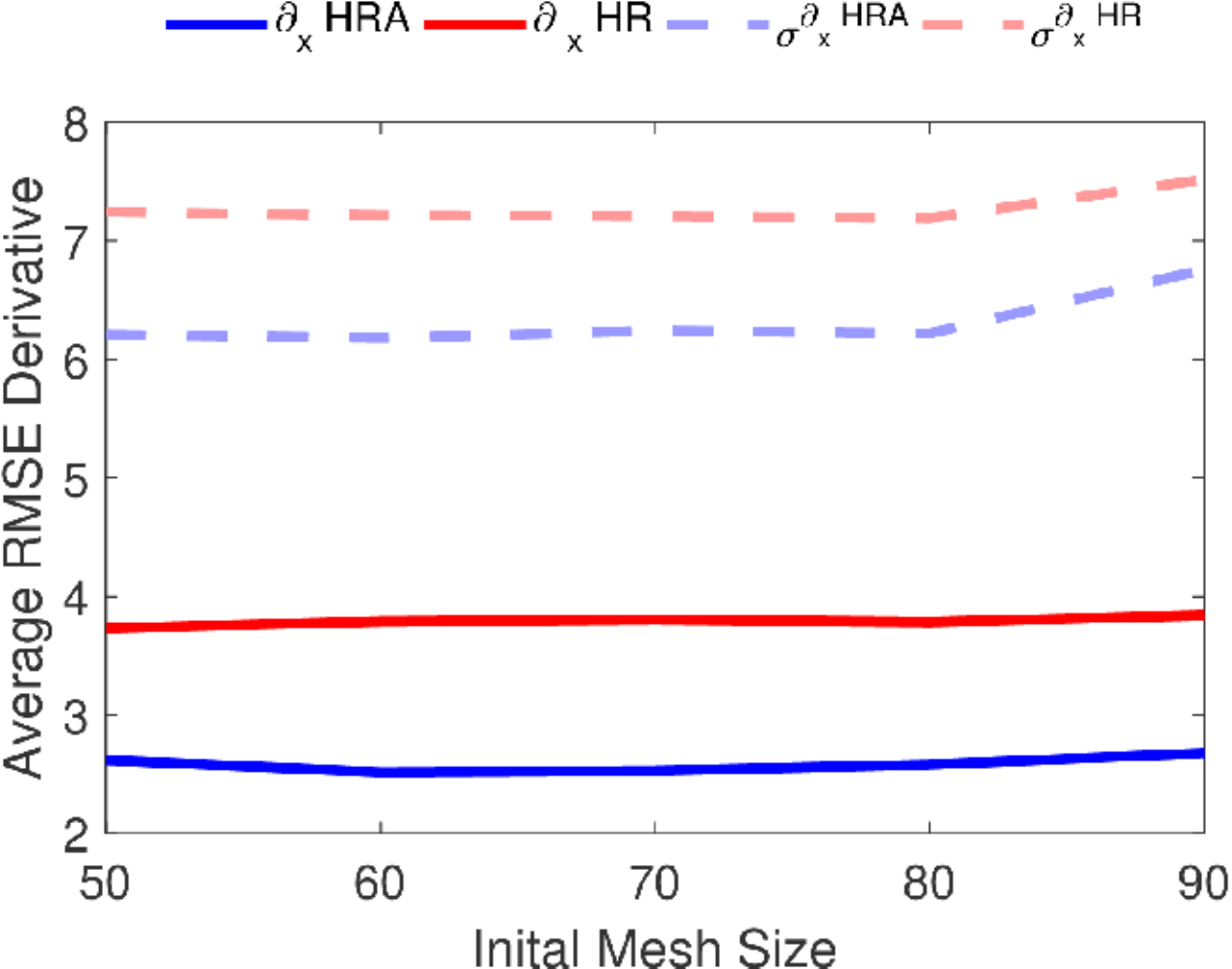}
        \label{INIoptimalderivativeKSM}
    \end{subfigure}
     \begin{subfigure}[b]{0.33\textwidth}
       \includegraphics[width=\textwidth]{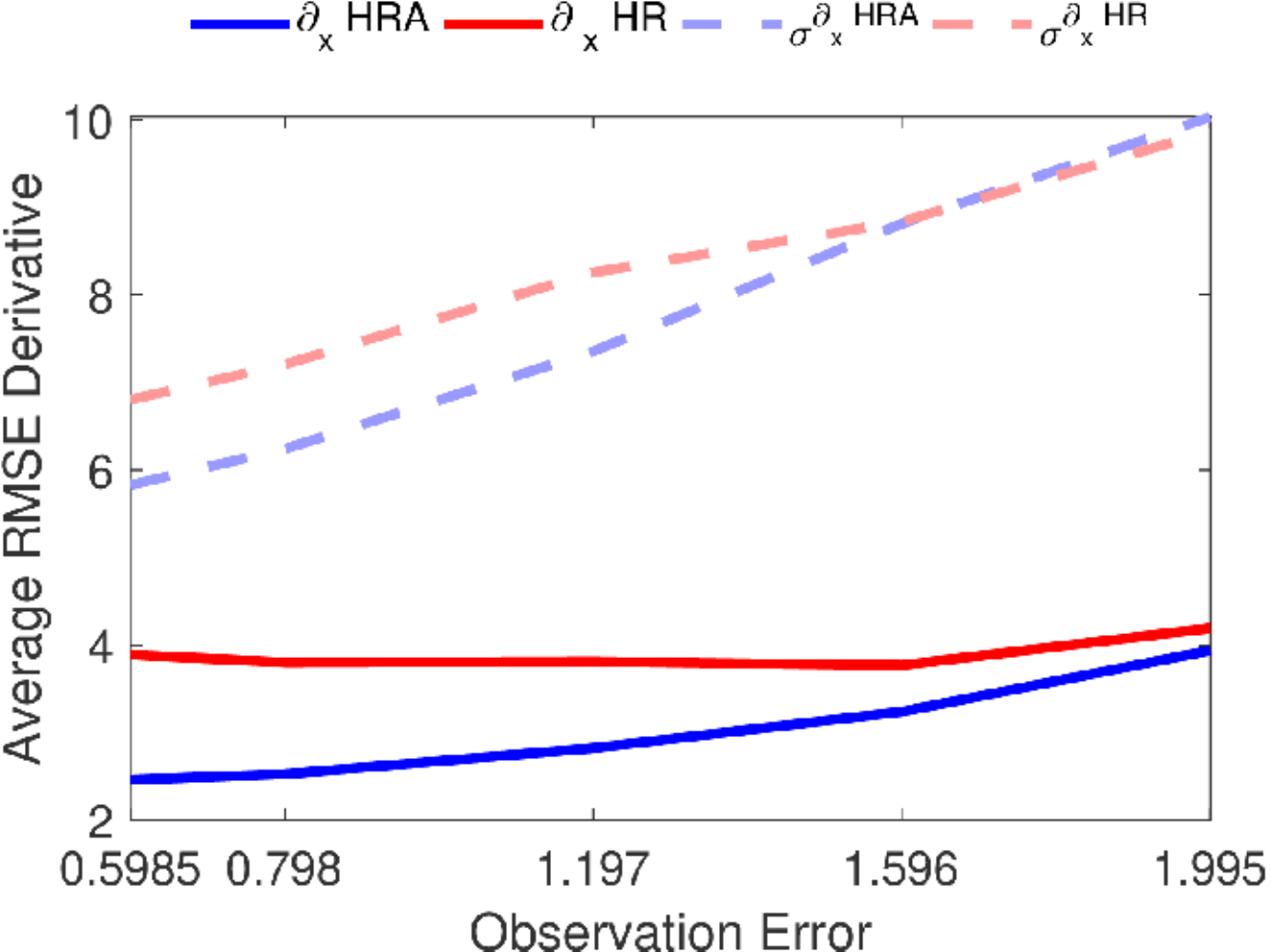}
        \label{OBSoptimalderivativeKSM}
    \end{subfigure}
   
   \caption{Same as Fig.~\ref{optimalBGM} for the KSM model.}\label{optimalKSM}
\end{figure*}

The comparison is carried out over ranges of the three key experimental parameters: the ensemble size, $N^{\rm e}$, the initial mesh size, $I_m$, and the observation error, $\sigma_{\rm o}$.
We study the performance of the methods by running experiments with two of them kept fixed while varying the other. 
For each parameter setting, the optimal jitter and inflation for each scheme are determined by running tuning experiments that identify the pair of values giving the lowest time averaged RMSE. This way we will compare the best possible configuration of each scheme. The values used in the experiments are given in Table~\ref{tab:sensitivitysettings}.  
Results are shown in Fig.~\ref{optimalBGM} and \ref{optimalKSM} for the BGM and KSM model respectively.

\begin{table}[ht]
\centering
\begin{tabular}{l| l l l| l l l}
\hline
   & \multicolumn{3}{c|}{BGM} & \multicolumn{3}{c}{KSM}\\
\hline
Experiment Type& $N^{\rm e}$   & $I_{\rm m}$   & $\sigma_{\rm o}$    & $N^{\rm e}$   & $I_{\rm m}$   & $\sigma_{\rm o}$    \\ \hline
 Varying $N^{\rm e}$   & [20-90] & 70      & 0.01   & [20-90] & 70      & 0.798      \\ 
 Varying $I_{\rm m}$ & 30      & [50-90] & 0.01 & 40      & [50-90] & 0.798     \\ 
 Varying $\sigma_{\rm0}$ & 30    & 70      & [0.01-0.07] & 40    & 70      & [0.60-2.0]\\  \hline
\end{tabular}
\caption{The settings used for the three sensitivity experiments for the BGM and KSM models. For all three experiment types the ranges of $\alpha$ and $\alpha_{\rm J}$ optimised over remained the same. The range of $\alpha_{\rm J}$ differed between the model types with a range of $[0-0.1]$ for the BGM model and $[0-0.5]$ for the KSM model. The range of $\alpha$ was $[0-1.6]$ for both models.}

\label{tab:sensitivitysettings}
\end{table}

First note that Fig.~\ref{optimalBGM} and \ref{optimalKSM} immediately reveal that the ensemble spread in the HR scheme is typically much larger than that of the HRA scheme even when performance is comparable, such as in the initial mesh size experiments. This is due to the inherent stochasticity of the HR scheme discussed in section~\ref{dimreturn}. 

By looking at the RMSE of the analysis mean, it is evident that the HRA scheme tends to out perform the HR scheme in general and particularly for smaller ensemble sizes. This is due to the extra information carried in the cross covariances between the physical values and the node locations.  
The RMSE of the spatial derivatives is also generally lower in the HRA scheme, except at ensemble size 50 for the BGM case. It is worth reiterating that we use parameters optimised for the solution itself and not the first derivatives here. We will discuss this behaviour more extensively later in this section together with other metrics used to understand this particular issue. Figures~\ref{optimalBGM} and \ref{optimalKSM} also highlight that only marginal improvements in time averaged RMSE are obtained after an ensemble size of 30 for the BGM model and 50 for the KSM model. This kind of behaviour can be observed when the ensemble size is larger than or equal to the dimension of the unstable, neutral subspace of the dynamics \citet{bocquet2017four}. 
For both the BGM and KSM models we do not see much dependence on the initial mesh size but with HRA performing comparably to or better than the HR scheme in the BGM case and out performing HR in the KSM case.  

We also make a comparison to the performance of each scheme with respect to increasing observation error. For the BGM case both schemes perform comparably but we see better results from the HRA scheme in the KSM case particularly with regard to the first spatial derivative of the solution. We would also like to remark that the clearer trends in the KSM experiments is likely a result of a longer time average of the RMSE available as the BGM model damps quickly limiting the experimental time window. When the observation error is large enough both models perform about the same suggesting that one might choose to accept extra computational cost of the HRA scheme when the observations are good enough to warrant doing so.

\begin{figure*}
    \centering
    \begin{subfigure}[b]{0.33\textwidth}
        \includegraphics[width=\textwidth]{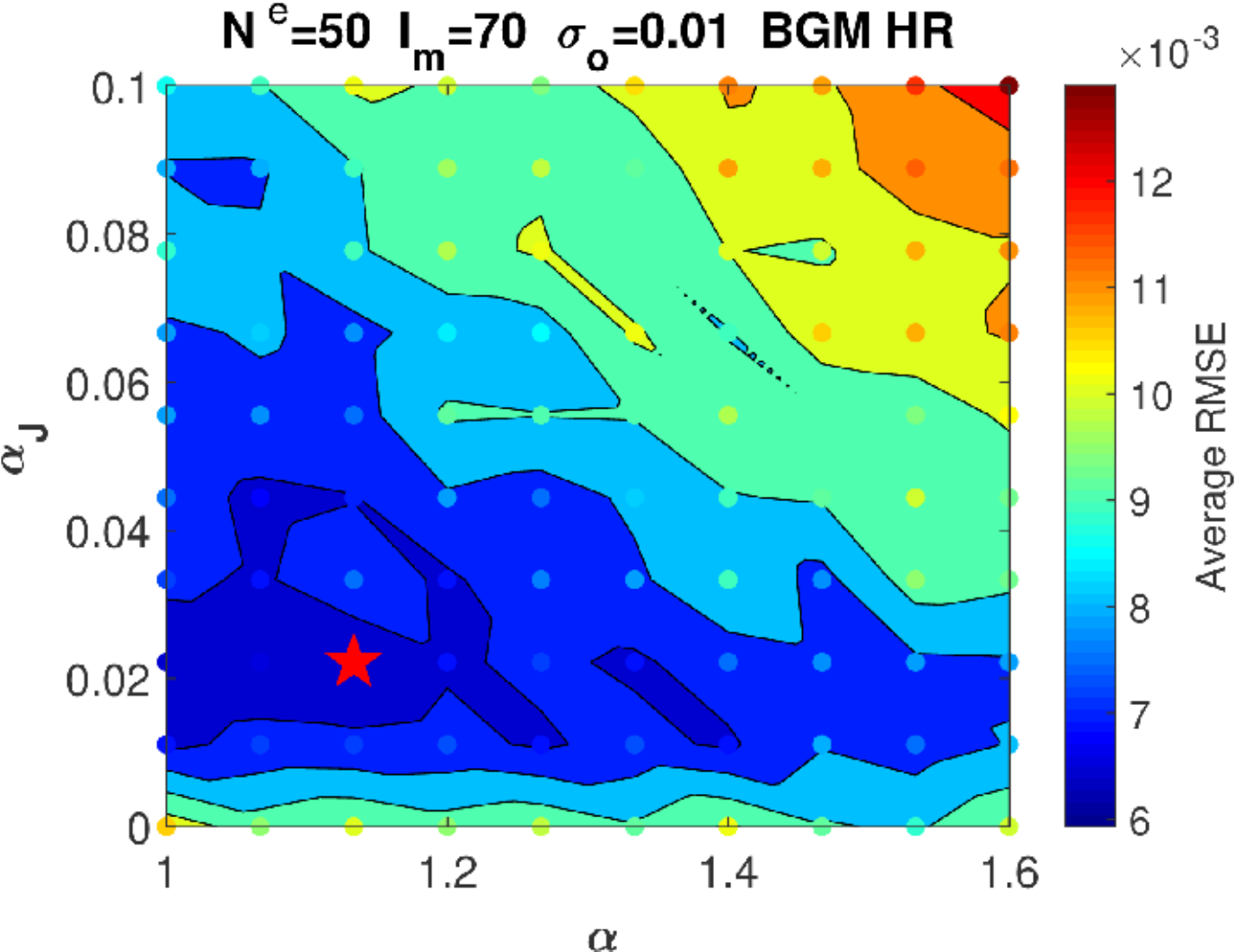}
         \label{ENSBGMHRSURF}
      \end{subfigure}
   \begin{subfigure}[b]{0.33\textwidth}
       \includegraphics[width=\textwidth]{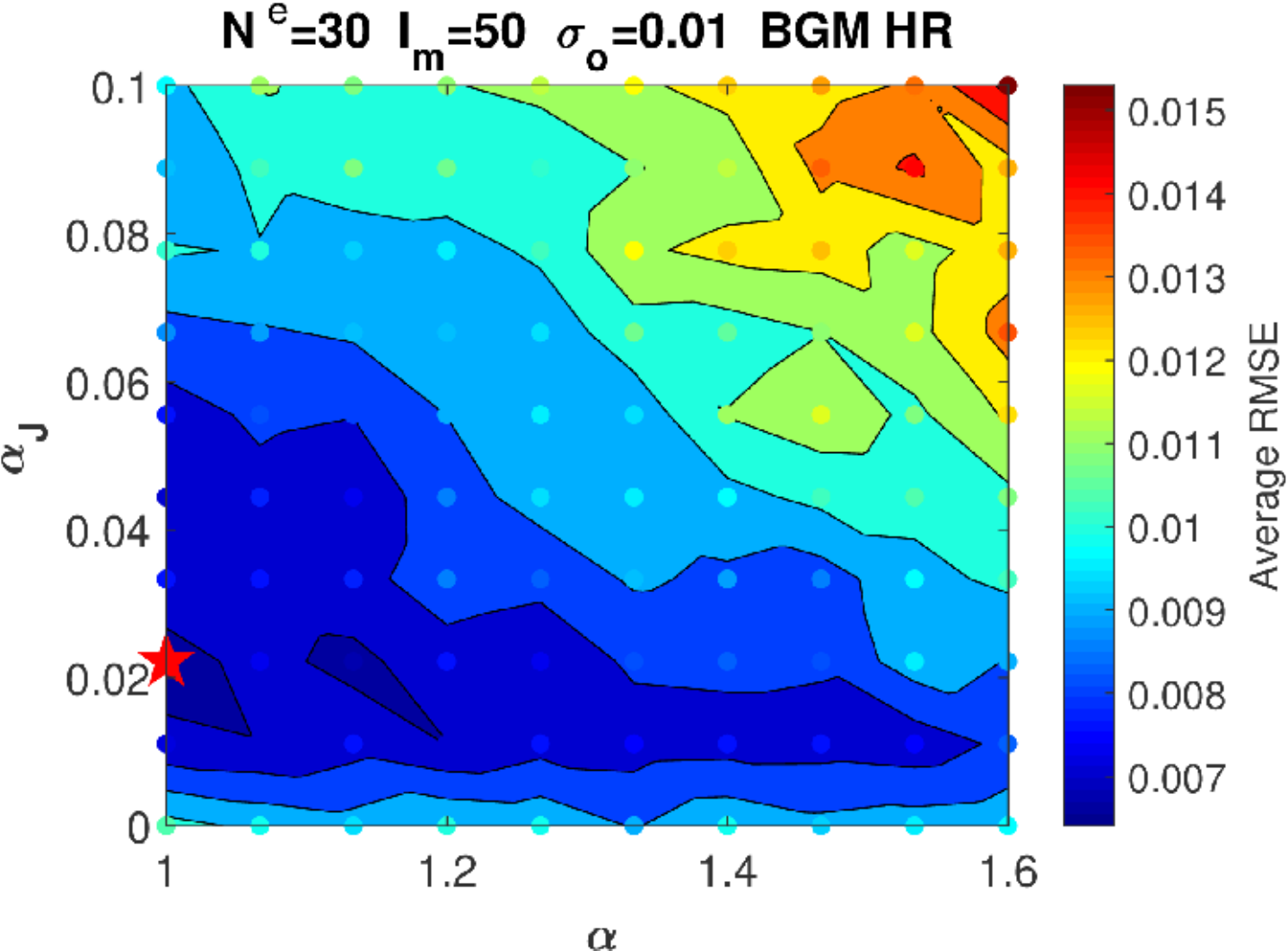}
        \label{INIBGMSHRURF}
    \end{subfigure}
     \begin{subfigure}[b]{0.33\textwidth}
       \includegraphics[width=\textwidth]{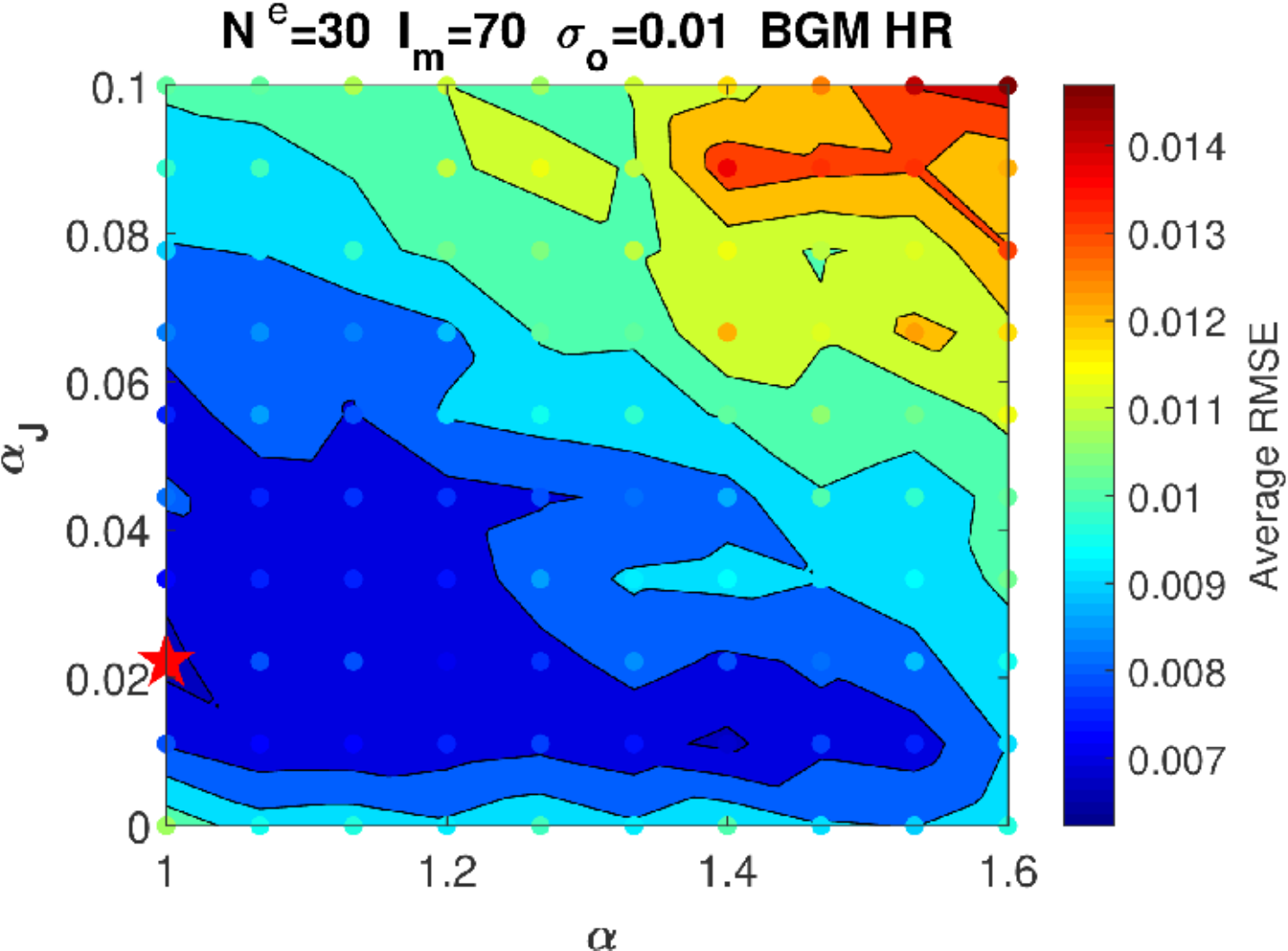}
        \label{OBSBGMHRSURF}
    \end{subfigure}
    
        \begin{subfigure}[b]{0.33\textwidth}
        \includegraphics[width=\textwidth]{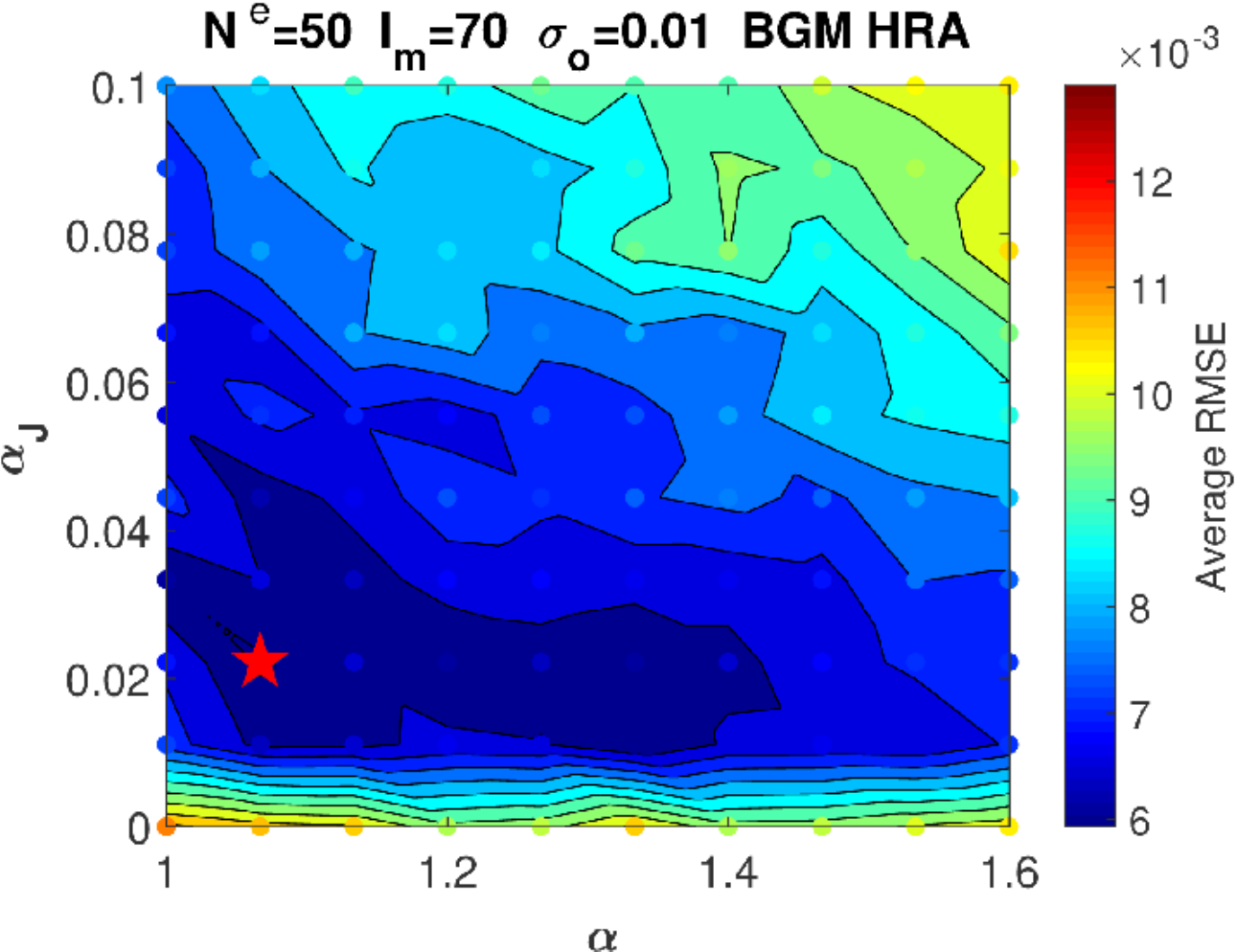}
         \label{ENSBGMHRASURF}
      \end{subfigure}
   \begin{subfigure}[b]{0.33\textwidth}
       \includegraphics[width=\textwidth]{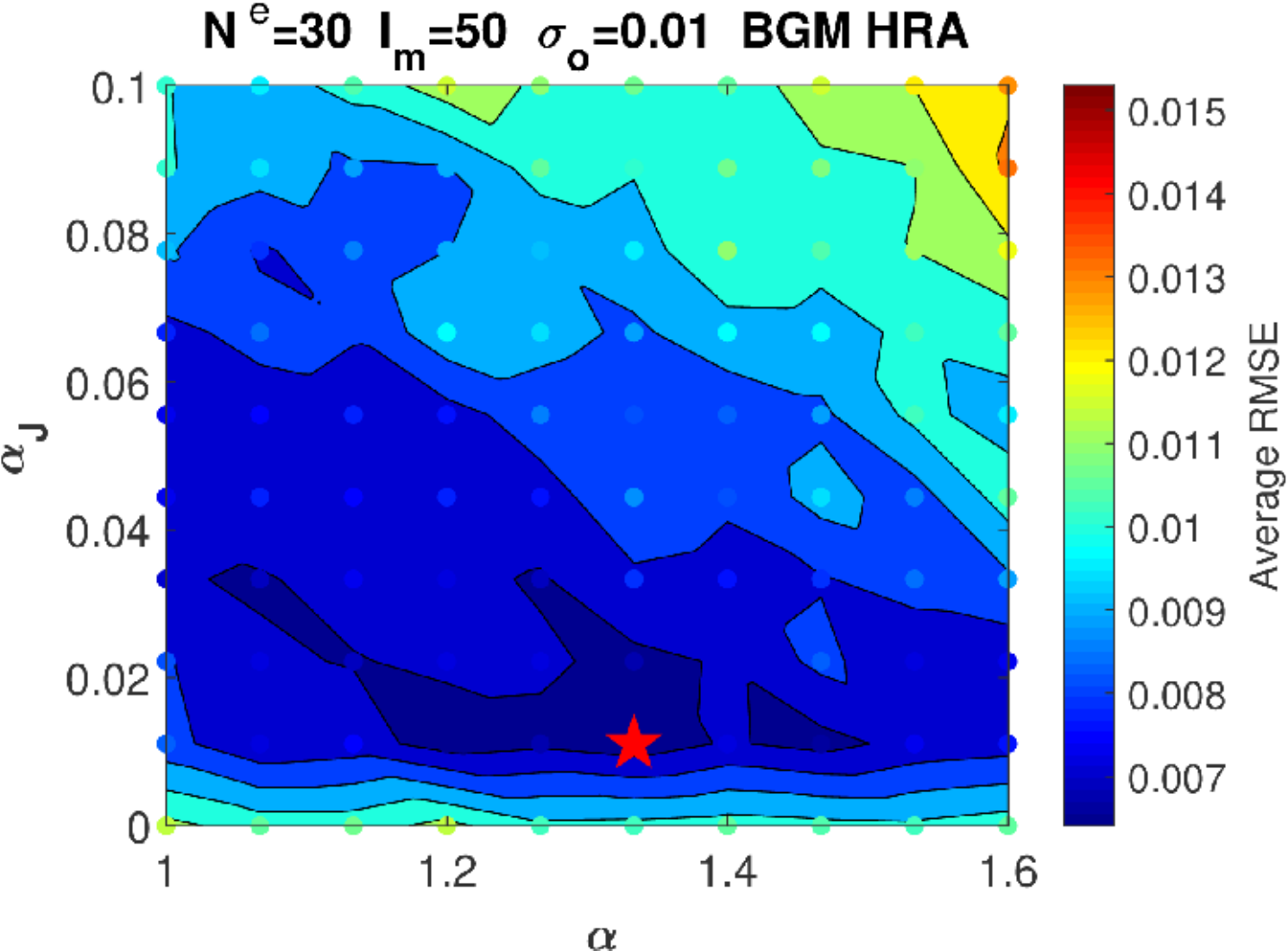}
        \label{INIBGMSHRAURF}
    \end{subfigure}
     \begin{subfigure}[b]{0.33\textwidth}
       \includegraphics[width=\textwidth]{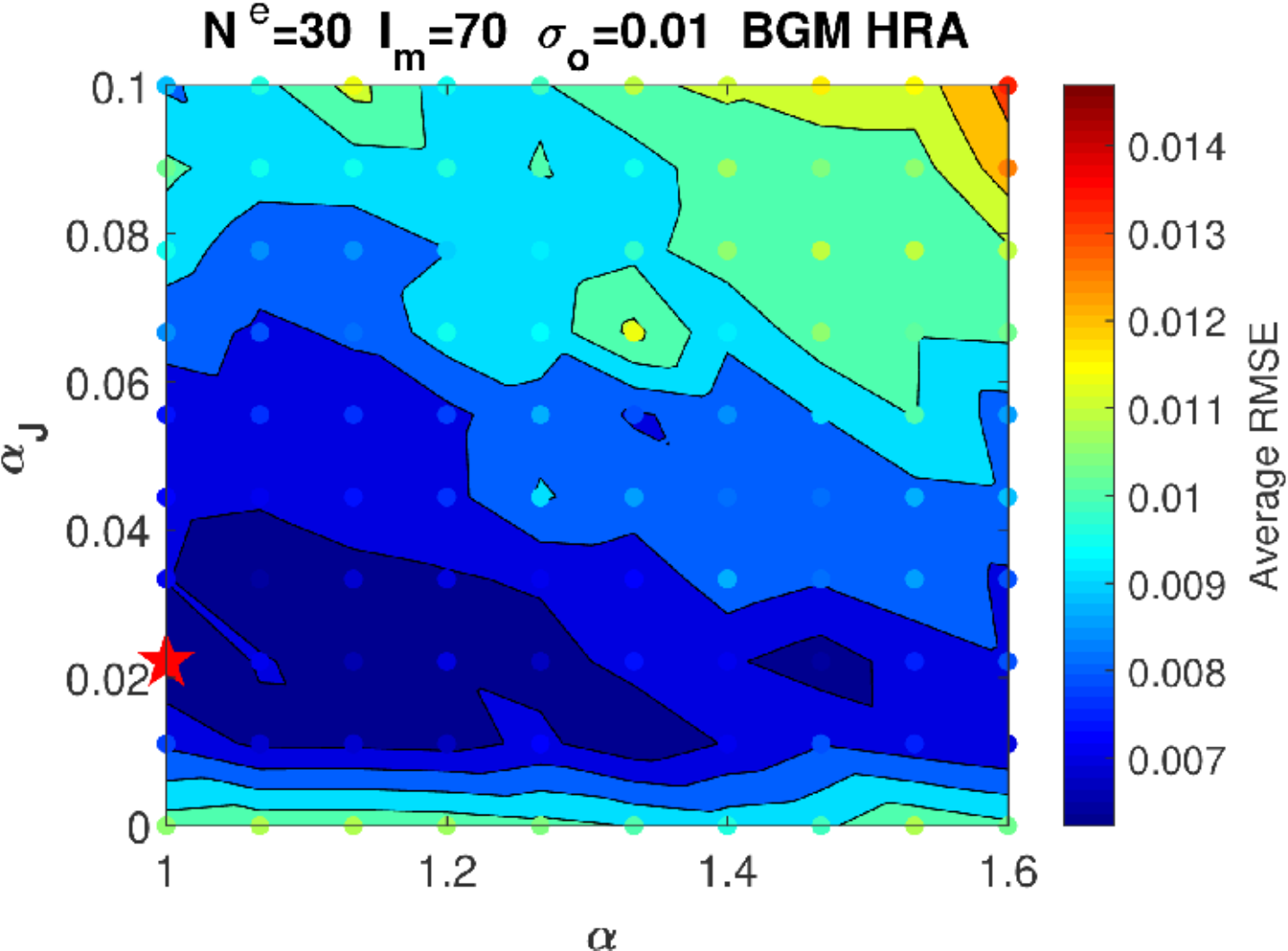}
        \label{OBSBGMHRASURF}
    \end{subfigure}

   \caption{Contour plots for the jitter and inflation calibrations for the BGM model for the HR (top row) and HRA (bottom row) schemes for the experimental parameters. The points in the plot represent the sample points used while the red star represents the jitter and inflation with the lowest time averaged RMSE. Of particular note is that the HRA scheme has its valley of low RMSE well away from the x-axis implying that some jitter is beneficial while this is not as strong a feature with the HR method due to the inherent stochasticity added during the dimension matching step.}\label{BGMsurf}
\end{figure*}

\begin{figure*}
    \centering
    \begin{subfigure}[b]{0.33\textwidth}
        \includegraphics[width=\textwidth]{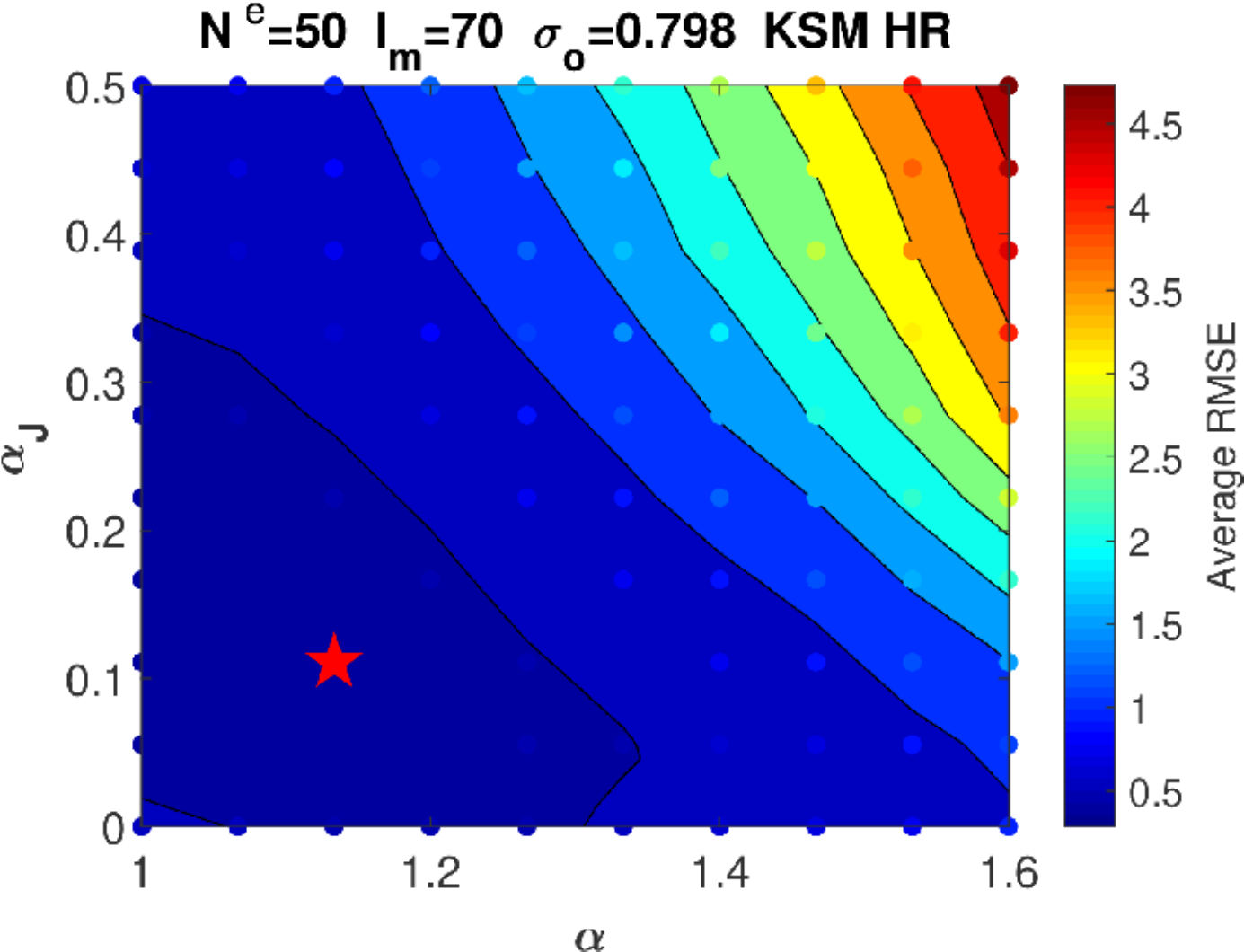}
         \label{ENSKSMHRSURF}
      \end{subfigure}
   \begin{subfigure}[b]{0.33\textwidth}
       \includegraphics[width=\textwidth]{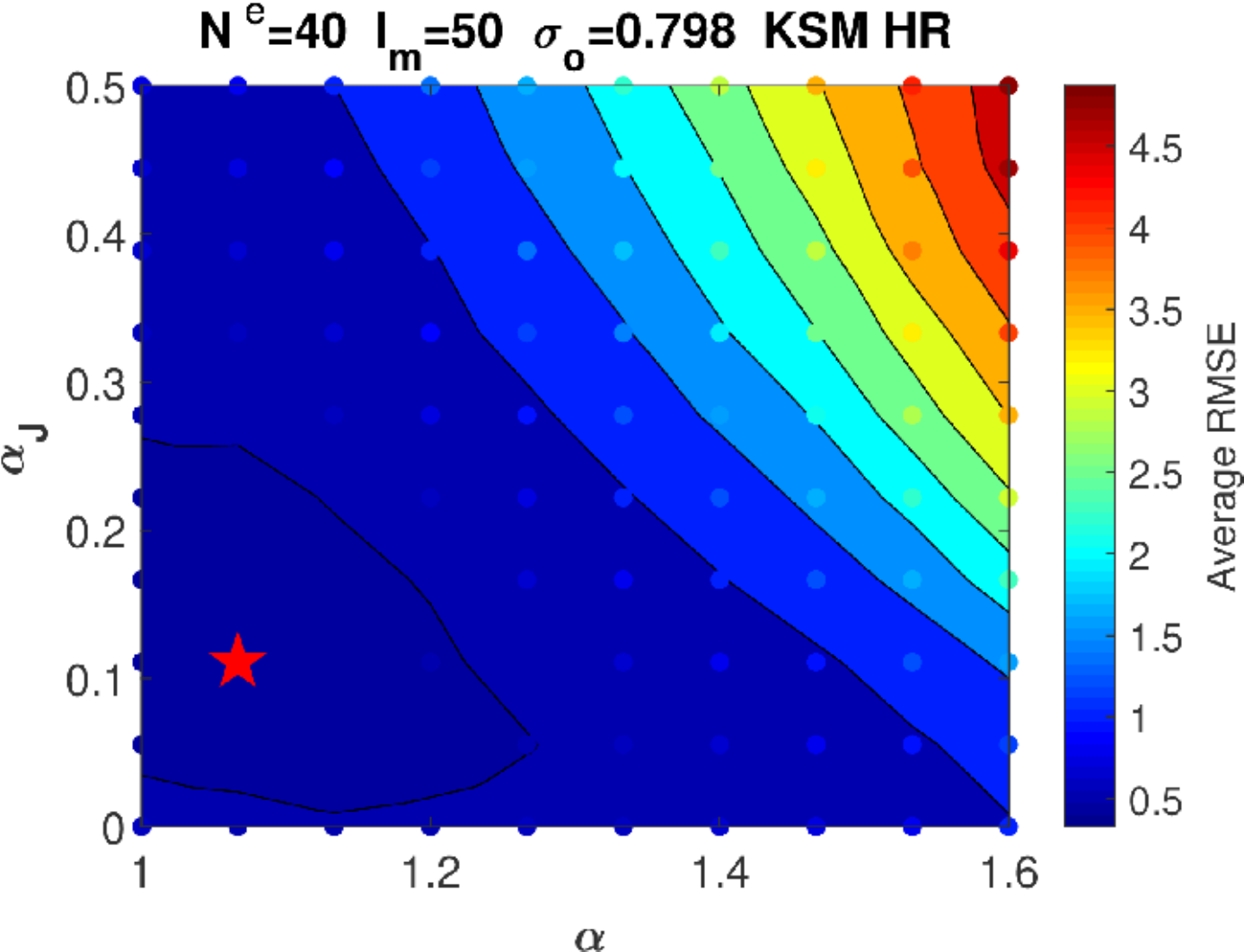}
        \label{INIKSMSHRURF}
    \end{subfigure}
     \begin{subfigure}[b]{0.33\textwidth}
       \includegraphics[width=\textwidth]{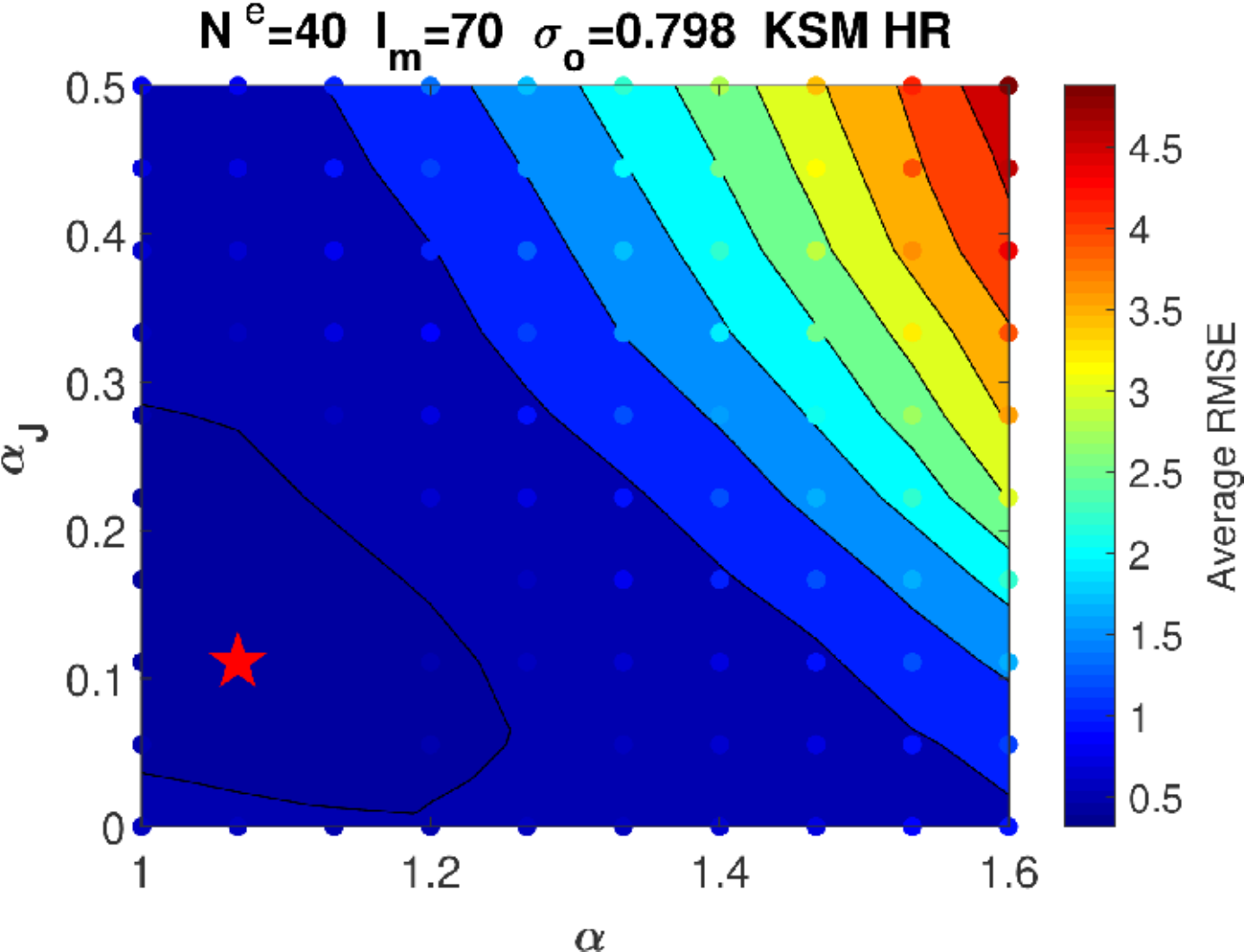}
        \label{OBSKSMHRSURF}
    \end{subfigure}
    
        \begin{subfigure}[b]{0.33\textwidth}
        \includegraphics[width=\textwidth]{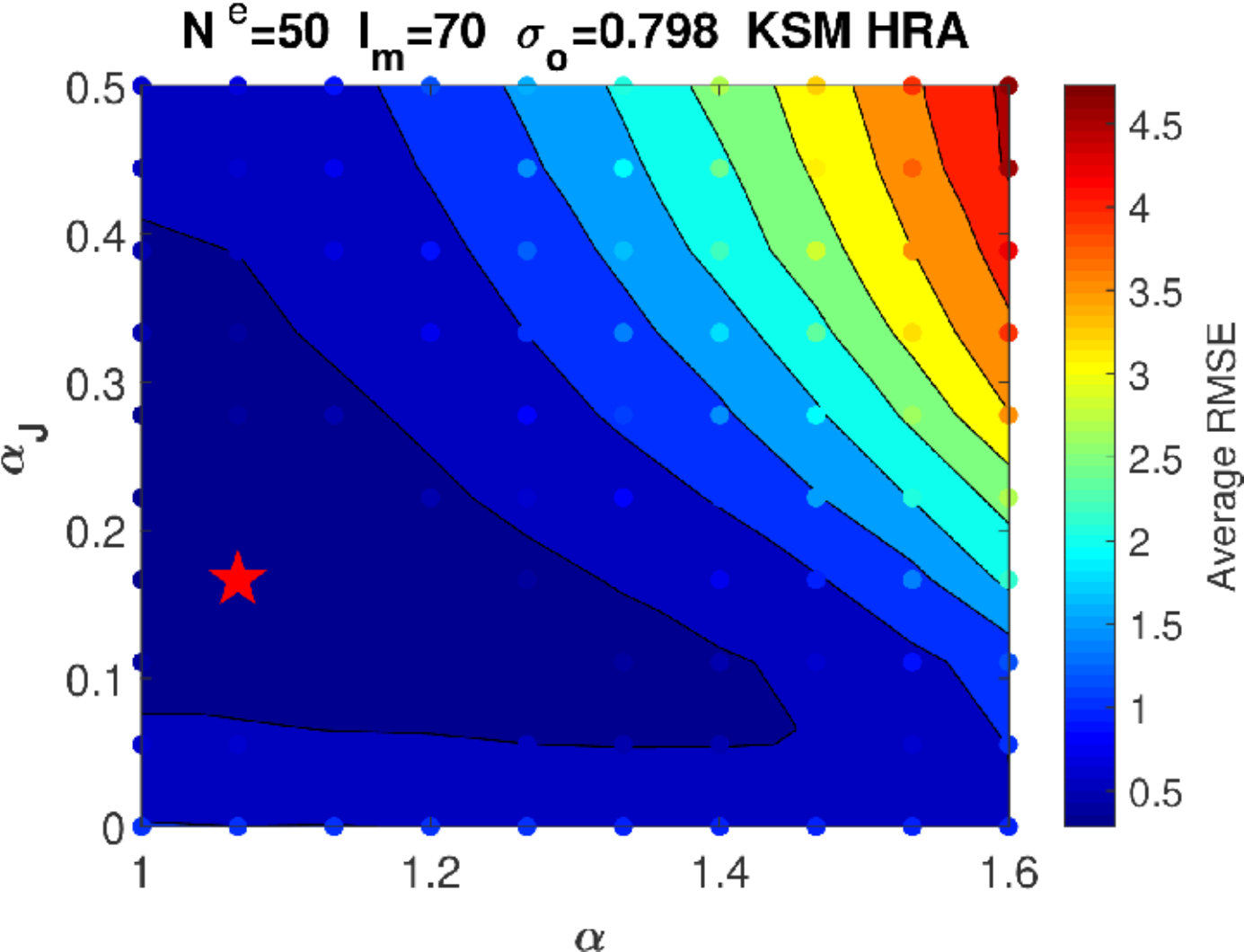}
         \label{ENSKSMHRASURF}
      \end{subfigure}
   \begin{subfigure}[b]{0.33\textwidth}
       \includegraphics[width=\textwidth]{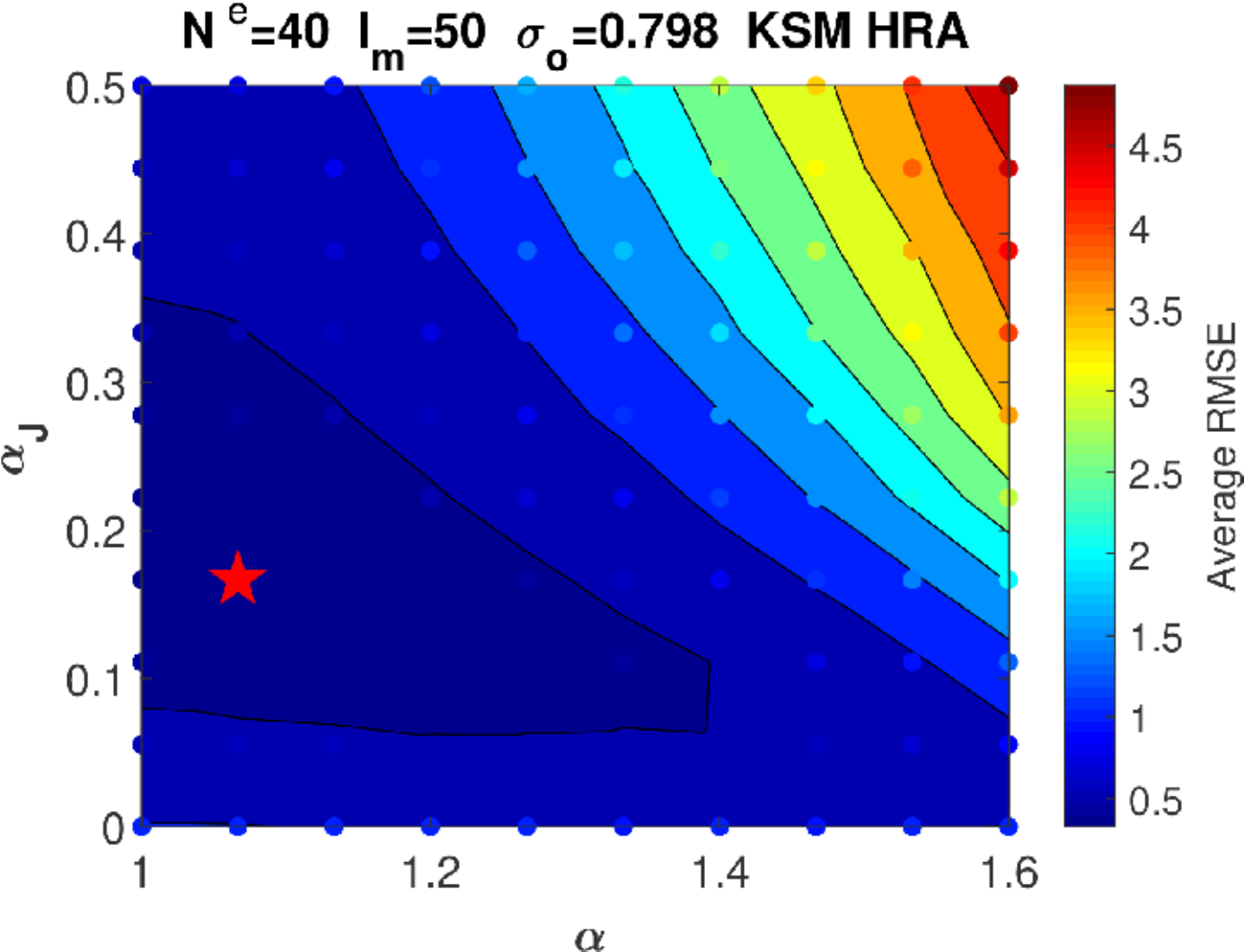}
        \label{INIKSMSHRAURF}
    \end{subfigure}
     \begin{subfigure}[b]{0.33\textwidth}
       \includegraphics[width=\textwidth]{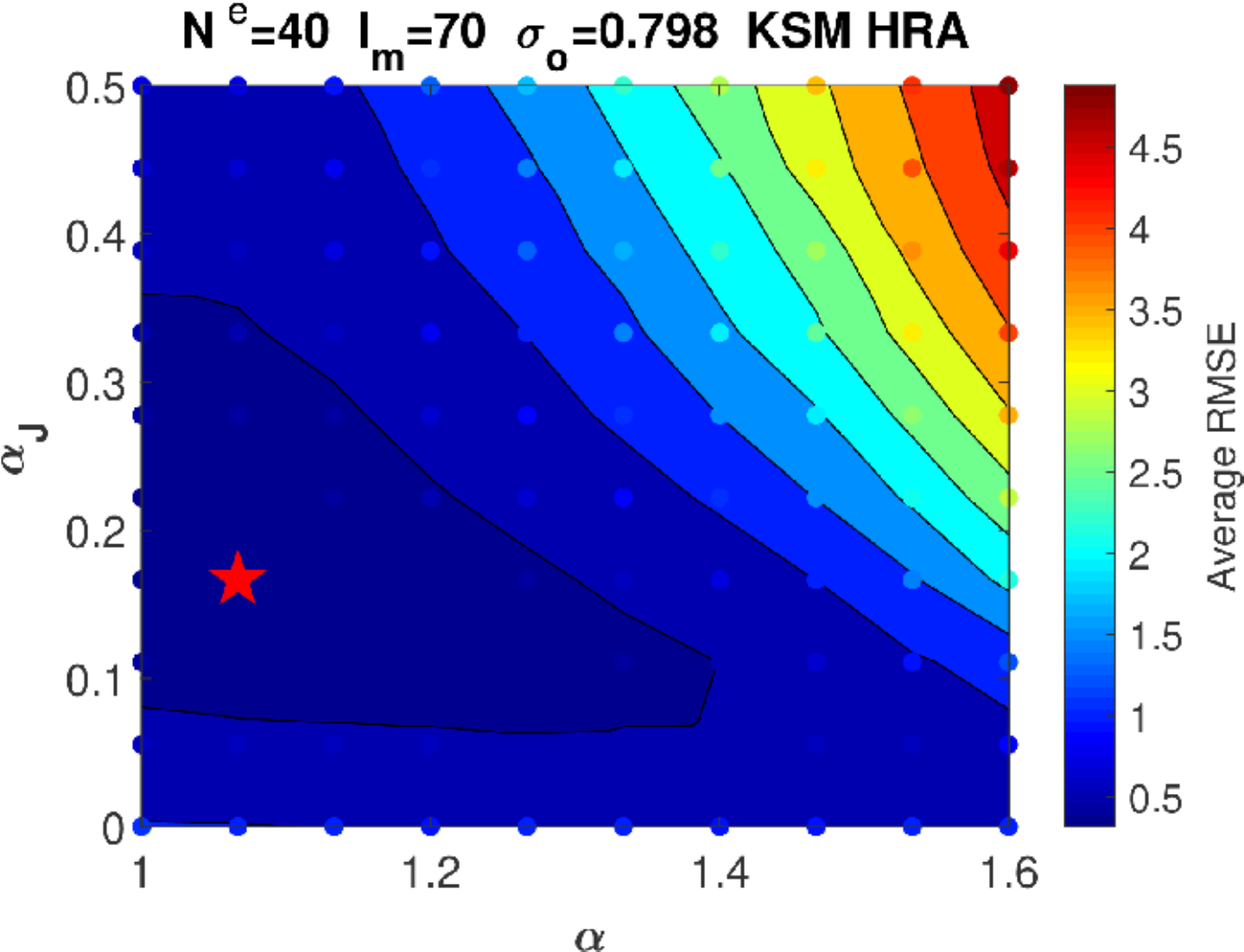}
        \label{OBSKSMHRASURF}
    \end{subfigure}
  
   \caption{Same as Fig.~\ref{BGMsurf} but for the KSM model.}\label{KSMsurf}
\end{figure*}

In Fig.~\ref{BGMsurf} and \ref{KSMsurf}, respectively for KSM and BGM, we show examples of the time averaged RMSE surfaces from the experiments described above as a function of inflation and jitter with the optimal pair of values marked by a red star. 
The difference in the smoothness of the contour plots arises from the aforementioned fact that we run the KSM model for a longer time than the BGM model which dissipates quickly due to the chosen viscosity term. The longer run provides a larger sample of RMSE values to average over producing a smoother surface. 

The need for some jitter in the HRA method (bottom panels) is highlighted by the fact that the time averaged RMSE error is higher near the x-axis ($\alpha_J=0$) for both models, but particularly with BGM). While this is also the case for the HR scheme with the BGM model the effect is less pronounced. For the HR method with the KSM model there is a region with $\alpha_J=0$ that the time averaged RMSE remains close to the one obtained using optimal jitter and inflation, this is likely achievable due to the chaos in the KSM model naturally increasing the spread.

\subsection{Error Covariance Structures}
\begin{figure*}
\centering
\includegraphics[width=\textwidth]{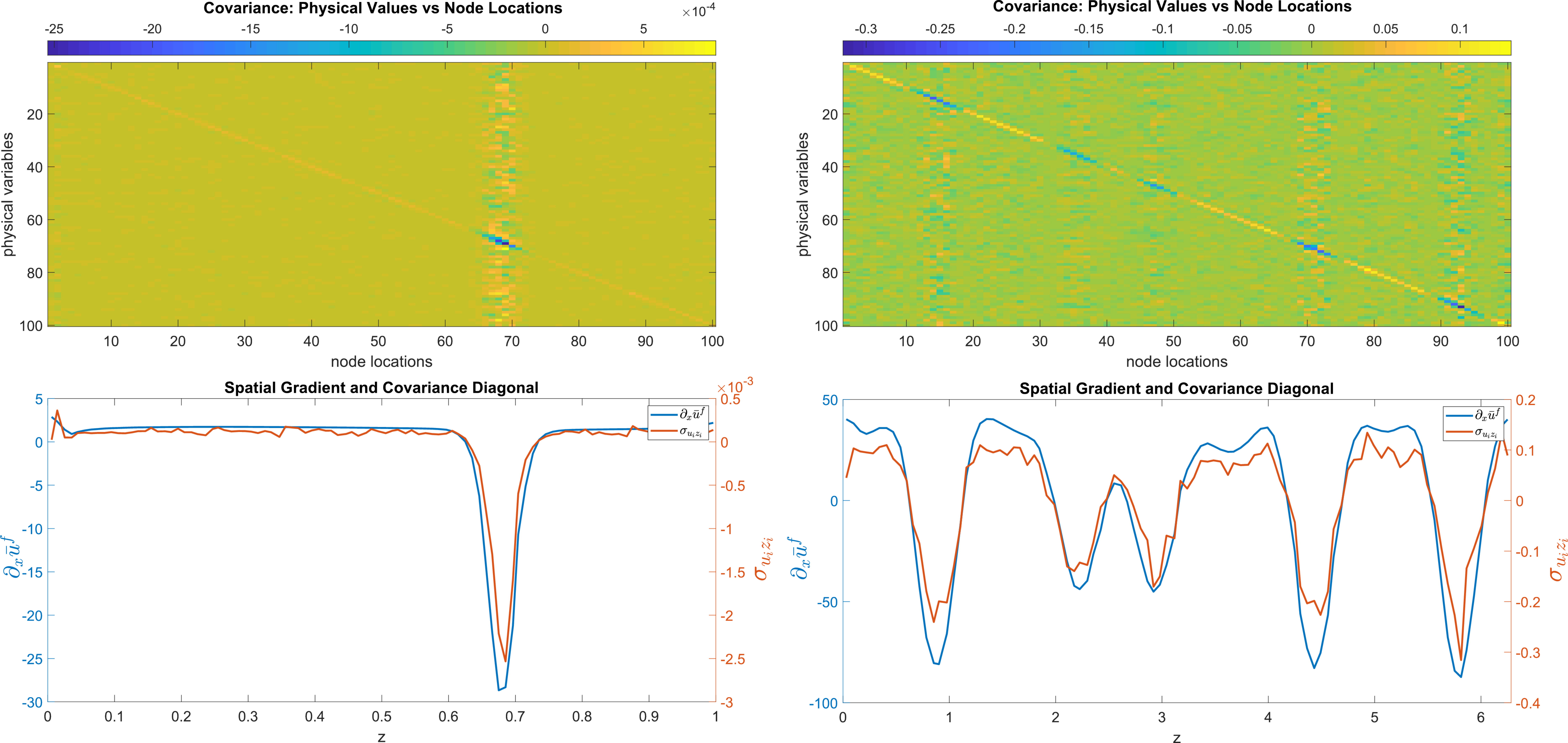}
\caption{ Top row: Examples showing the forecast covariances between the physical values and the node locations in the HRA method for the BGM and KSM models right before the 10th and 20th assimilation steps respectively. Bottom row: The spatial gradient of the forecast mean and covariance between $u_i$ and $z_i$ corresponding to the covariance matrices above, this highlights the extra information encoded into the Kalman gain when using the HRA scheme.}\label{covuz}
\end{figure*}

\begin{figure*}
    \centering
    {\bf BGM }\par \medskip
    \begin{subfigure}[b]{0.33\textwidth}
        \includegraphics[width=\textwidth]{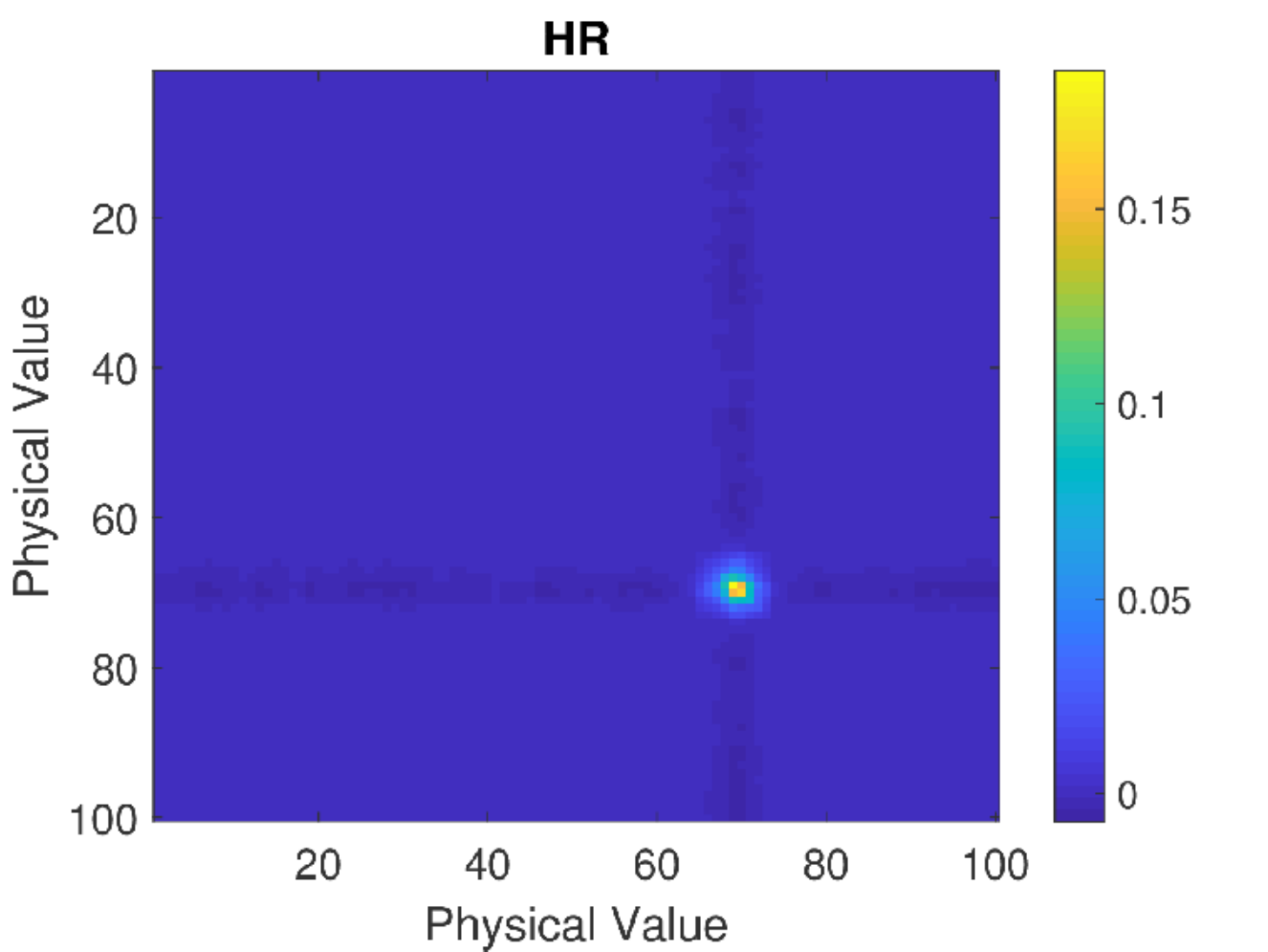}
         \label{BGMUUHR}
      \end{subfigure}
   \begin{subfigure}[b]{0.33\textwidth}
       \includegraphics[width=\textwidth]{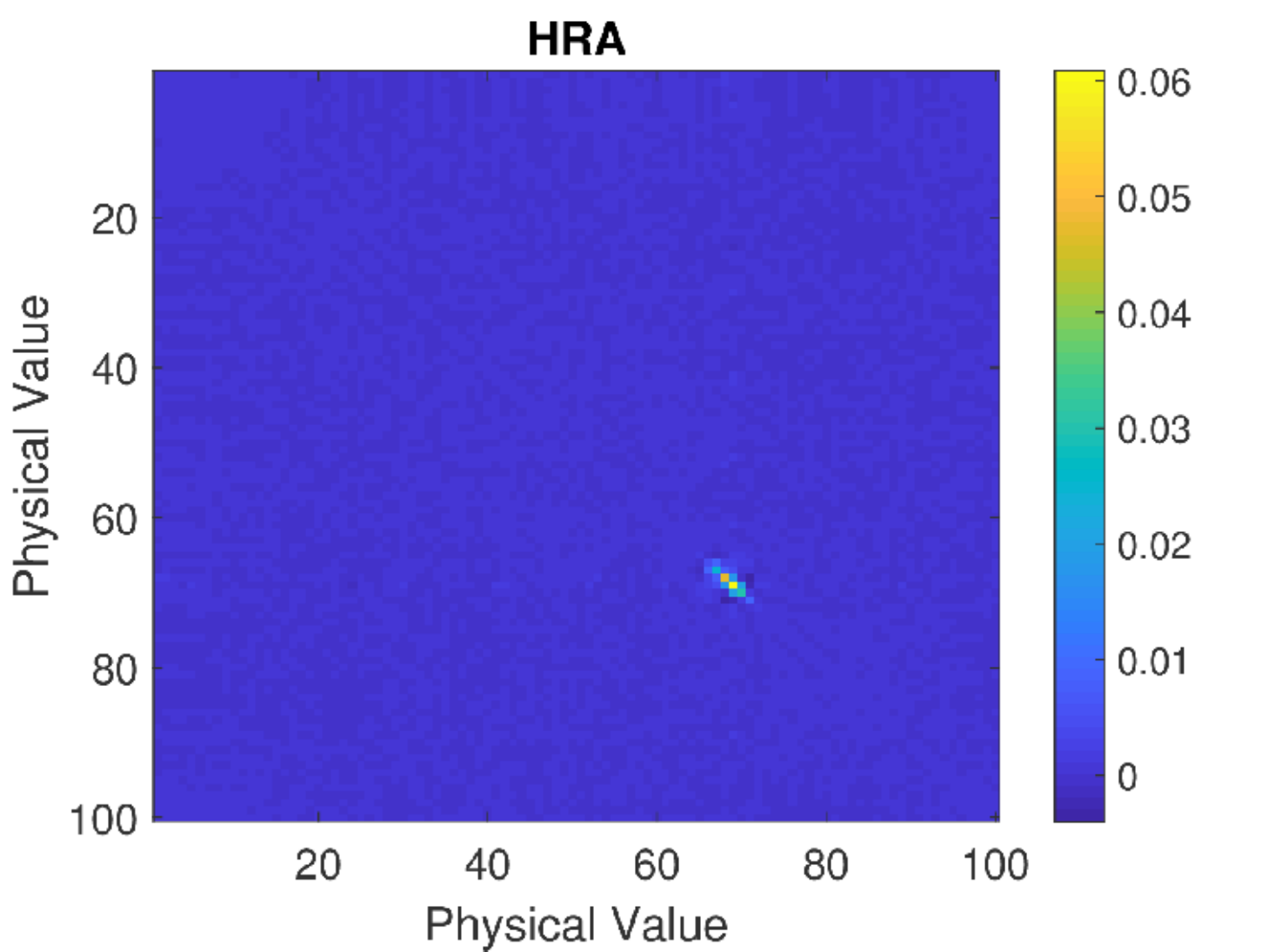}
        \label{BGMUU}
    \end{subfigure}
    \begin{subfigure}[b]{0.33\textwidth}
       \includegraphics[width=\textwidth]{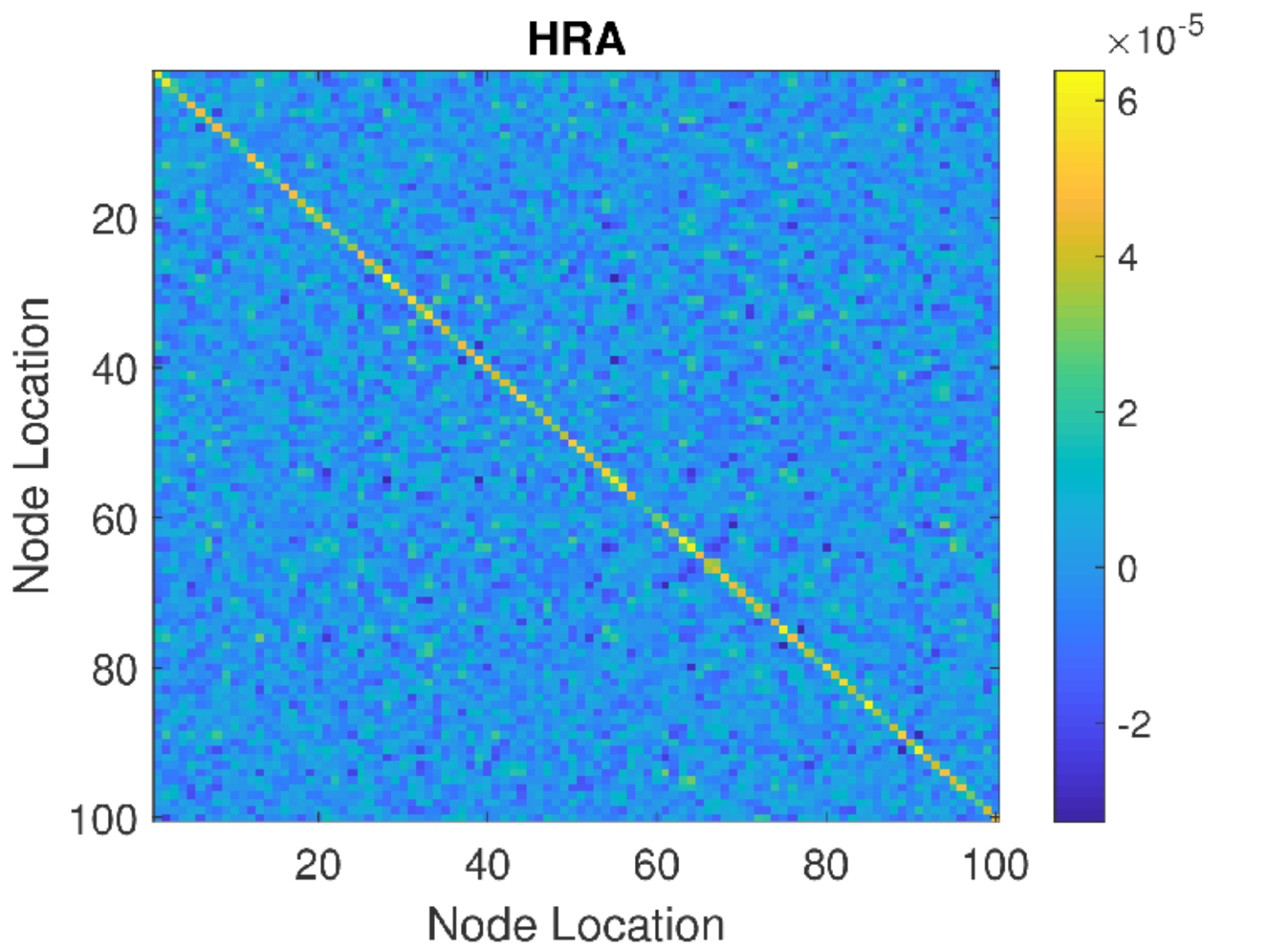}
        \label{BGMZZ}
    \end{subfigure}
   {\bf KSM }\par \medskip
      \begin{subfigure}[b]{0.33\textwidth}
        \includegraphics[width=\textwidth]{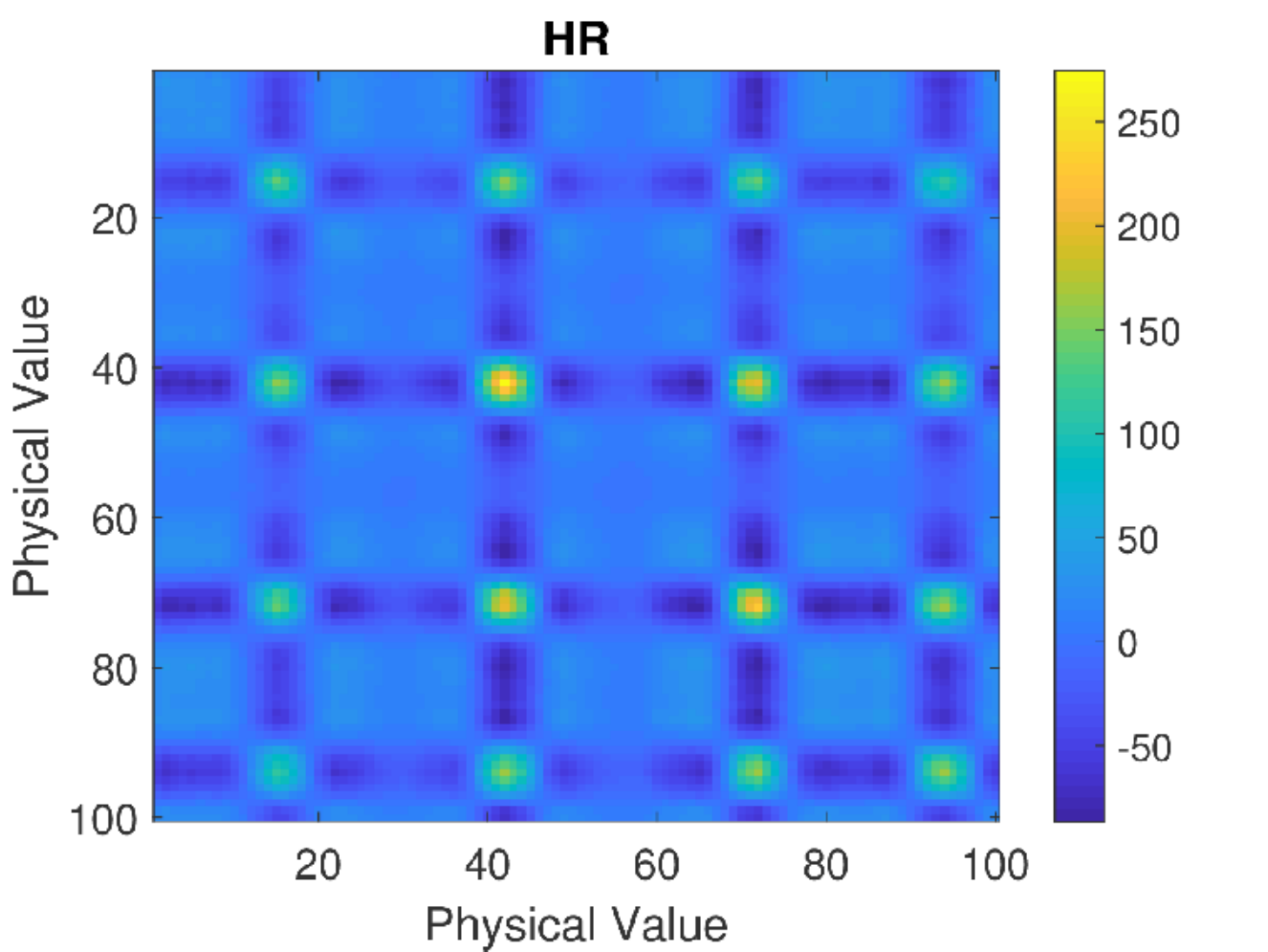}
         \label{KSMUUHR}
      \end{subfigure}
   \begin{subfigure}[b]{0.33\textwidth}
       \includegraphics[width=\textwidth]{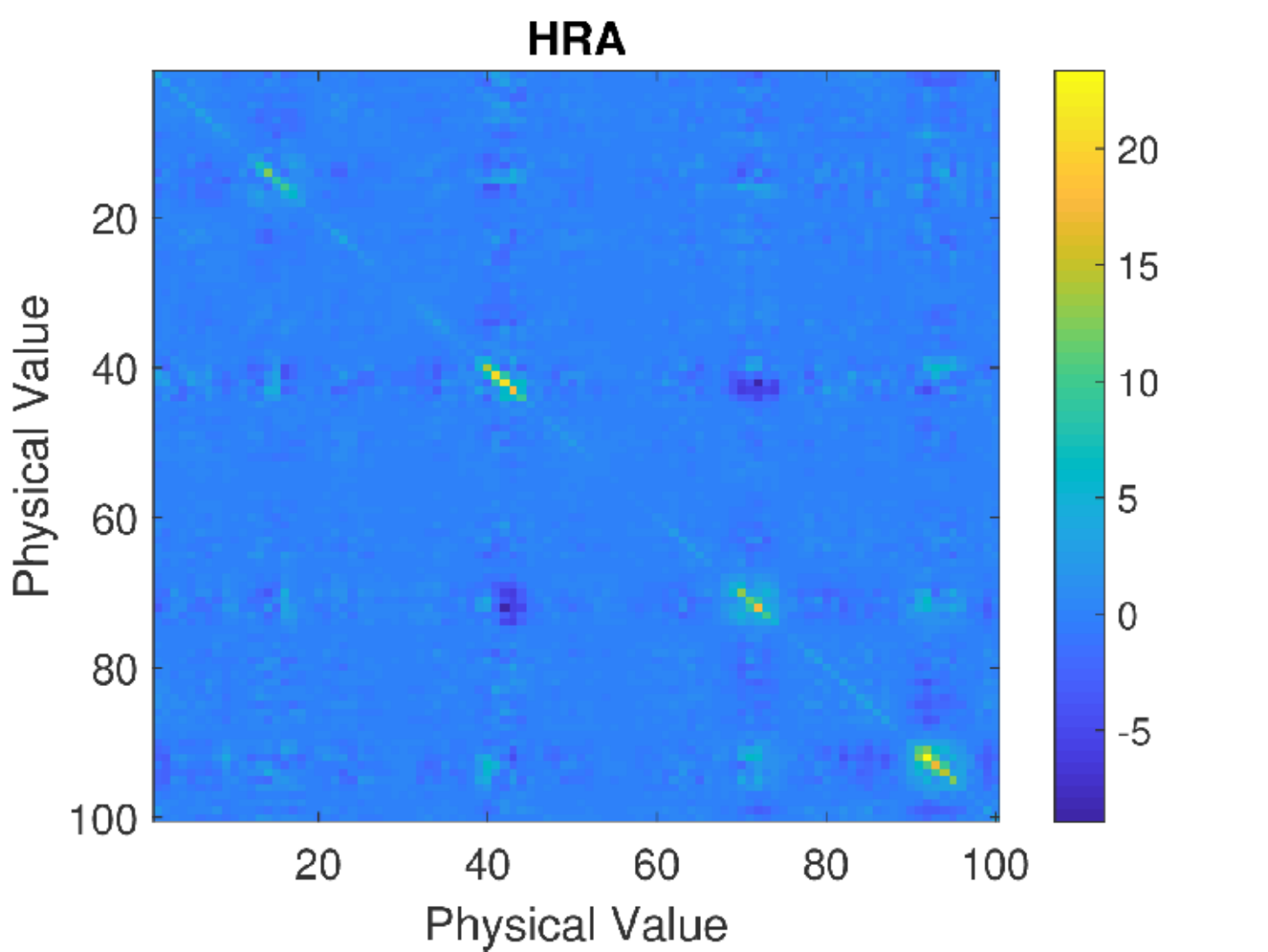}
        \label{KSMUU}
    \end{subfigure}
    \begin{subfigure}[b]{0.33\textwidth}
       \includegraphics[width=\textwidth]{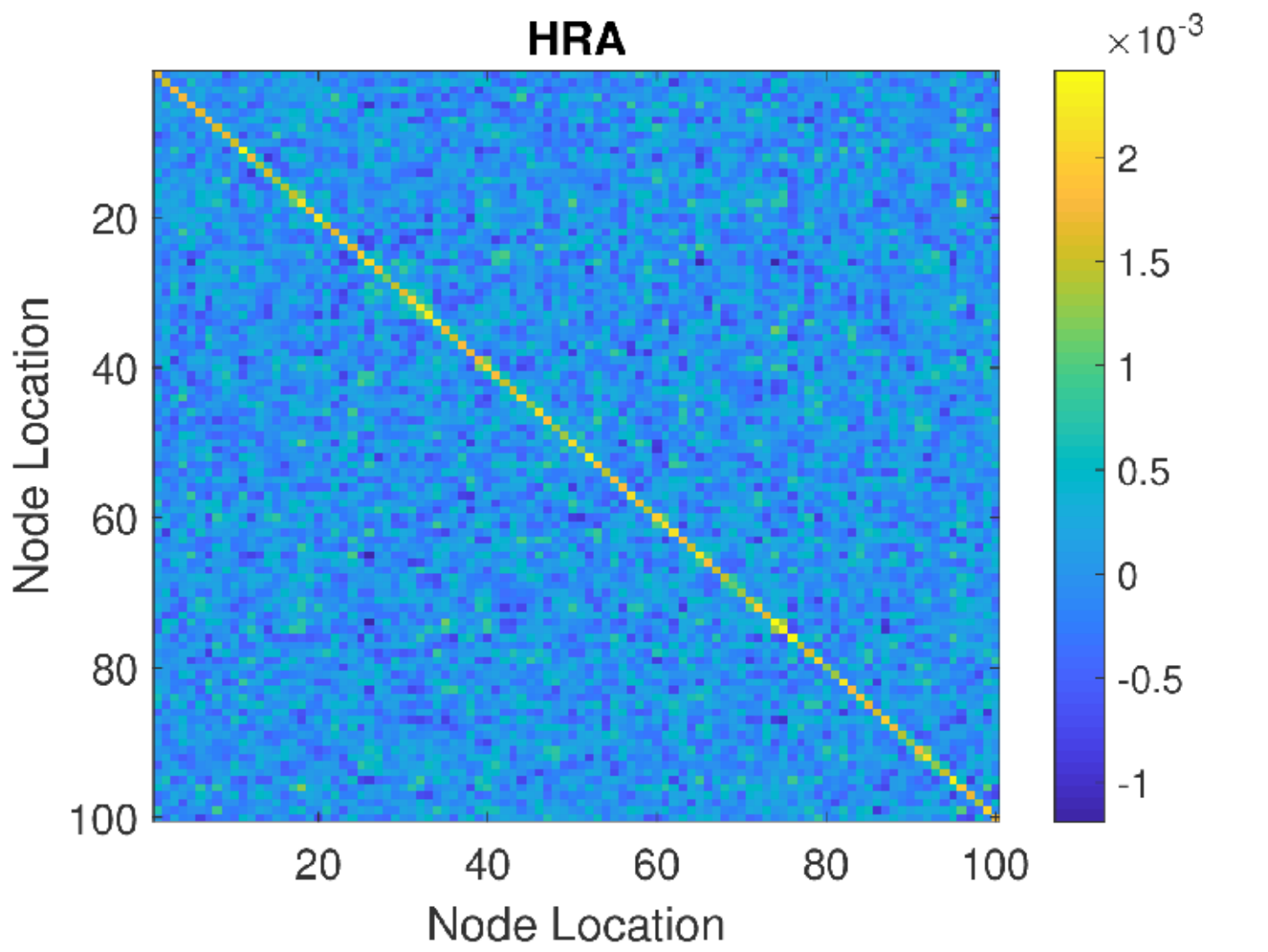}
        \label{KSMZZ}
    \end{subfigure}
    \caption{Examples of the error covariance matrix before the update step for the HR and HRA methods. The top row corresponds to the BGM model while the bottom row corresponds to the KSM model. Each row shows $\sigma_{u_i u_j}$ (left and middle columns) for the HR and HRA schemes respectively while the right column shows $\sigma_{z_i z_j}$ which only exists for the HRA scheme. }\label{covuuzz}
\end{figure*}

We study here the ensemble-based forecast error covariance matrices, ${\bf X}^{\rm f} {\bf X}^{{\rm f}^{\rm T}}$, with ${\bf X}^{\rm f}$ defined in Eq.~\eqref{errorcov}. The structure of the matrices is shown in Fig.~\ref{covuz} and~\ref{covuuzz}. The size of ${\bf X}^{\rm f}{\bf X}^{{\rm f}^{\rm T}}$ for the HRA method in this case is $200\times200$ while that for the HR case is $100\times100$. With the HR method we only have covariances between physical values themselves in contrast to HRA where we have covariances between physical values, physical values and node locations and the node locations themselves. In Fig.~\ref{covuz} we show the forecast covariances between the physical values and node locations for the HRA method with both the BGM and KSM models just before the 10th and 20th assimilation steps respectively.
These error covariance matrices correspond to no jitter or inflation in an effort to understand the intrinsic differences between the methods. Also shown is the gradient of the corresponding forecast mean and the associated covariance between the physical values $u_i$ and their node location $z_i$, {\it i.e.} the diagonal of the matrix. This is done to highlight that the largest covariances occur at sharp gradients and have the same sign. This is natural since a negative gradient would imply a negative correlation between a physical value and its independent variable, likewise for a positive gradient. In fact, the shape of the diagonal closely matches that of the gradient demonstrating that including the node locations in the state vector encodes a deeper level of information into the Kalmain gain matrix. 

 In Fig.~\ref{covuuzz} we show the covariances for the physical values for  HR and HRA ({\it i.e.} the top left $100\times100$ block) as well as the HRA covariances of the node locations ({\it i.e.} the bottom right $100\times100$ block) for both the BGM and KSM models. Typically the HR covariances are higher in magnitude than that of the HRA scheme, this is because the ensemble members are compared on the same mesh in conjunction with the effect of the intrinsic stochasticity caused by the mapping to and from the reference mesh. The shock is immediately identifiable in the physical value covariances for the BGM model as a bright spot near the sharp gradient ({\it cf} Fig.~\ref{covuz}). The sharp gradients of the KSM model are also apparent in the physical error covariances. For the KSM model, there is a strong, albeit regular structure in the matrix for the HR method resulting form the fixed mesh with some long distance cross-correlations. Those long distance correlations are greatly reduced in the HRA. The correlations between the node locations themselves in the HRA scheme (rightmost panels) are very small due to the fact that they are not very far from each other since the intervals themselves are very small. This means that the extra contribution to the innovation in the HRA scheme is mainly coming from the correlations between the physical values and the node locations as opposed to the node locations themselves. This is indeed desirable since we need to inject new nodes in the embedding process and would prefer to avoid incidental biases. 

\subsection{Ensemble Member Fidelity}
As discussed earlier the addition of jitter can disrupt the shape of the ensemble members while still improving the analysis mean. This may be problematic if the ensemble members are used to feed information to another model component.  

As an example where ensemble member fidelity may be important, we consider the Heterogeneous Multiscale Method (HMM) described for various applications in \citet{e2011principles}. In a general setting, the HMM method connects a macro scale model with parameters dependent on micro scale variables to a model of this micro variables in order to simulate a physical process. Typically one has a macro scale model $F(U,D)$ where $U$ is the physical macro scale variable and $D$ the data needed in order for the macro scale model to be complete, a stress tensor for example. Paired with the macro scale model we have a micro scale model $f(u,d)=0$ and $d=d(U)$ where $d$ is the data needed to set up the micro scale model and is dependent on the macro scale state. 
Typically the HMM process proceeds as follows.
\begin{enumerate}
    \item Given the current state of the macro variables, initialise the micro variables using the needed micro model data $d=d(U)$.
    \item Evolve the micro variables for some micro model time steps.
    \item Through the appropriate method, calculate $D$, needed for the macro model.
    \item Evolve the macro variables using the macro-solver. 
\end{enumerate}
In this setting if an ensemble member is solution of either the macro or micro models disruption in their fidelity will naturally cause a problem with steps 1. and 3.  through an inaccurate calculation of $d$ or $D$ and likely propagating such errors in the evolution steps. An example of a system like this can be found in Cloud-Resolving Convection Parameterization (CRCP) \citet{Grabowski2001}. There, a macro model solving inviscid moist equations is coupled with a micro model representing sub grid scale cloud physics.  

\begin{figure*}
    \centering
   {\bf HR Scheme}\par \medskip
    \begin{subfigure}[b]{0.33\textwidth}
        \includegraphics[width=\textwidth]{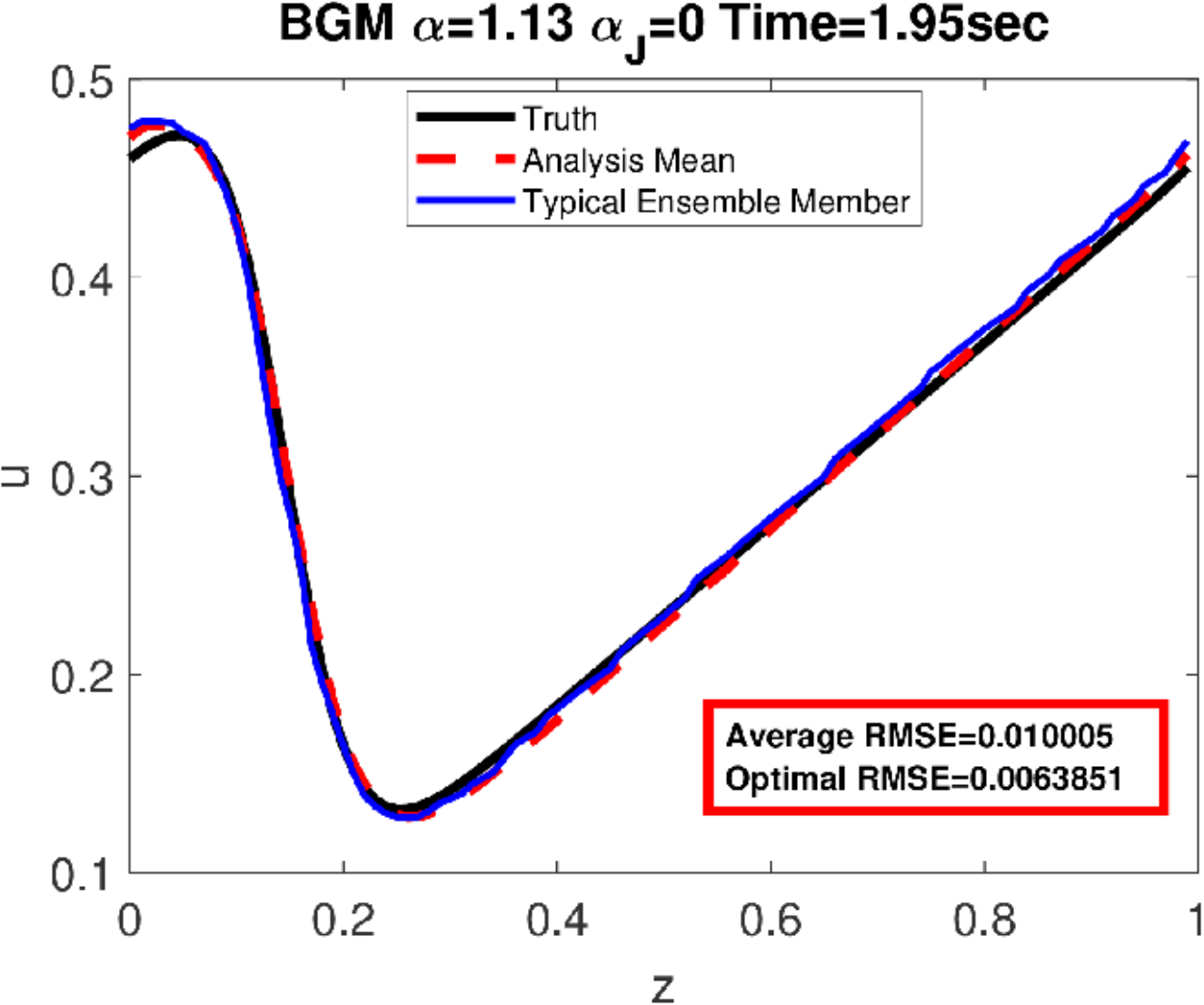}
         \label{BGMEXJ_0HR}
      \end{subfigure}
   \begin{subfigure}[b]{0.33\textwidth}
       \includegraphics[width=\textwidth]{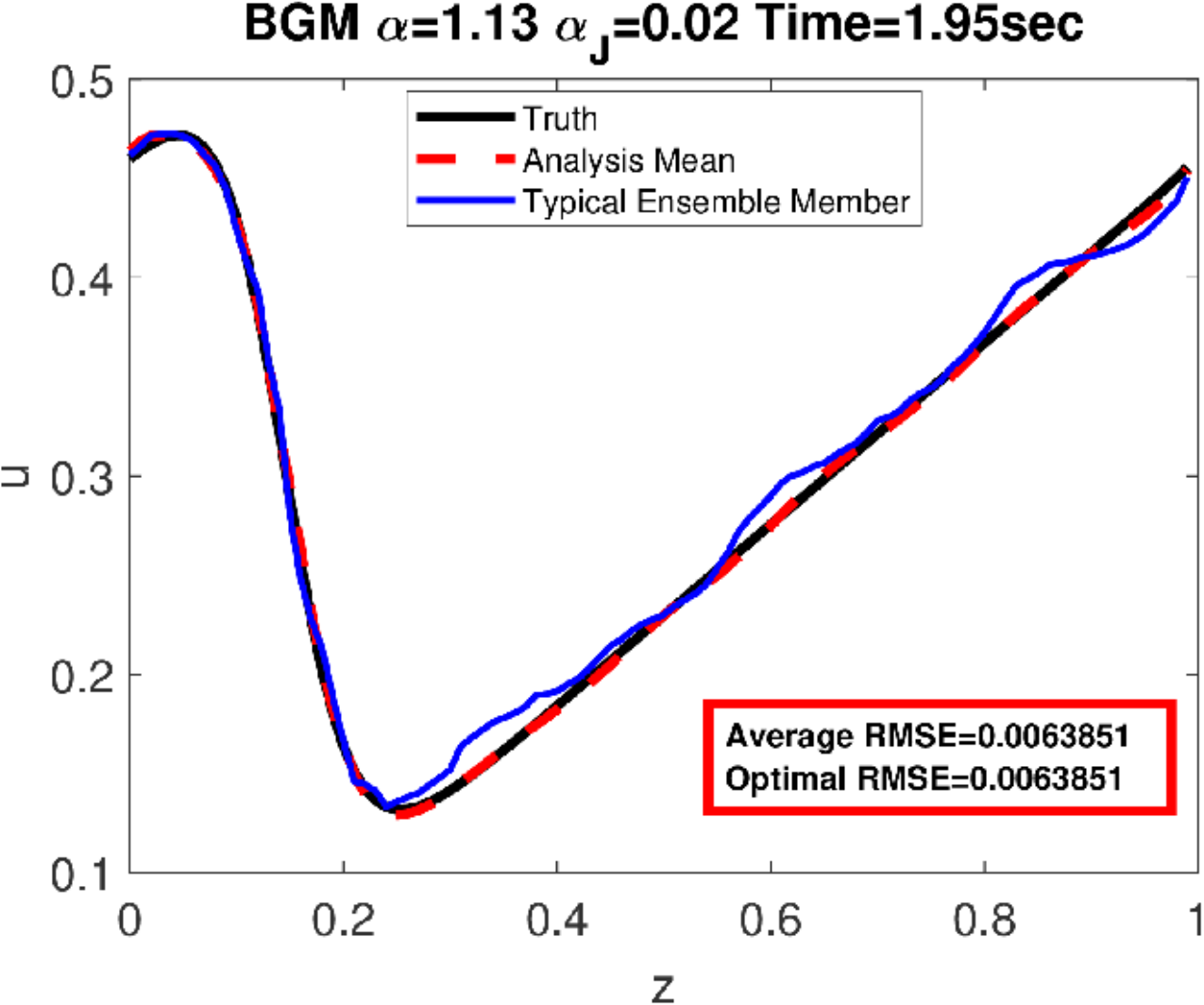}
        \label{BGMEXJ_0_02HR}
    \end{subfigure}
     \begin{subfigure}[b]{0.33\textwidth}
       \includegraphics[width=\textwidth]{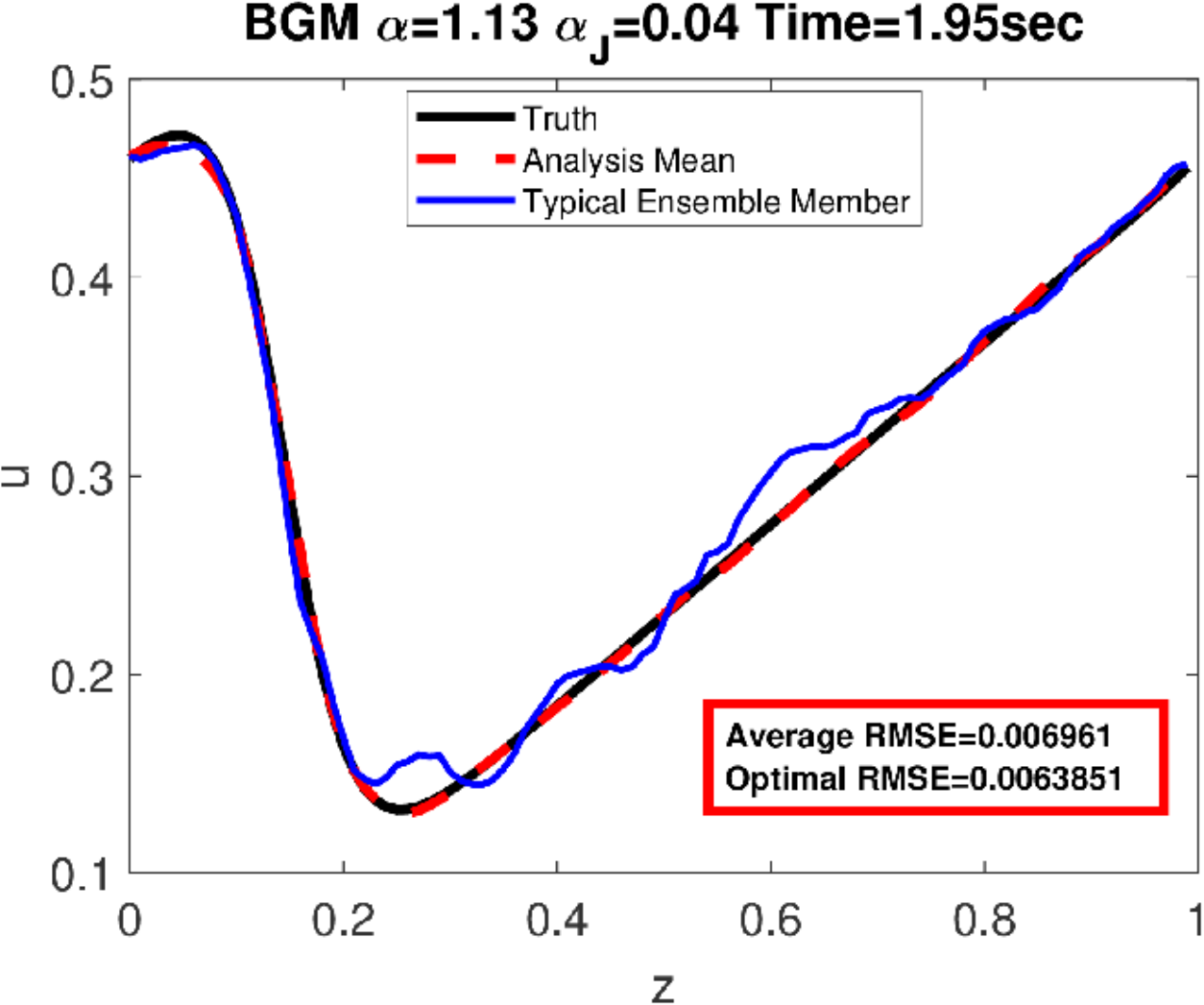}
        \label{BGMEXJ_0_04HR}
    \end{subfigure}
    
   {\bf HRA Scheme}\par \medskip
        \begin{subfigure}[b]{0.33\textwidth}
        \includegraphics[width=\textwidth]{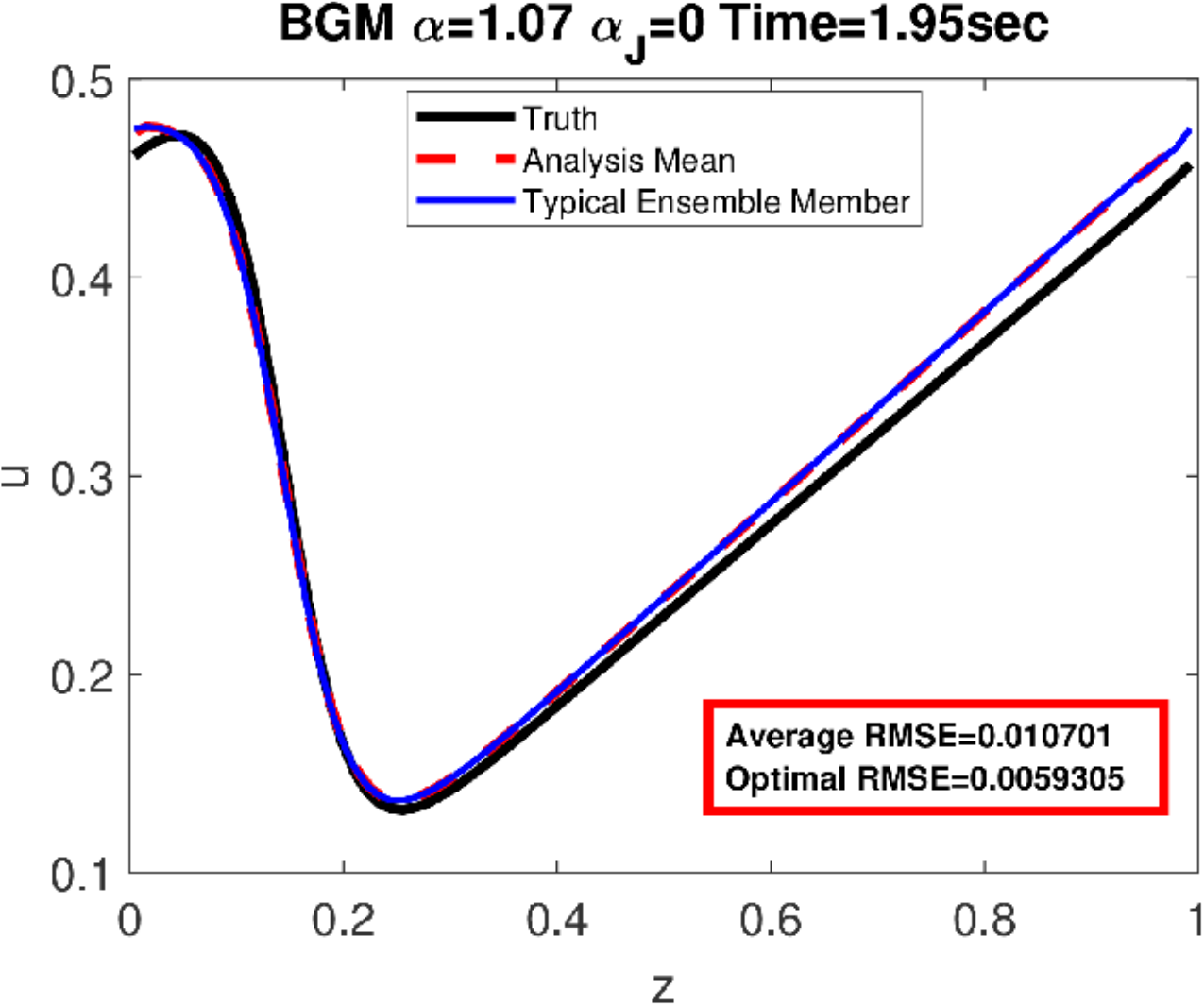}
         \label{BGMEXJ_0HRA}
      \end{subfigure}
   \begin{subfigure}[b]{0.33\textwidth}
       \includegraphics[width=\textwidth]{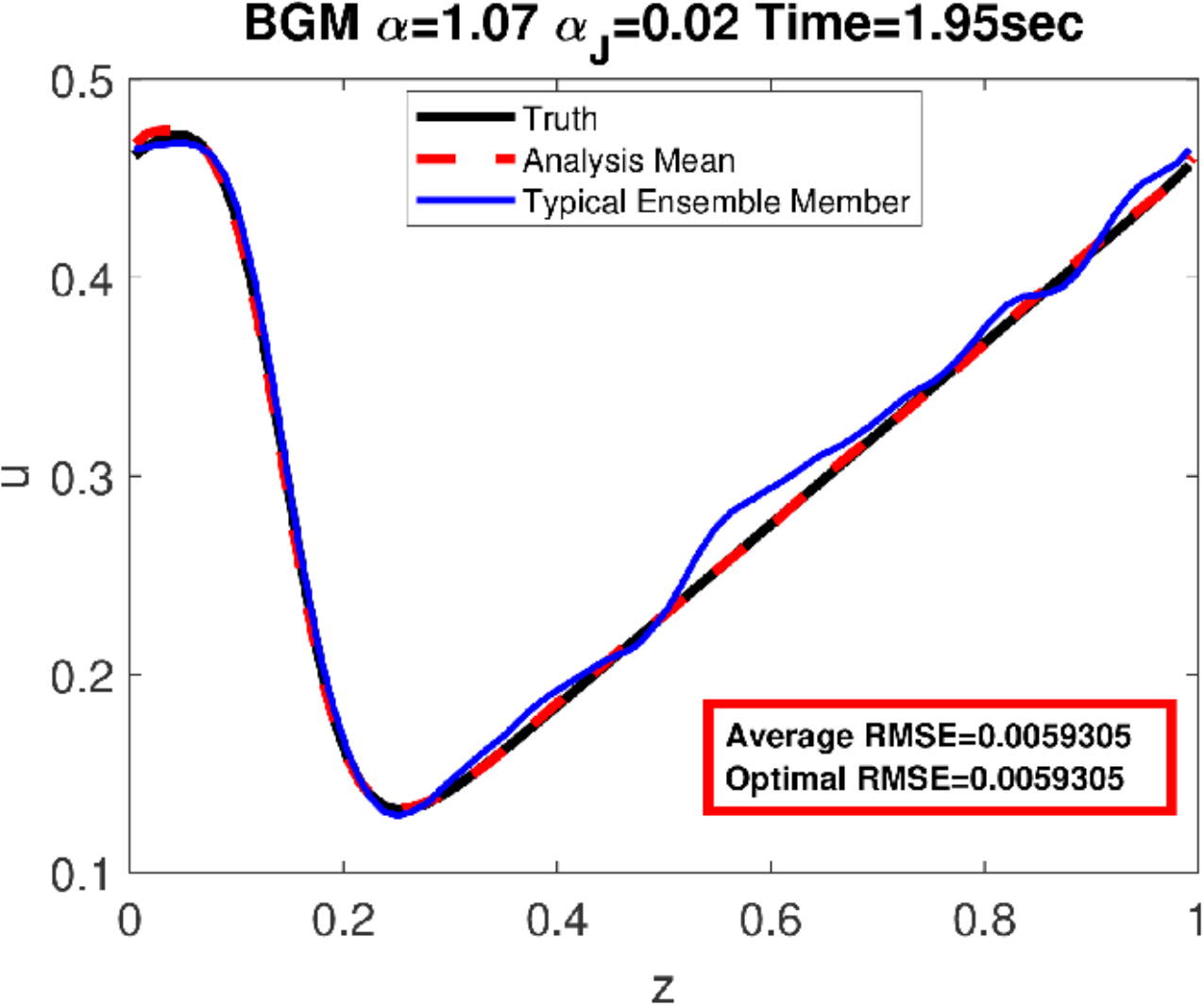}
        \label{BGMEXJ_0_02HRA}
    \end{subfigure}
     \begin{subfigure}[b]{0.33\textwidth}
       \includegraphics[width=\textwidth]{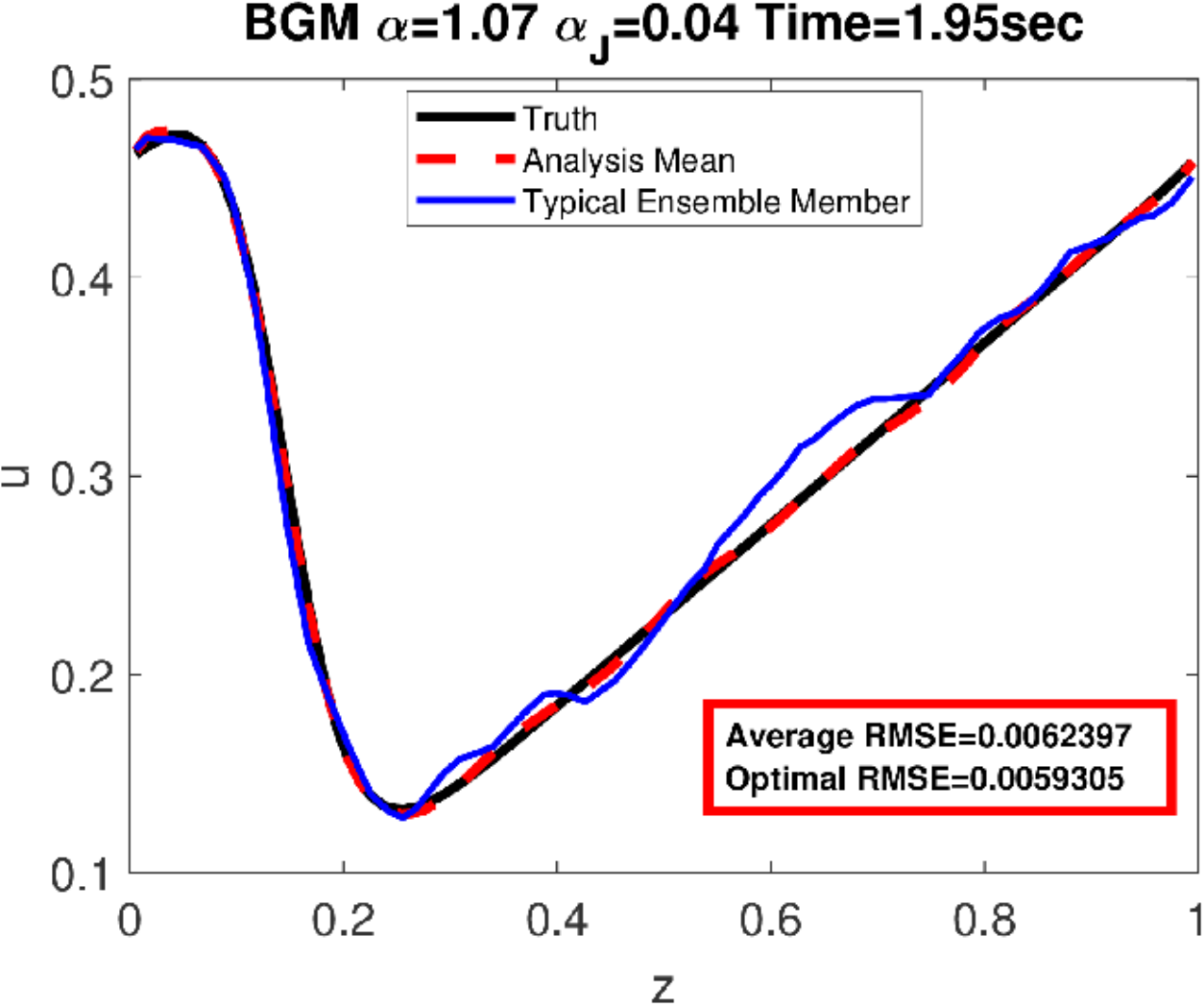}
        \label{BGMEXJ_0_04HRA}
    \end{subfigure}

   \caption{Examples of BGM analysis and an ensemble member for different inflation and jitter choices. Top HR and bottom HRA. The RMSE of the analysis mean can be low even when the ensemble members themselves represent unrealistic solutions to the PDE. }\label{wiggles}
\end{figure*}

We saw in Fig.~\ref{BGMsurf} and
\ref{KSMsurf} that there are regions of low analysis mean RMSE for relatively high values of $\alpha_J$, suggesting that the mean is smoothing out the added noise from jitter. In Fig.~\ref{wiggles} we show examples of the analysis mean, truth, and a typical ensemble member of the BGM model for a fixed inflation and three values of $\alpha_J$. The inflation chosen corresponds to the optimal value found for an ensemble size of 50. We show $\alpha_J=0$, the optimal $\alpha_J$ (in terms of lowest time averaged RMSE), and a larger $\alpha_J$ for which the ensemble mean still has low time averaged RMSE. The figure clearly shows that applying jitter to the ensemble members has the potential to disrupt them (see the waving profile of the displayed arbitrarily chosen ensemble member), especially if your scheme requires you to act on each ensemble member as we do here with dimension matching and return. Depending on the application, such as a model using the HMM framework, it may be better to sacrifice a small amount of analysis accuracy to preserve the fidelity of each ensemble member in terms of representing a valid solution to the underlying PDE. In other applications, that may not matter quite so much.

\begin{figure*}
\centering
{\bf Kurtosis Examples}\par \medskip

      \begin{subfigure}[b]{0.45\textwidth}
        \includegraphics[width=\textwidth]{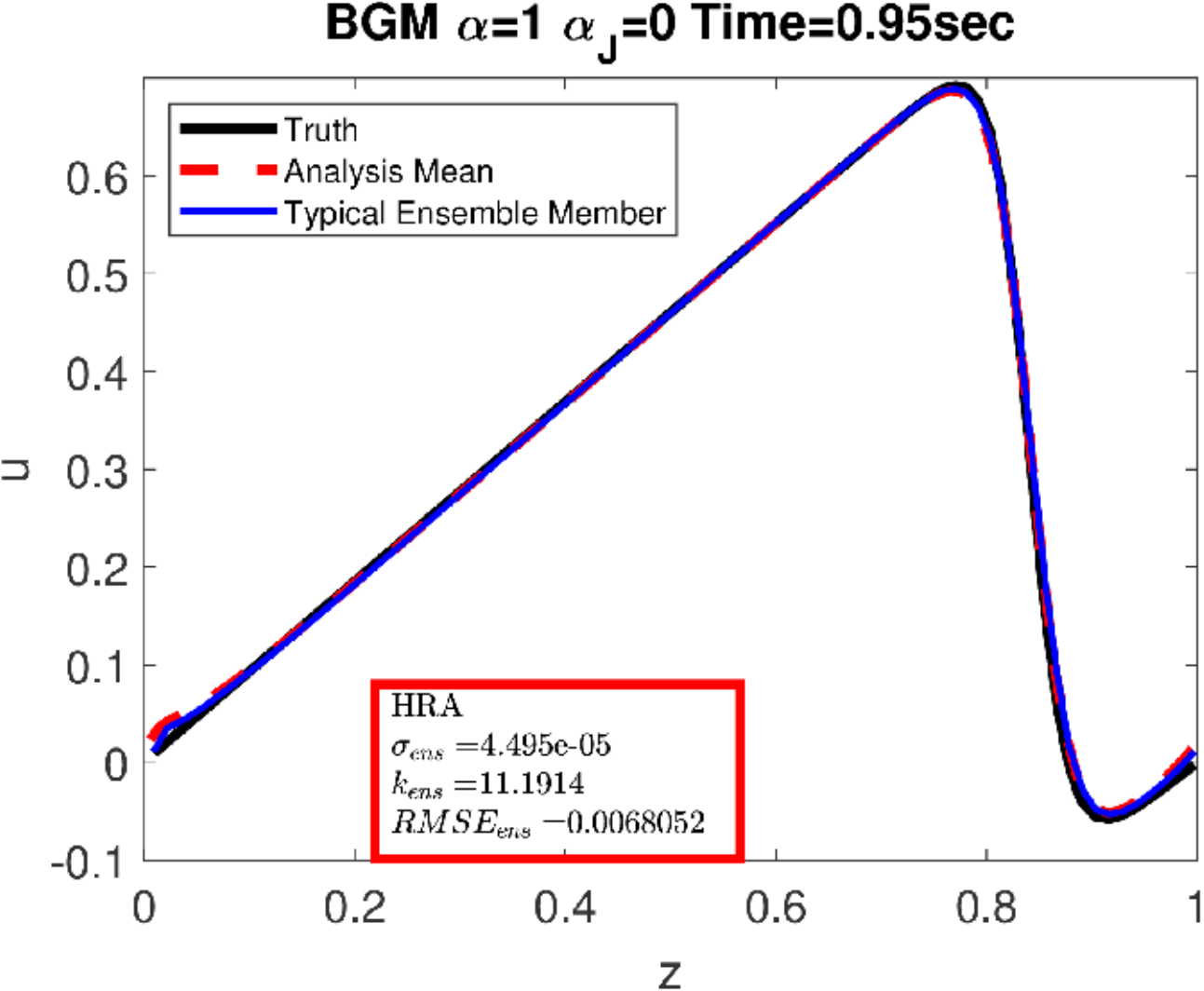}
         \label{HRALOWKURT}
      \end{subfigure}
     \begin{subfigure}[b]{0.45\textwidth}
       \includegraphics[width=\textwidth]{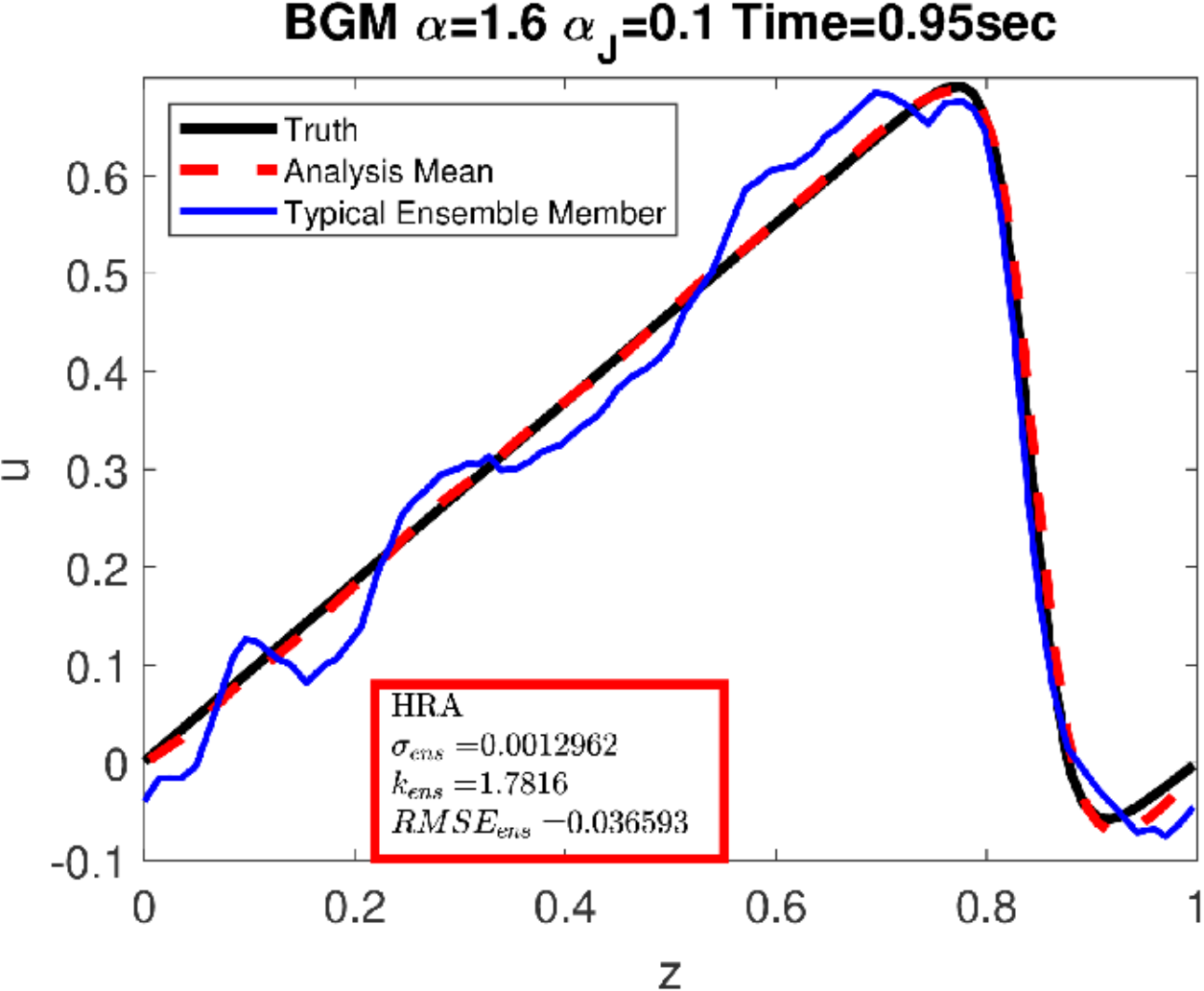}
        \label{HRAHIGHKURT}
    \end{subfigure}

  \caption{Examples of the three statistical measures we use to measure curve distortion.}\label{KURTEX}
\end{figure*}

\begin{figure*}
\centering
 {\bf Extra Metrics Forecast Members (BGM)}\par\medskip
\begin{subfigure}[b]{0.33\textwidth}
\includegraphics[width=\textwidth]{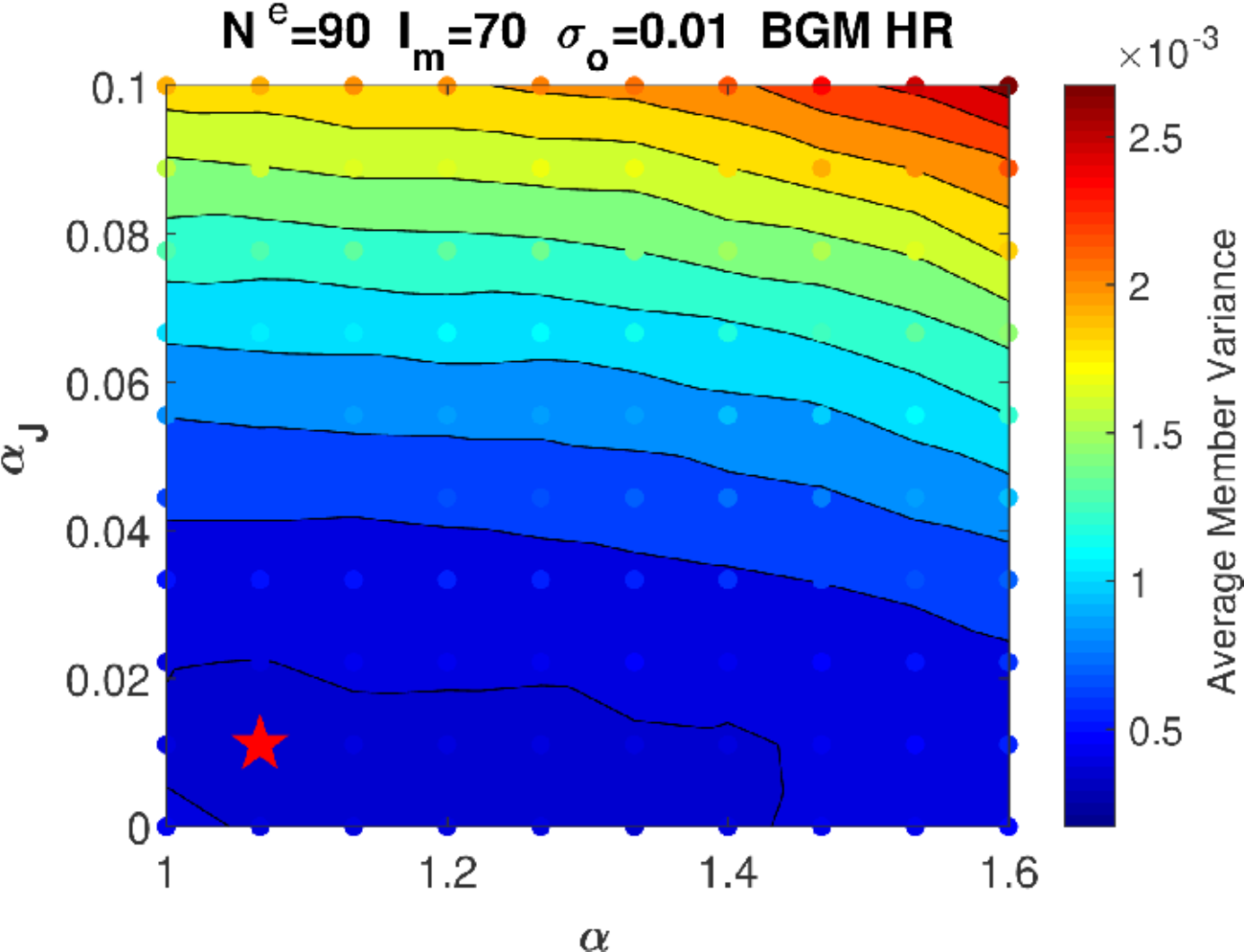}
         \label{VARBGMHRSURFFC}
      \end{subfigure}
   \begin{subfigure}[b]{0.33\textwidth}
       \includegraphics[width=\textwidth]{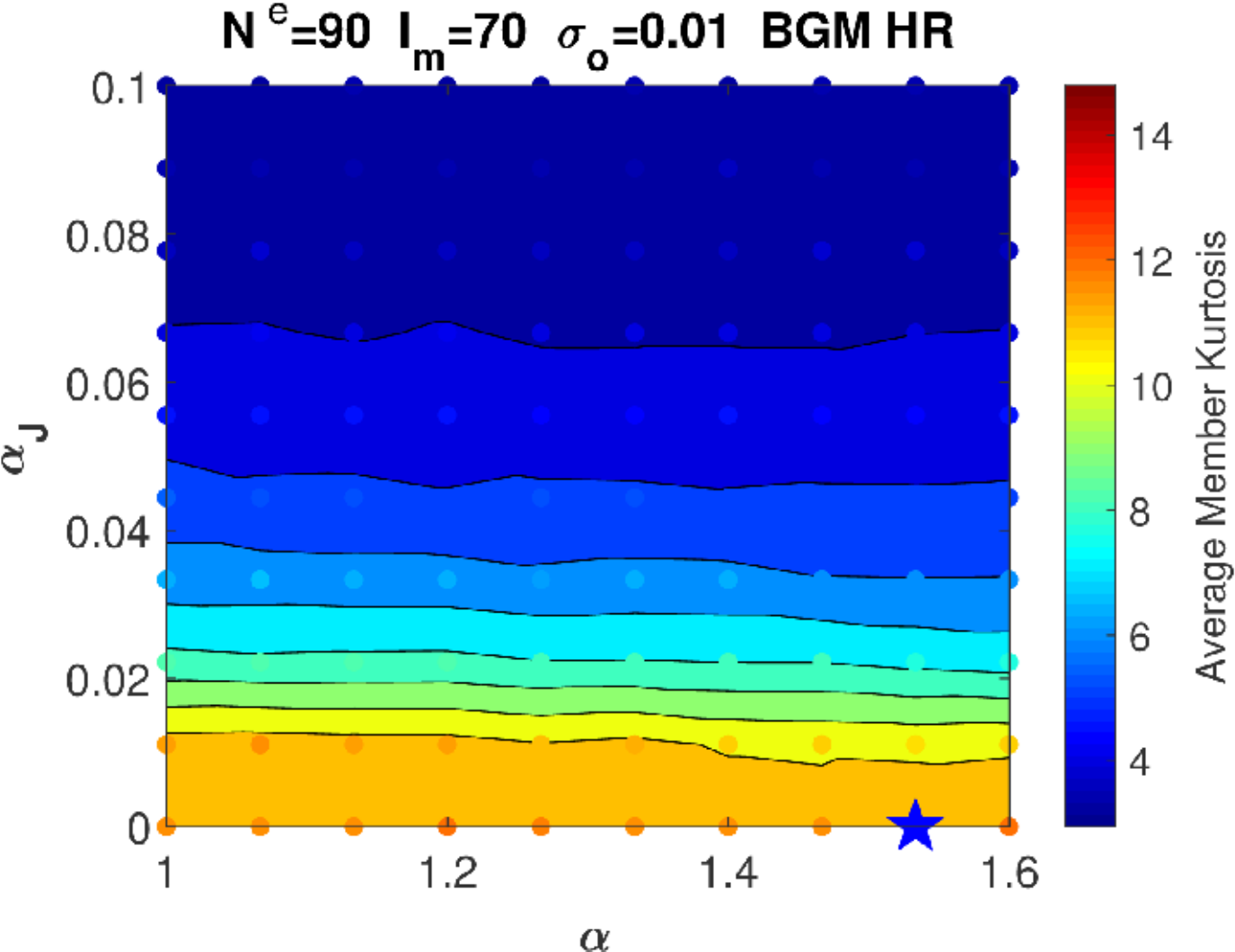}
        \label{KURTBGMSHRURFFC}
    \end{subfigure}
     \begin{subfigure}[b]{0.33\textwidth}
       \includegraphics[width=\textwidth]{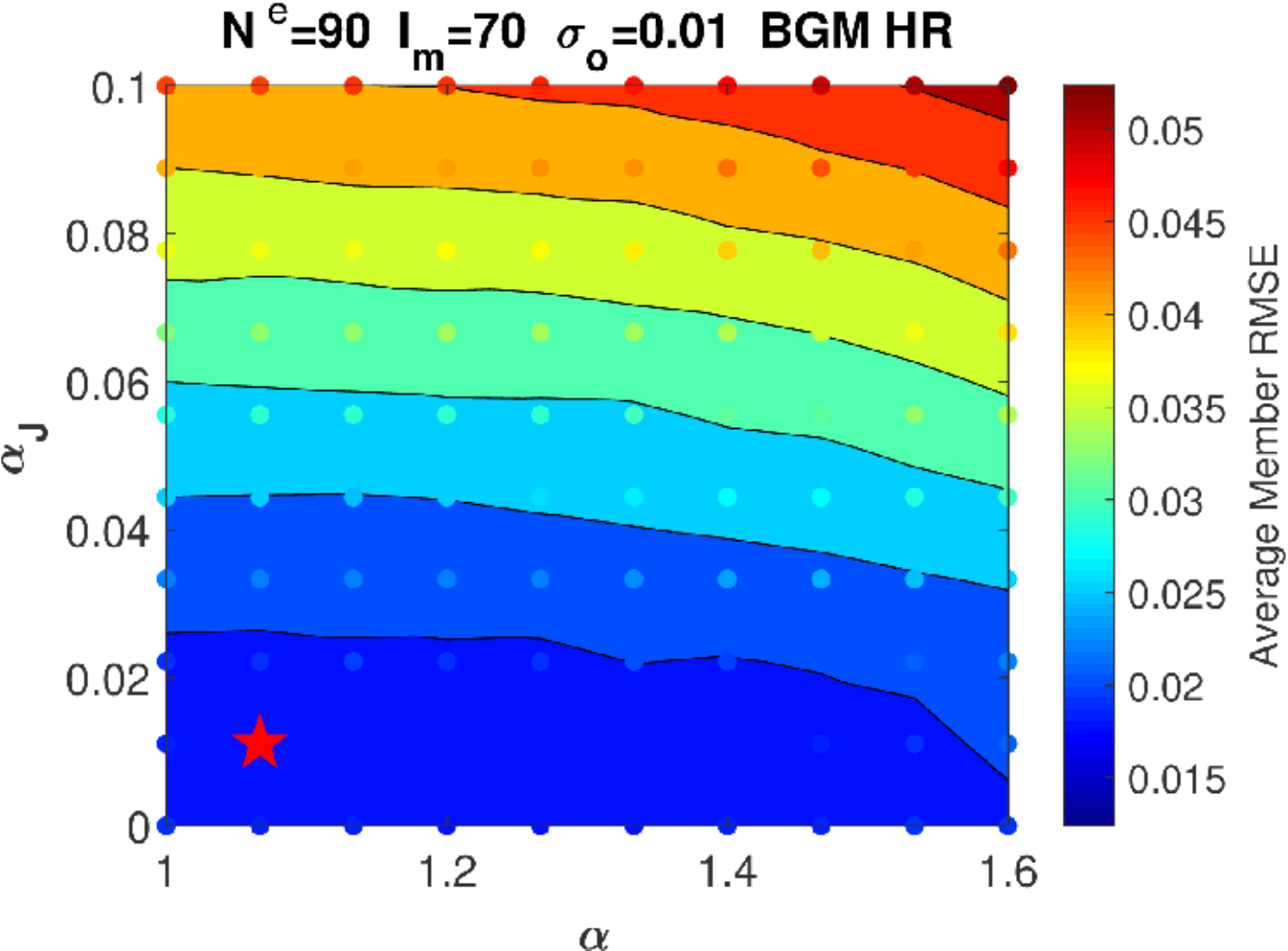}
        \label{RMSEBGMHRSURFFC}
    \end{subfigure}
    
        \begin{subfigure}[b]{0.33\textwidth}
        \includegraphics[width=\textwidth]{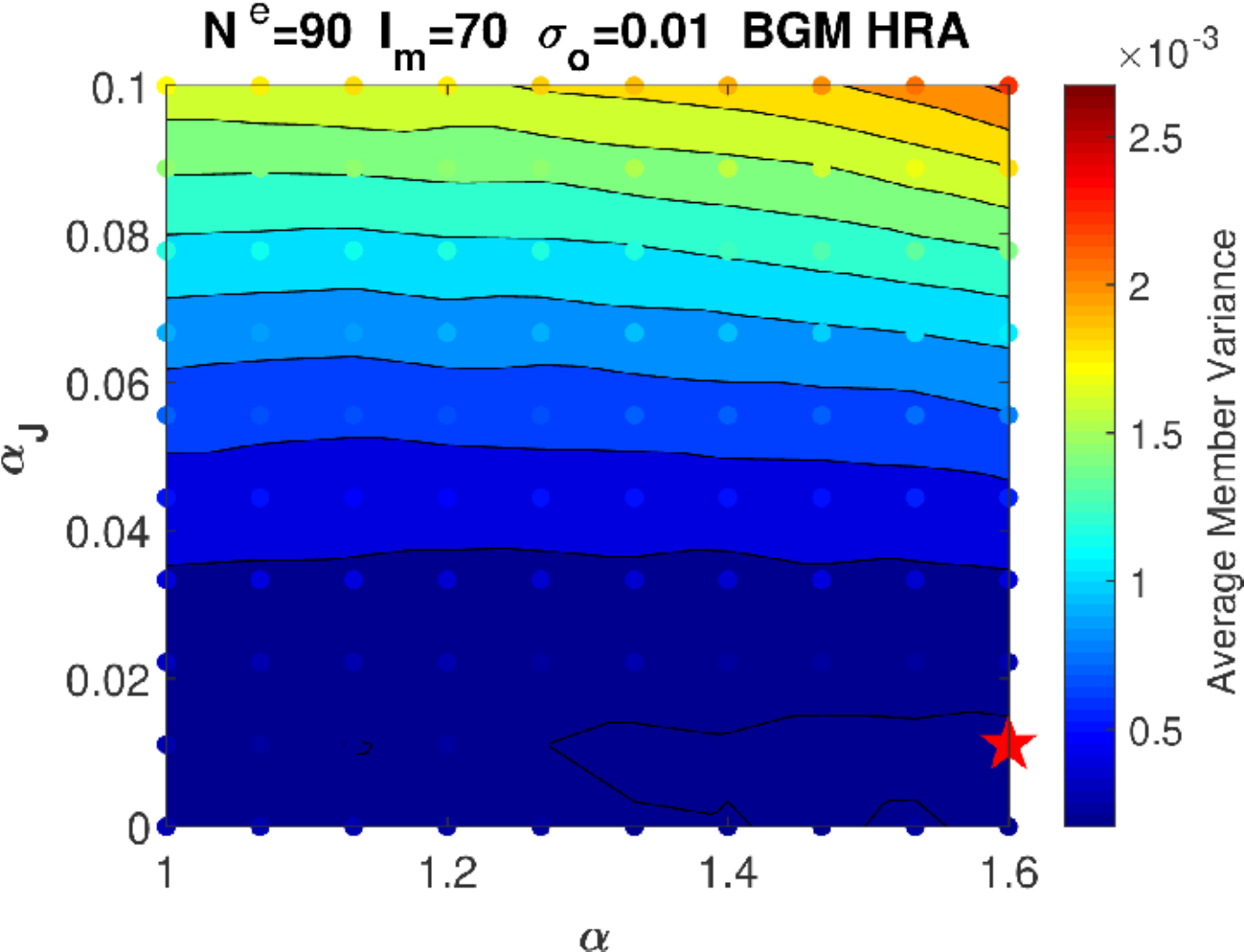}
         \label{VARBGMHRASURFFC}
      \end{subfigure}
   \begin{subfigure}[b]{0.33\textwidth}

       \includegraphics[width=\textwidth]{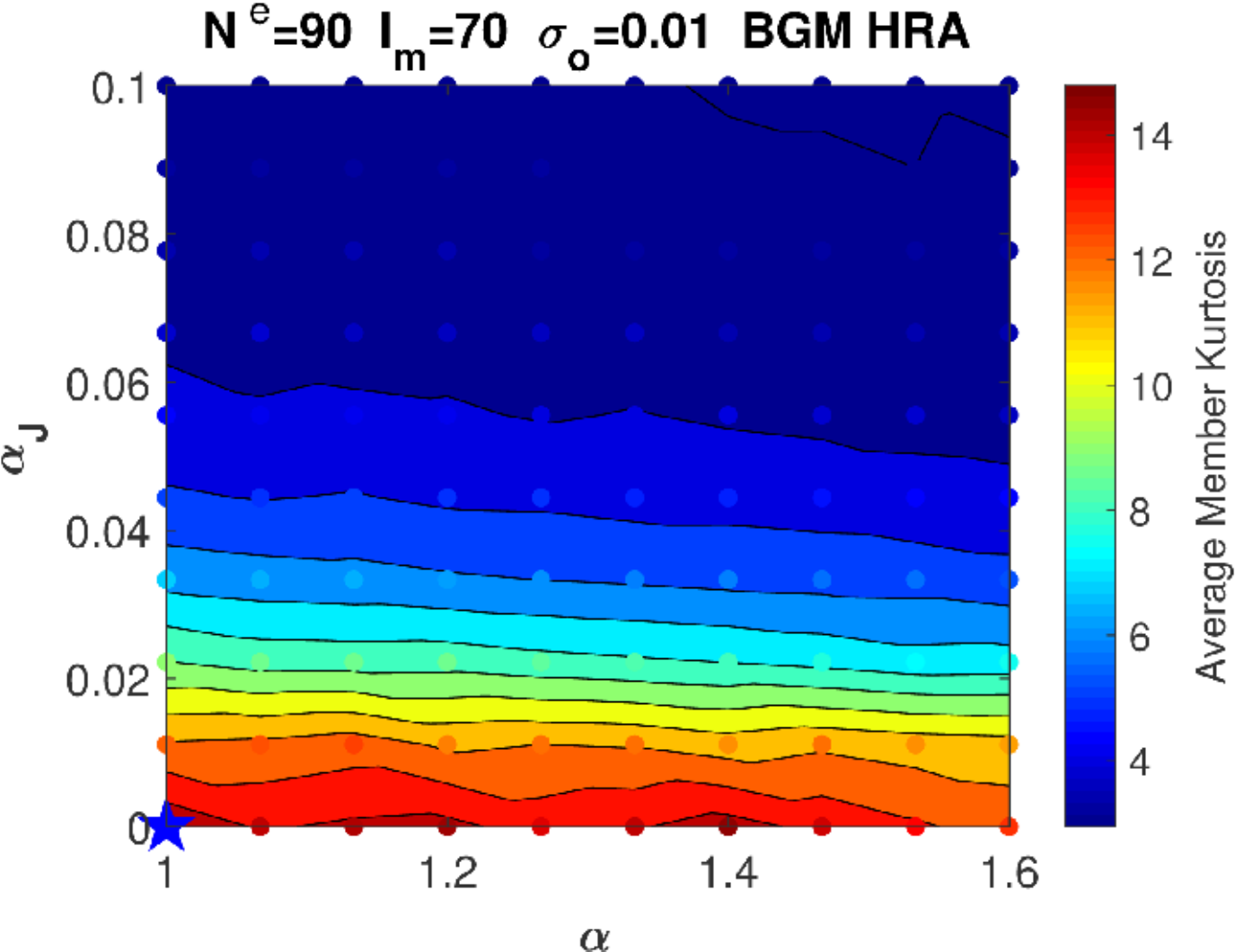}
        \label{VARBGMSHRAURFFC}
    \end{subfigure}
     \begin{subfigure}[b]{0.33\textwidth}
       \includegraphics[width=\textwidth]{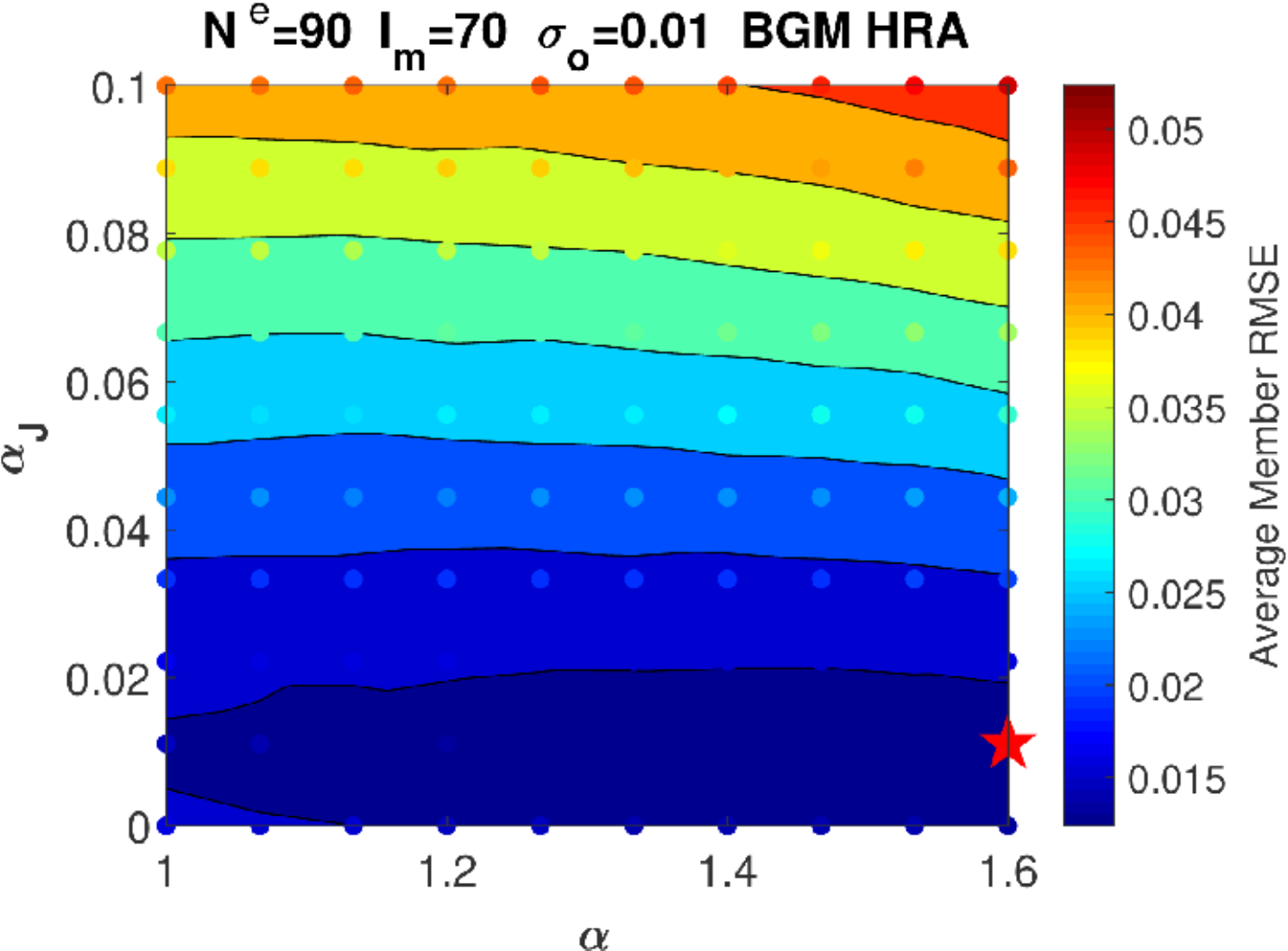}
        \label{RMSEBGMHRASURFFC}
    \end{subfigure}
     \caption{Surfaces for $\sigma_{ens}$, $K_{ens}$ and RMSE$_{ens}$ as a function of jitter and inflation for the BGM model using the forecast ensemble members, right before update, for HR (top row) and HRA (bottom row). It is notable that both $\sigma_{ens}$ and RMSE$_{ens}$ increase primarily as a function of jitter ($\alpha_J$) with less dependence on multiplicative inflation ($\alpha$) while the same is true for decreasing kurtosis evidenced by horizontal contours. This identifies the jitter as the primary source of ensemble member distortion. The red stars represent lowest values and blue stars highest values.}\label{BGMextraFC}
\end{figure*}

\begin{figure*}
    \centering
     {\bf Extra Metrics Analysis Members (BGM)}\par\medskip
    \begin{subfigure}[b]{0.33\textwidth}
        \includegraphics[width=\textwidth]{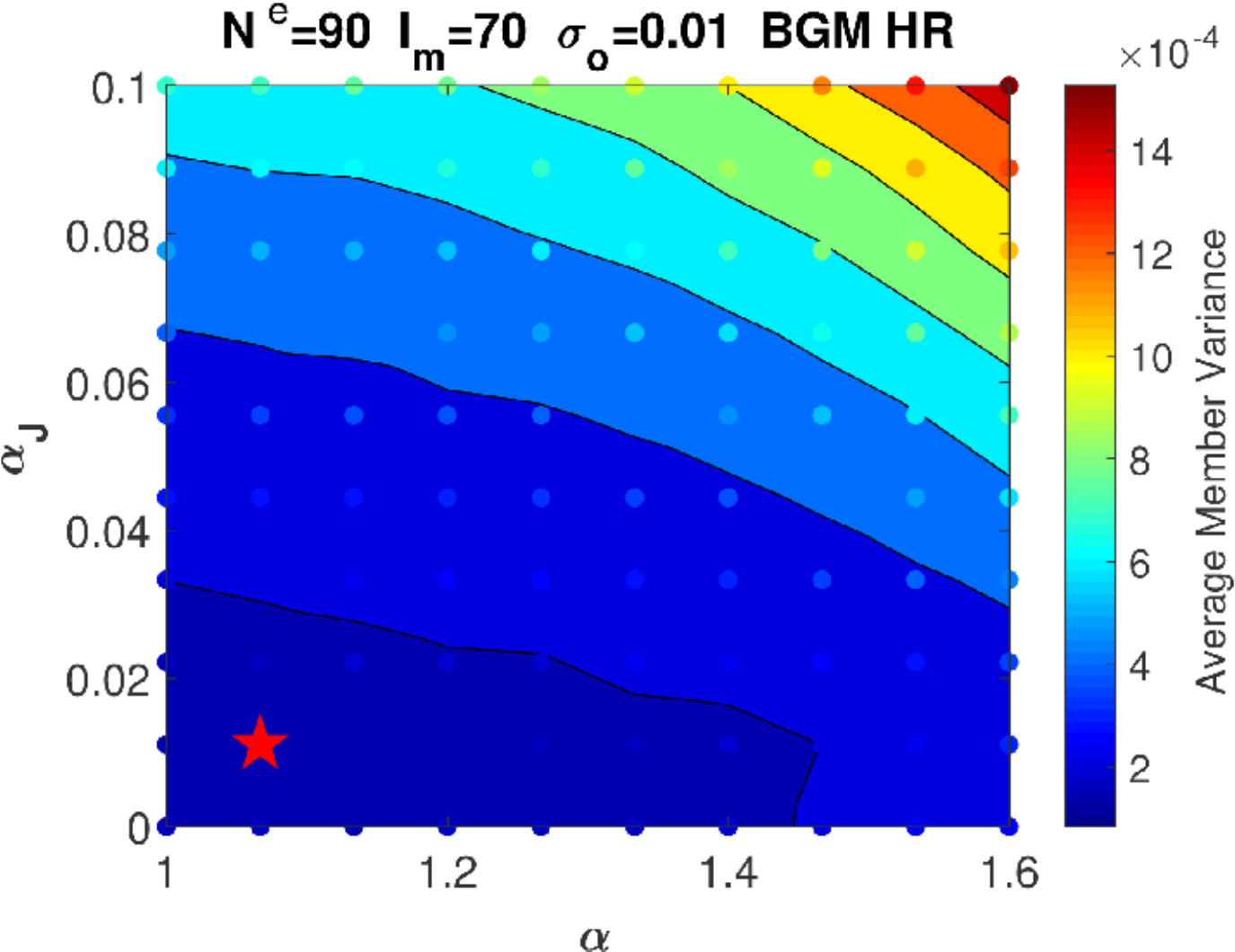}
         \label{VARBGMHRSURFAN}
      \end{subfigure}
   \begin{subfigure}[b]{0.33\textwidth}
       \includegraphics[width=\textwidth]{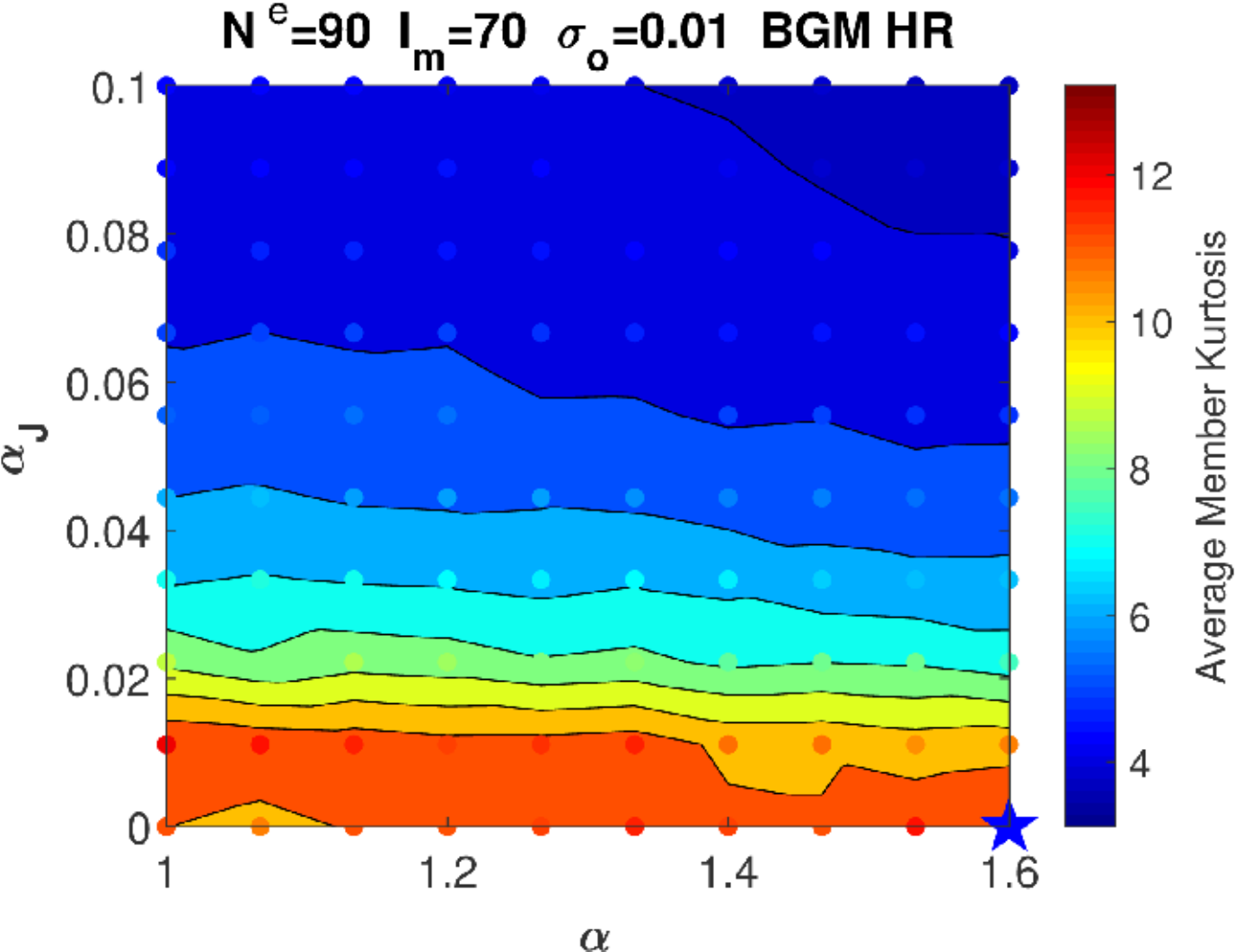}
        \label{KURTBGMSHRURFAN}
    \end{subfigure}
     \begin{subfigure}[b]{0.33\textwidth}
       \includegraphics[width=\textwidth]{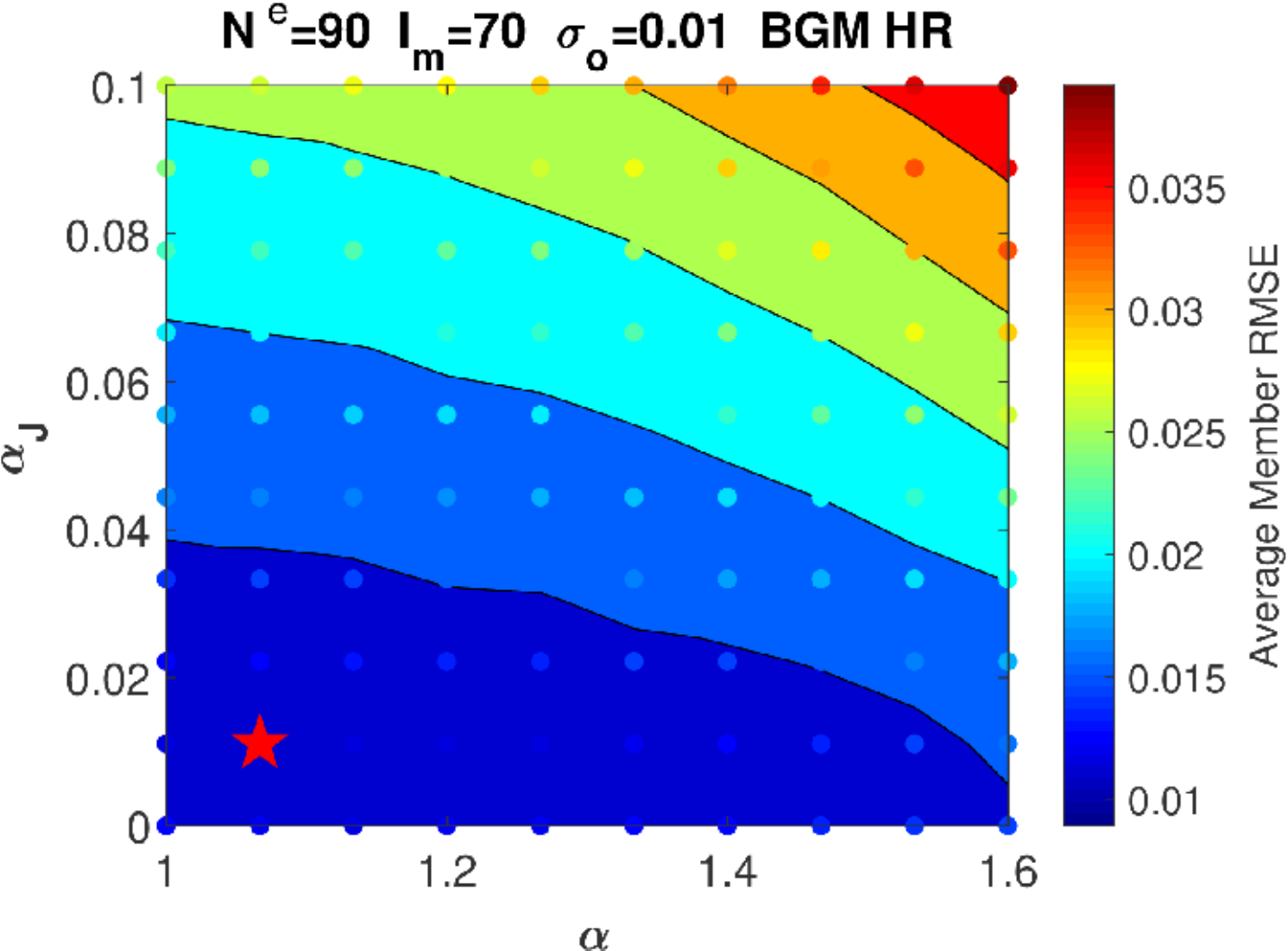}
        \label{RMSEBGMHRSURFAN}
    \end{subfigure}
    
        \begin{subfigure}[b]{0.33\textwidth}
        \includegraphics[width=\textwidth]{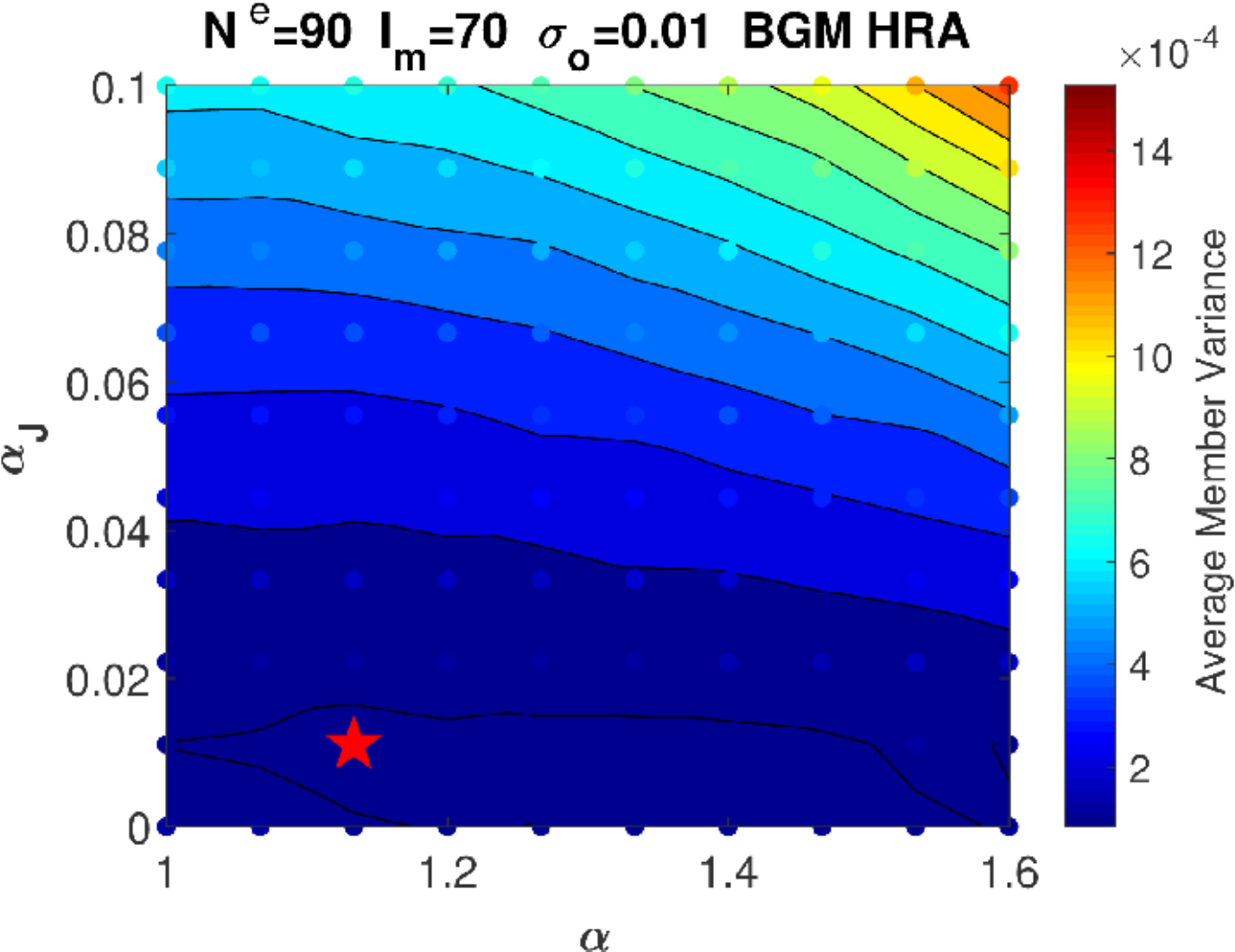}
         \label{VARBGMHRASURFAN}
      \end{subfigure}
   \begin{subfigure}[b]{0.33\textwidth}
       \includegraphics[width=\textwidth]{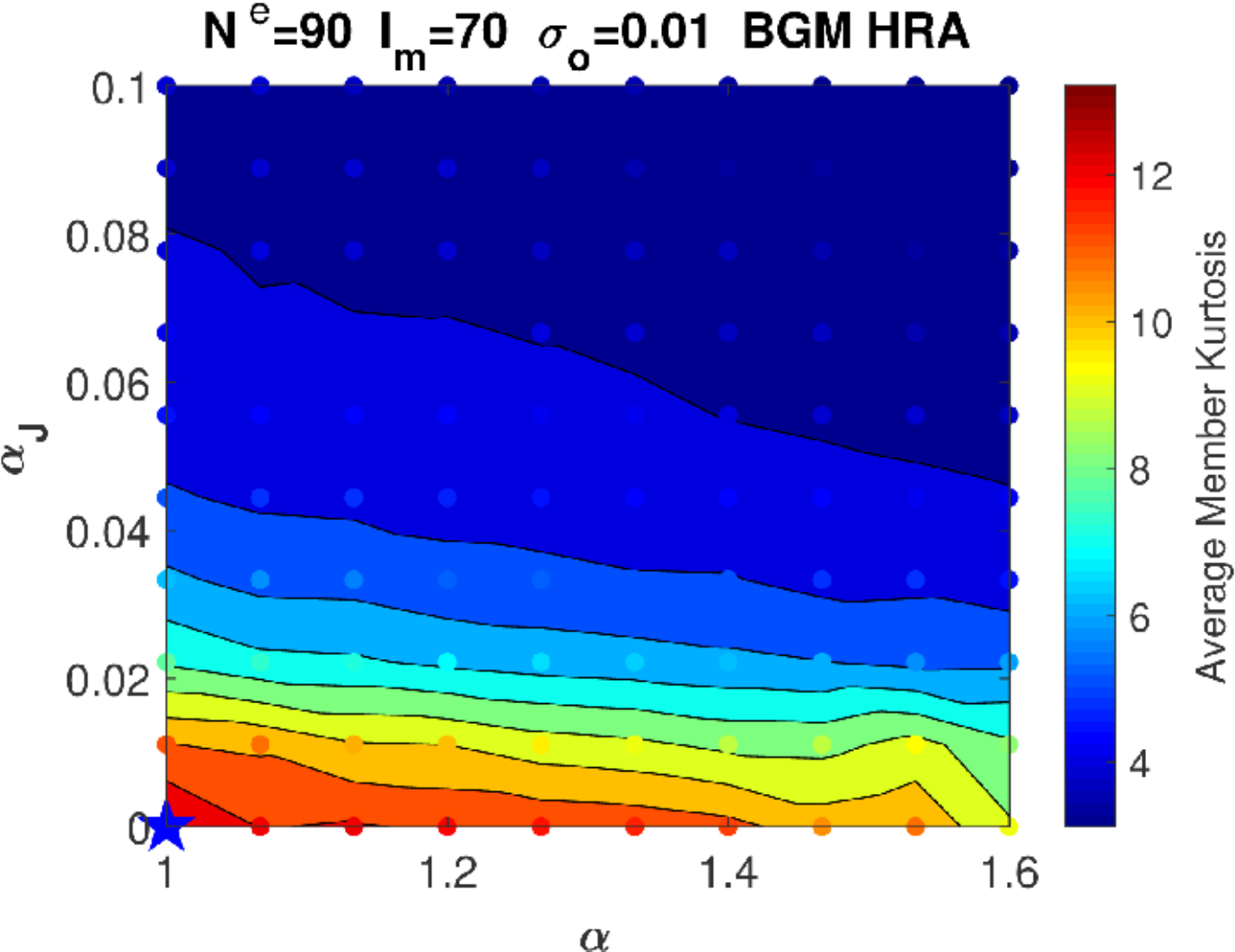}
        \label{VARBGMSHRAURFAN}
    \end{subfigure}
     \begin{subfigure}[b]{0.33\textwidth}
       \includegraphics[width=\textwidth]{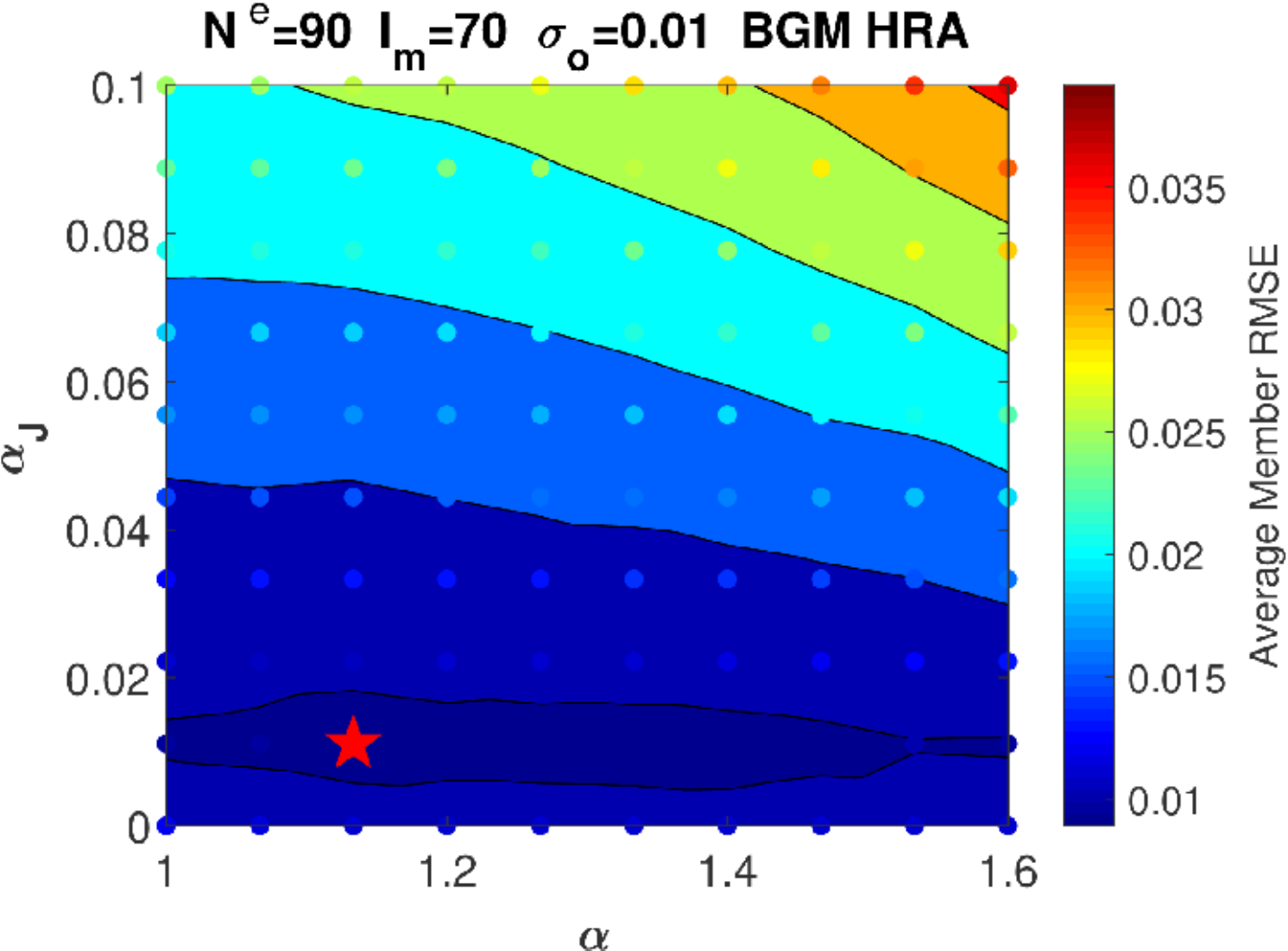}
        \label{RMSEBGMHRASURFAN}
    \end{subfigure}
     \caption{Surfaces for $\sigma_{ens}$, $K_{ens}$ and RMSE$_{ens}$ as a function of jitter and inflation for the BGM model using the analysis ensemble members for HR (top row) and HRA (bottom row). It is notable that horizontal contours shown in the same analysis for the forecast members are now slanted with improvements in $\sigma_{ens}$ and RMSE$_{ens}$ for lower values of inflation ($\alpha$). This is due to the update step, however the Kurtosis remains relatively unchanged implying that the members are still distorted for larger values of $\alpha_J$ and only the scale of the errors has been reduced. The red stars represent lowest values and blue stars highest values.}\label{BGMextraAN}
\end{figure*}

\begin{figure*}
    \centering
     {\bf Extra Metrics Forecast Members (KSM)}\par\medskip
    \begin{subfigure}[b]{0.33\textwidth}
        \includegraphics[width=\textwidth]{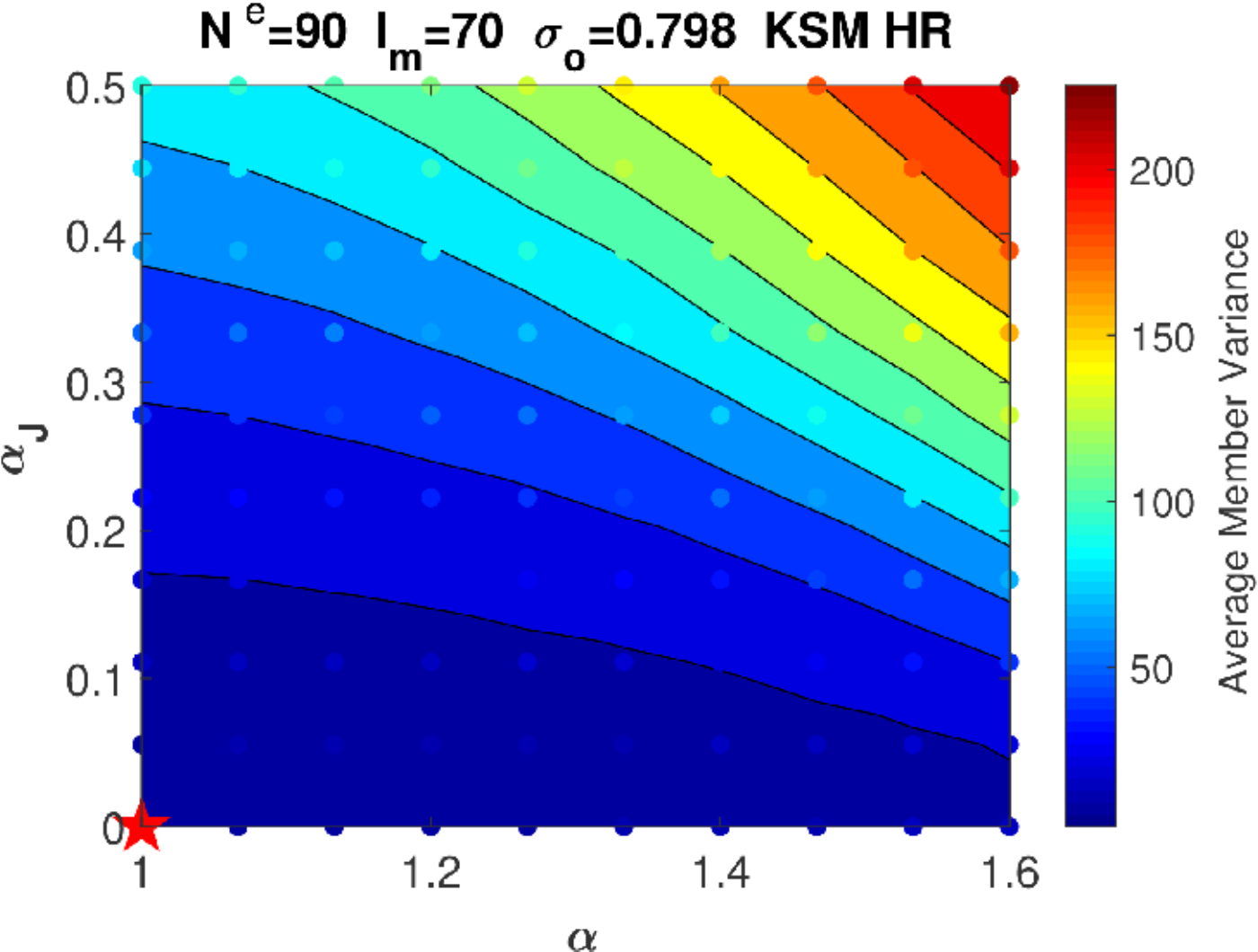}
         \label{VARKSMHRSURFFC}
      \end{subfigure}
   \begin{subfigure}[b]{0.33\textwidth}
       \includegraphics[width=\textwidth]{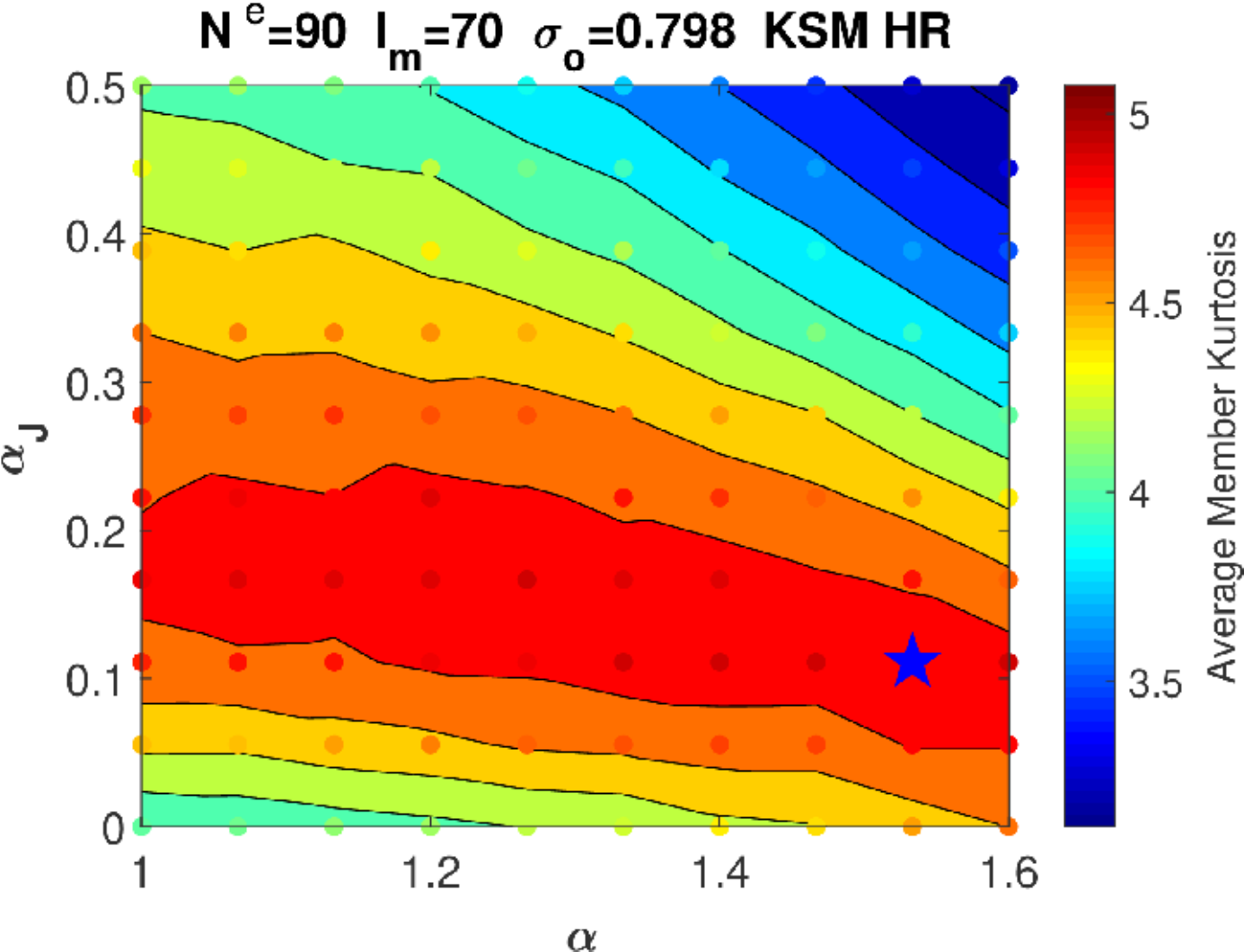}
        \label{KURTKSMSHRURFFC}
    \end{subfigure}
     \begin{subfigure}[b]{0.33\textwidth}
       \includegraphics[width=\textwidth]{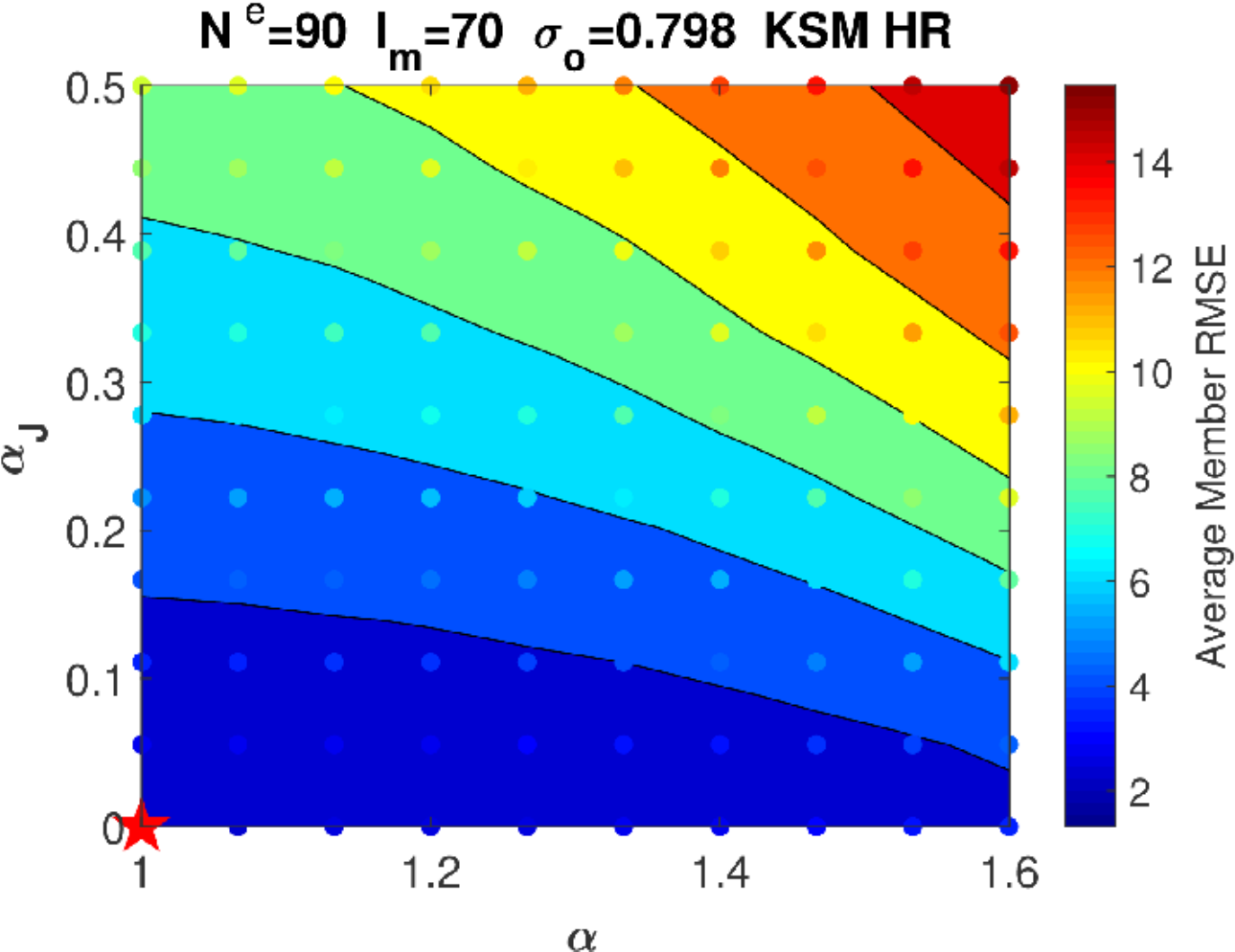}
        \label{RMSEKSMHRSURFFC}
    \end{subfigure}
    
        \begin{subfigure}[b]{0.33\textwidth}
        \includegraphics[width=\textwidth]{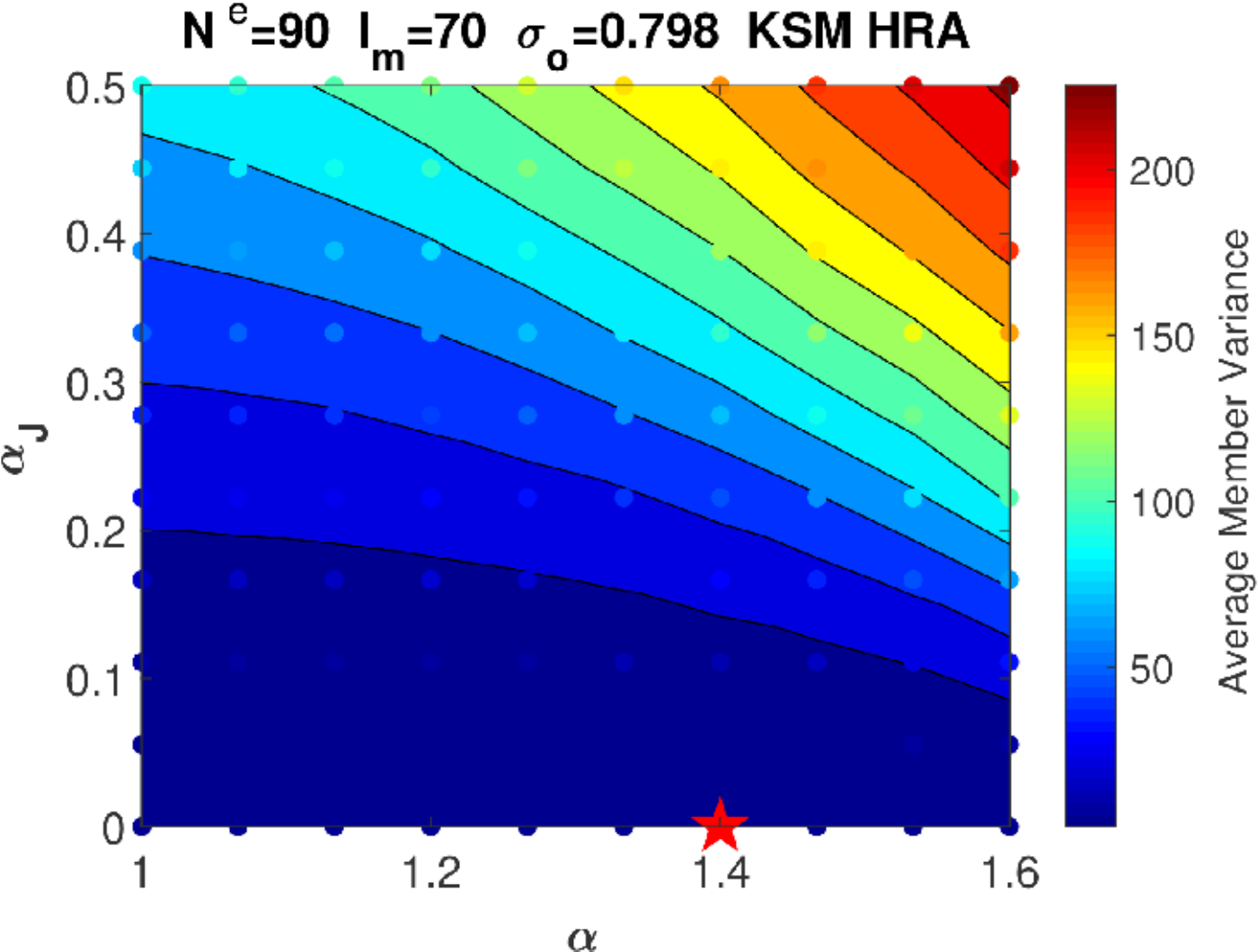}
         \label{VARKSMHRASURFFC}
      \end{subfigure}
   \begin{subfigure}[b]{0.33\textwidth}
       \includegraphics[width=\textwidth]{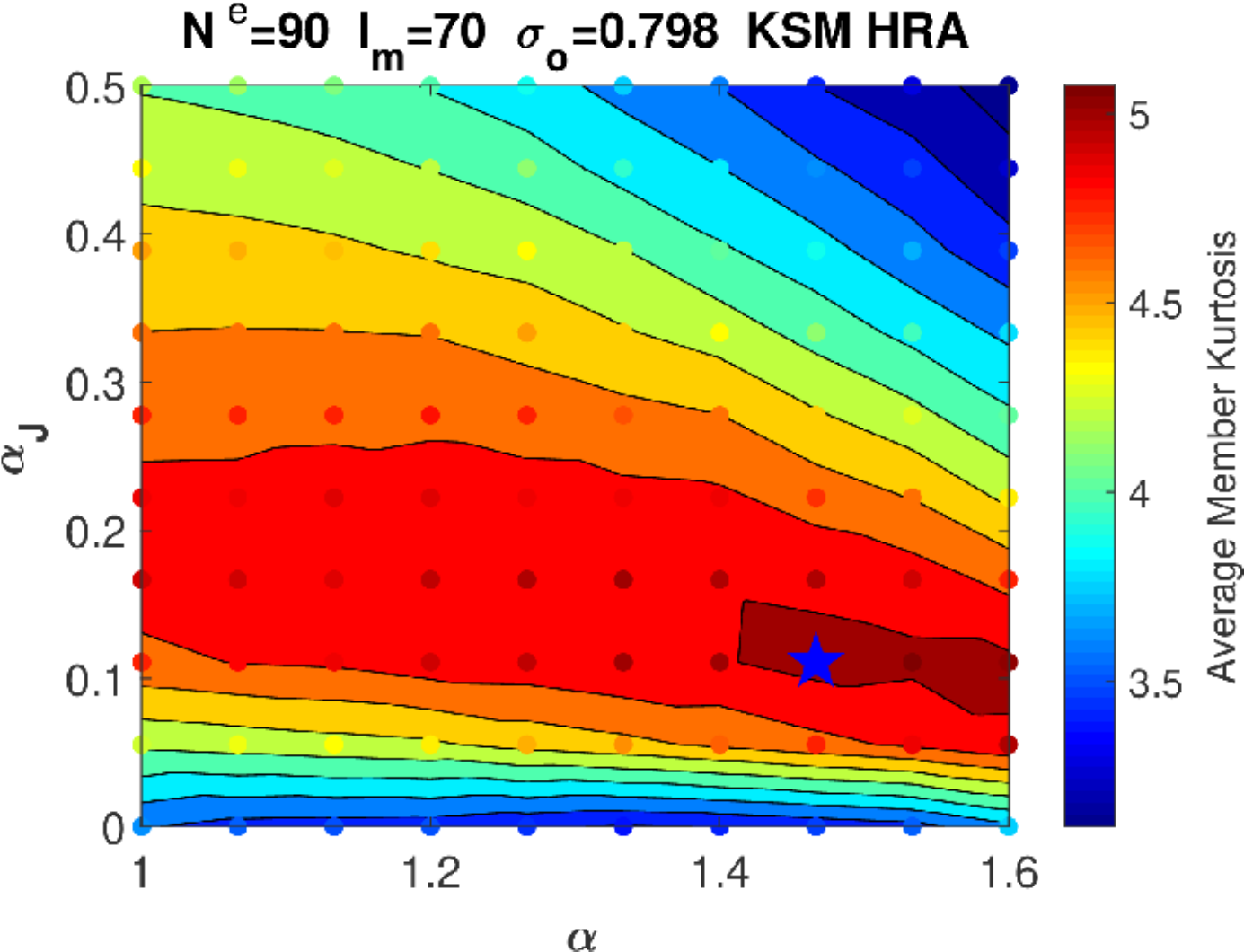}
        \label{VARKSMSHRAURFFC}
    \end{subfigure}
     \begin{subfigure}[b]{0.33\textwidth}
       \includegraphics[width=\textwidth]{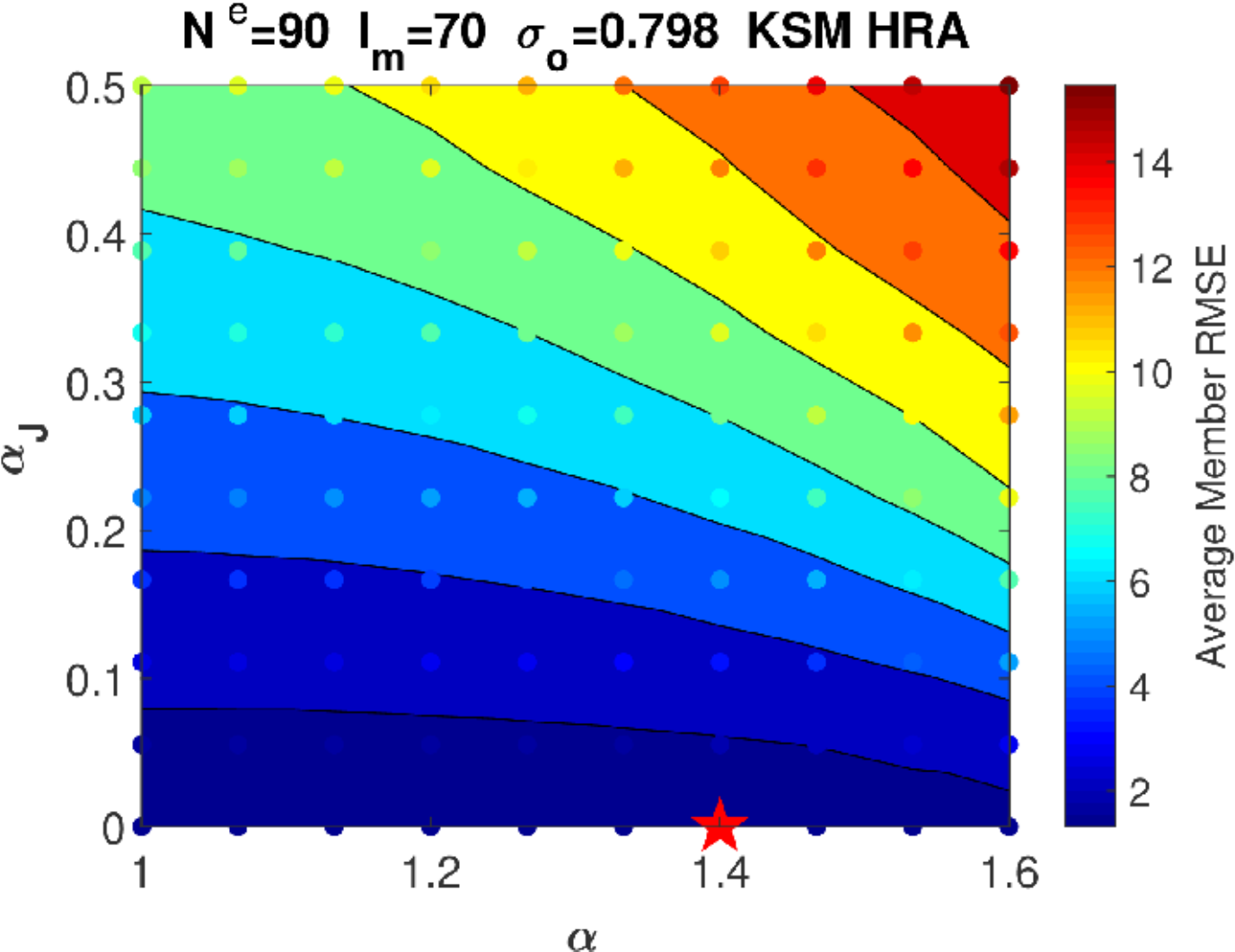}
        \label{RMSEKSMHRASURFFC}
    \end{subfigure}

  \caption{Surfaces for $\sigma_{ens}$, $K_{ens}$ and RMSE$_{ens}$ as a function of jitter and inflation for the KSM model using the forecast ensemble members, right before update, for HR (top row) and HRA (bottom row). It is notable that both $\sigma_{ens}$ and RMSE$_{ens}$ increase primarily as a function of jitter ($\alpha_J$) with less dependence on multiplicative inflation ($\alpha$) while the same is true for decreasing kurtosis evidenced by horizontal contours. This identifies the jitter as the primary source of ensemble member distortion. The red stars represent lowest values and blue stars highest values.}\label{KSMextraFC}
\end{figure*}

\begin{figure*}
    \centering
     {\bf Extra Metrics Analysis Members (KSM)}\par\medskip
    \begin{subfigure}[b]{0.33\textwidth}
        \includegraphics[width=\textwidth]{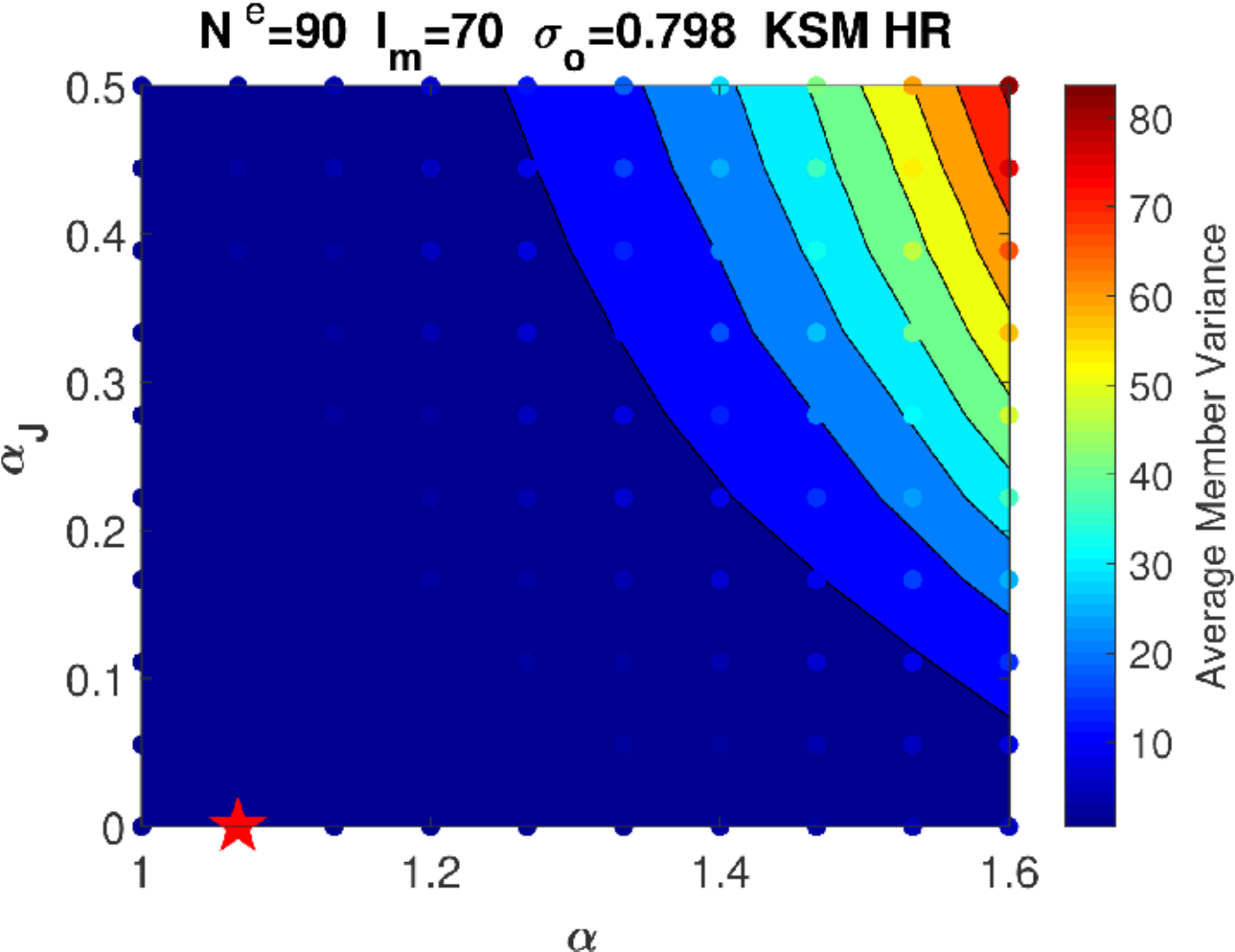}
         \label{VARKSMHRSURFAN}
      \end{subfigure}
   \begin{subfigure}[b]{0.33\textwidth}
       \includegraphics[width=\textwidth]{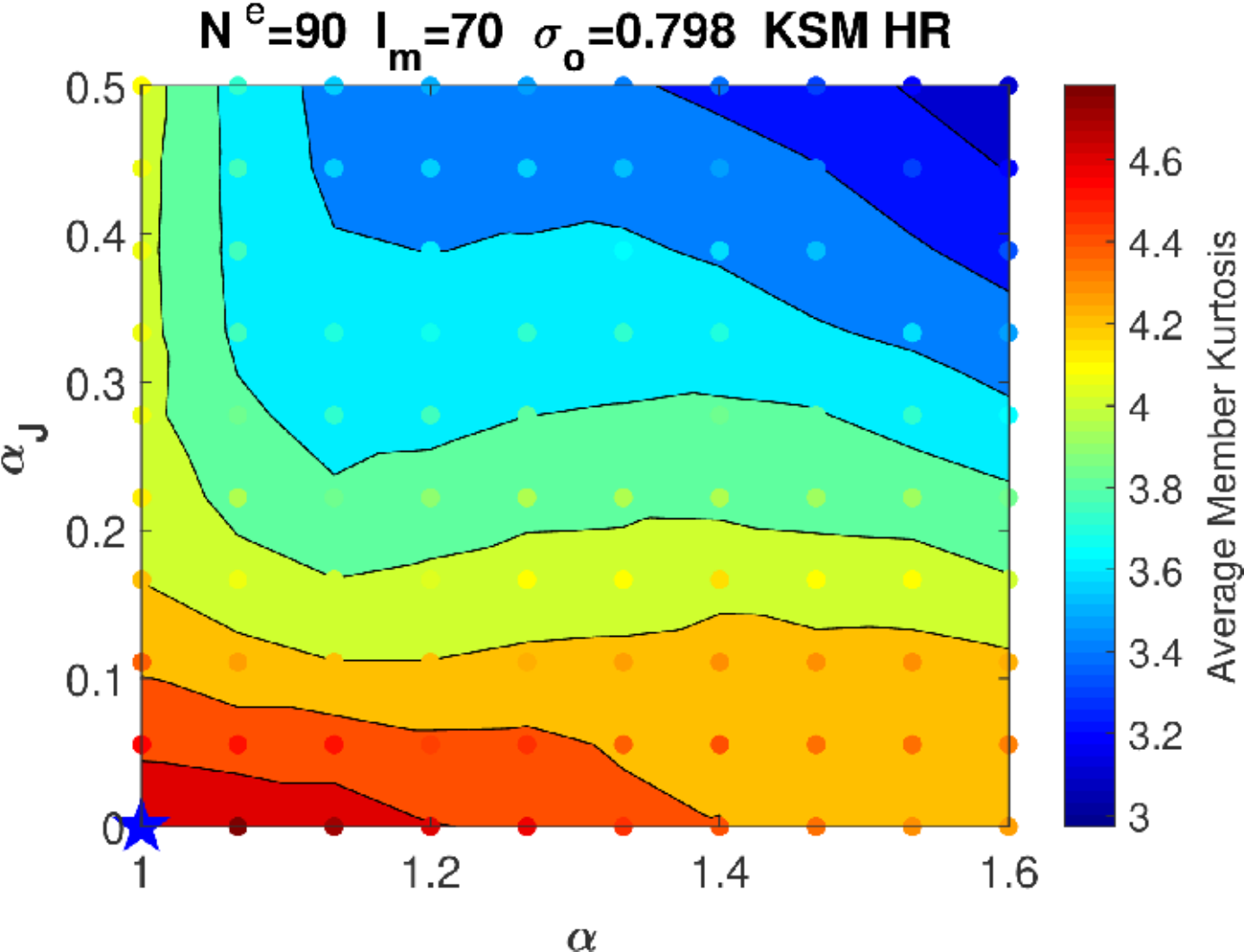}
        \label{KURTKSMSHRURFAN}
    \end{subfigure}
     \begin{subfigure}[b]{0.33\textwidth}
       \includegraphics[width=\textwidth]{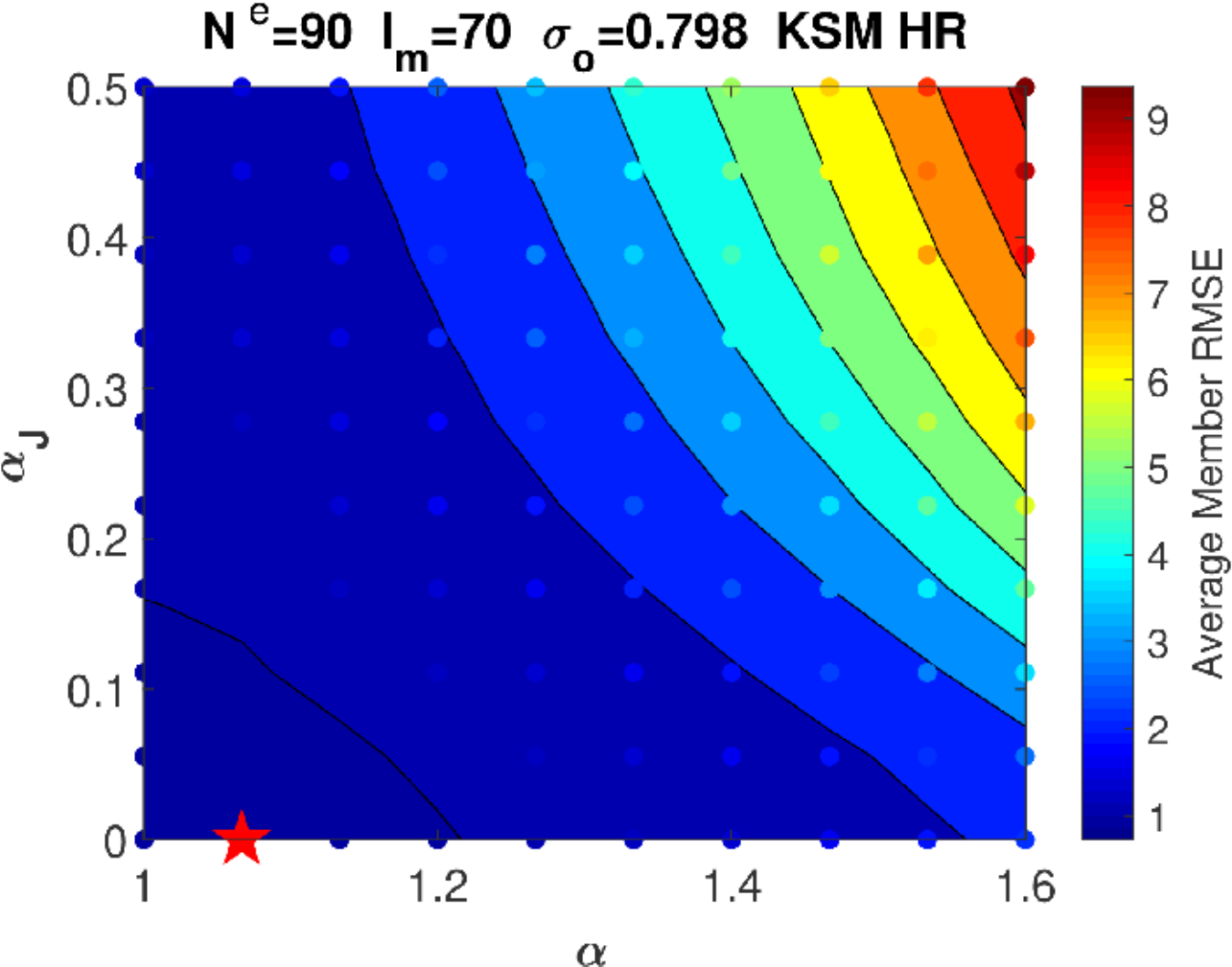}
        \label{RMSEKSMHRSURFAN}
    \end{subfigure}
    
        \begin{subfigure}[b]{0.33\textwidth}
        \includegraphics[width=\textwidth]{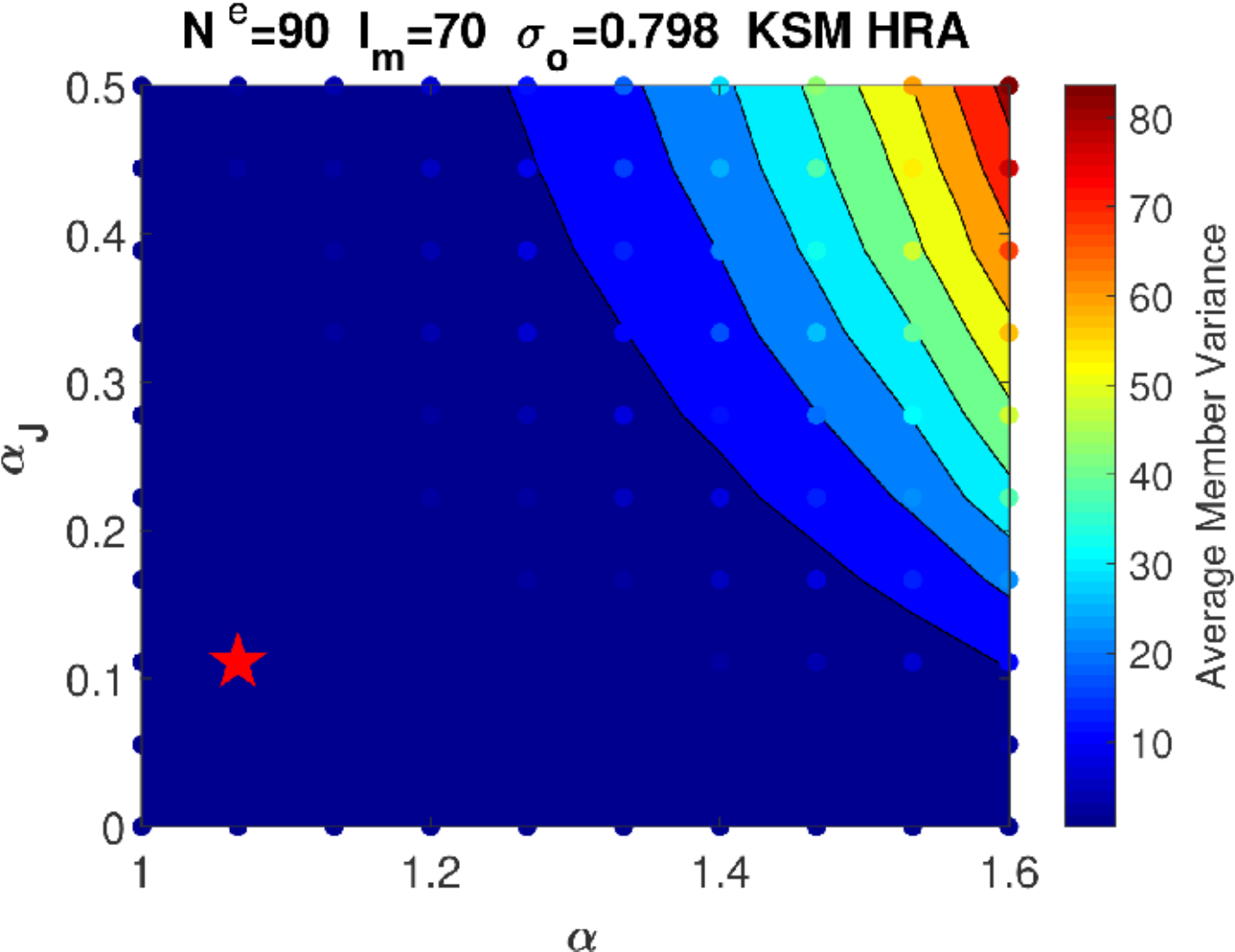}
         \label{VARKSMHRASURFAN}
      \end{subfigure}
   \begin{subfigure}[b]{0.33\textwidth}
       \includegraphics[width=\textwidth]{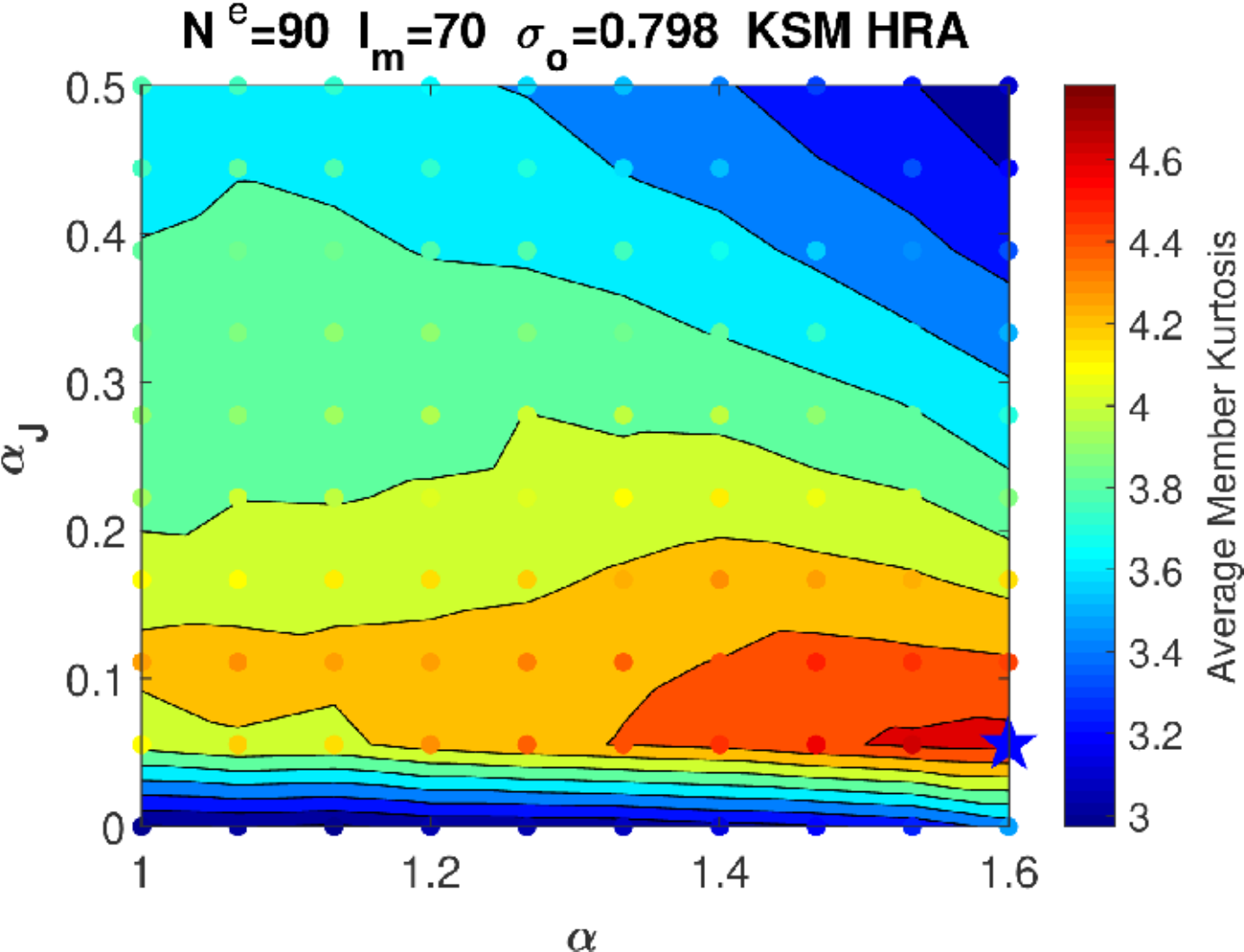}
        \label{VARKSMSHRAURFAN}
    \end{subfigure}
     \begin{subfigure}[b]{0.33\textwidth}
       \includegraphics[width=\textwidth]{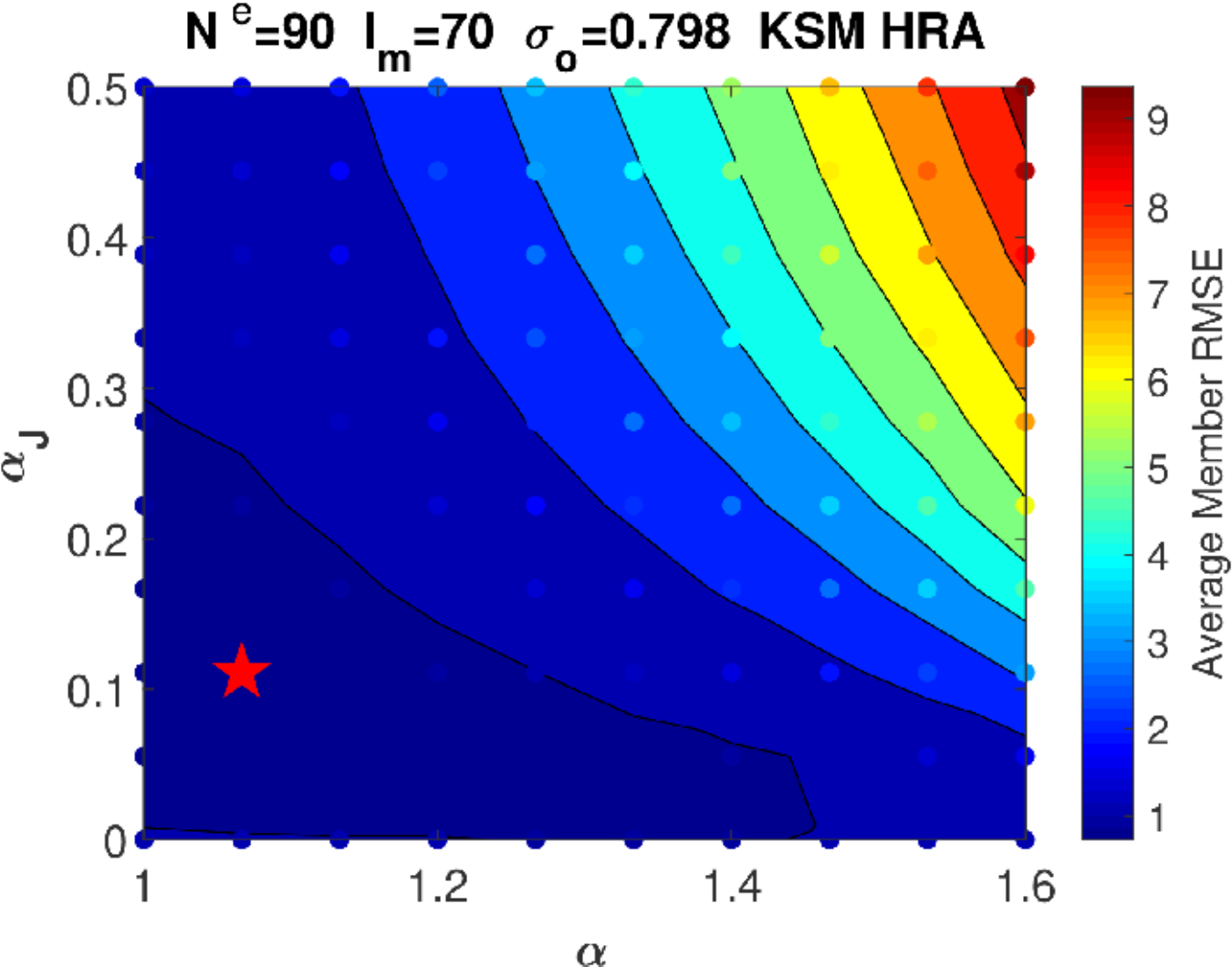}
        \label{RMSEKSMHRASURFAN}
    \end{subfigure}

   \caption{Surfaces for $\sigma_{ens}$, $K_{ens}$ and RMSE$_{ens}$ as a function of jitter and inflation for the KSM model using the analysis ensemble members for HR (top row) and HRA (bottom row). It is notable that horizontal contours shown in the same analysis for the forecast members are now slanted with improvements in $\sigma_{ens}$ and RMSE$_{ens}$ for lower values of inflation ($\alpha$). 
   The red stars represent lowest values and blue stars highest values.}\label{KSMextraAN}
\end{figure*}

To better quantify the effect that adding jitter and inflation may have on the PDE fidelity of the ensemble members we look at three different metrics. The average of the time averaged variance of the difference between the ensemble members  and the truth at each node ($\sigma_{{\rm ens}}$), the kurtosis of the same difference ($k_{{\rm ens}}$), and the average of the time averaged RMSE errors of the ensemble members (RMSE$_{{\rm ens}}$). If ${ \bf d}_i^{\tau_j}=\left({ \bf u}_i^{\tau_j}-{ \bf u}_T^{\tau_j}\right) \in \mathbb{R}^M$  is the difference between the $i^{th}$ ensemble member and the truth at assimilation time $\tau_j$ then we can define these quantities as,

\begin{align}
\sigma_{ens}=\frac{1}{N^{\rm e}}\sum_{i=0}^{N^{\rm e}} \frac{1}{N_{an}}\sum_{\tau_1}^{\tau_n} \frac{1}{M} \sum_{k=0}^M \left(d_{i_{k}}^{\tau_j}-E\left[ d_{i_{k}}^{\tau_j}\right] \right)^2, \label{sigens} \\
k_{ens}=\frac{1}{N^{\rm e}}\sum_{i=0}^{N^{\rm e}} \frac{1}{N_{an}}\sum_{\tau_1}^{\tau_n} \frac{\frac{1}{M} \sum_{k=0}^M \left(d_{i_{k}}^{\tau_j}-E\left[ d_{i_{k}}^{\tau_j}\right] \right)^4}{\left(\frac{1}{M} \sum_{k=0}^M \left(d_{i_{k}}^{\tau_j}-E\left[ d_{i_{k}}^{\tau_j}\right] \right)^2 \right)^2}, \label{kurtens} \\ 
RMSE_{ens}=\frac{1}{N^{\rm e}}\sum_{i=0}^{N^{\rm e}} \frac{1}{N_{an}}\sum_{\tau_1}^{\tau_n} \frac{1}{M} \sqrt{ \sum_{k=0}^M \left(d_{i_{k}}^{\tau_j}\right)^2}, \label{rmseens}
\end{align}
where $N_{an}$ is the number of assimilation steps completed.
If the ensemble members have a low $\sigma_{ens}$ this would suggest they don't deviate from the mean error along the domain axis which can suggest that the shape of the curve is consistent with the true solution to the PDE and had not been overly distorted by the inflation or jitter. The kurtosis can give us a measure of how concentrated around the mean error the errors are. A low value for the kurtosis suggests a more uniform distribution with the normal distribution having a kurtosis of 3. Kurtosis above 3 would suggest either that the probability mass is concentrated around the mean and values far from the mean are rare, or that the probability mass is concentrated in the tails. In this particular case, a high $k_{ens}$ likely implies the ensemble members are not overly distorted by jitter and inflation with large error occurring infrequently along the domain. A low $k_{ens}$ signals that the ensemble members are distorted by jitter and inflation with larger deviations from the mean occurring more uniformly. However, one may have a large $k_{ens}$ with ensemble members that have high PDE fidelity but are far apart from each other and or far from the truth. Nevertheless, in this analysis we are looking at a {\it long time average} and expect the ensemble to converge around the true solution in time. Examples of ensemble members with low and high kurtosis at a specific time are shown in Fig.~\ref{KURTEX}. Ideally one would hope for each ensemble member to have a low variance, low RMSE and high kurtosis calculated from ${ \bf d}_i^{\tau_j}$. In Fig.~\ref{BGMextraFC},~\ref{BGMextraAN},~\ref{KSMextraFC} and~\ref{KSMextraAN} we show $\sigma_{ens}$, $k_{ens}$ and RMSE$_{ens}$ as a function of $\alpha$ and $\alpha_J$ for ensemble members before and after the update step. The lowest values for $\sigma_{ens}$ and RMSE$_{ens}$ are denoted by a red star while the largest value of $k_{ens}$ is denoted by a blue star. We calculate these metrics for the both the forecast ensemble, right before the update step, and the analysis ensemble after. 

    

When comparing between the forecast and analysis metric surfaces for the BGM model there is a detectable change in structure for $\sigma_{ens}$ and RMSE$_{ens}$ before and after the update. Before the update, both metrics tend to increase with increasing $\alpha_J$ somewhat independently of $\alpha$ and after the update the metrics are significantly reduced for lower values of $\alpha$. This is evidenced by the relatively horizontal contours in the forecast surfaces, which changes after the update. However, the general structure of the kurtosis remains similar before and after the update, implying that the update is reducing the size of the error, but that the distortions remain among the ensemble members for larger values of $\alpha_J$. The analysis mean is anyway smoothed through averaging still providing low RMSE$_{ens}$. Comparing the Kurtosis surfaces between the HR and HRA schemes also again shows that some jitter is desirable in the HRA scheme as the larger $k_{ens}$ values are away from the x-axis. This may be caused by the ensemble members collapsing around a solution and then deviating from the truth in time due to model instability and over confidence in the model solutions. It is notable that the lowest values for $\sigma_{ens}$ and RMSE$_{ens}$ occur for almost the same value of $\alpha_J$ at which we have our maximum $k_{ens}$ for the analysis metric surfaces in the HRA case. 

When making similar considerations between the forecast and analysis metric surfaces for the KSM model we see a similar change in structure between the $\sigma_{ens}$ and RMSE$_{ens}$ surfaces as that of the BGM case. However, the contours in the KSM forecast surfaces are less horizontal implying that multiplicative inflation alone can increase the average errors. This does not necessariliy mean that the solutions are of low fidelity given that the chaos exhibited by the KSM model can simply produce ensemble members which are further from the truth but still viable solutions of the underlying PDE. In fact, they do seem to be corrected at the analysis step. Interestingly the Kurtosis surface structure changes more for the HR scheme than for HRA between forecast and analysis. It is important to note though that there is a significant difference in the range of $k_{ens}$ between the BGM (1-15) and KSM (3-5) models. This is likely due to the presence of chaos in the KSM model naturally increasing the spread of the ensemble members causing some to have little distortion and thus more a more normal distribution of errors. The small range in $k_{ens}$ likely makes this a less informative measure for the KSM model.

How one would choose $\alpha$ and $\alpha_J$ would depend on the problem at hand, minimizing only the time averaged RMSE may be the desired outcome but if ensemble member fidelity is important considering other metrics such as those presented for the forecast ensemble may also be important. For example, with the BGM model there is a drop in $k_{ens}$ from 10 to 8 when going from $\alpha_J=0.01$ to $\alpha_J=0.02$, the optimal value suggested when using 50 ensemble members ({\it cf} Fig.~\ref{BGMsurf}), for the forecast ensemble with almost no trade off in the time averaged RMSE between the two values. Depending on the sensitivity another model component may have on the fidelity of the ensemble members a more careful choice of inflationary parameters may be warranted.

\section{Conclusions}\label{conclusions}
Adaptive mesh solvers have the potential to greatly improve model skill and predictions but present difficulties for traditional data assimilation methods such as the EnKF. We consider here the particular case of a non conservative adaptive moving mesh for which each member of an ensemble will have different numbers of nodes in different locations. The key steps in an EnKF scheme for models of this sort are {\it dimension matching} often involving interpolation with a sub step of paring state vector components should they be in different locations, and {\it dimension return}. Dimension return involves removing added points in the matching step, if they were, or if points were removed whether or not to add points back in. Building on the work presented in \citet{Aydodu2019npg} we develop an EnKF scheme for a non-conservative adaptive moving mesh solver in 1-d using an augmented state vector that includes the locations of the nodes, locations that are also updated in the analysis step. Dimension matching is done using the properties of the adaptive mesh scheme itself, via a partition of the domain with intervals of the same size as the proximity tolerance $\delta_1$ which guarantees each interval will have at most one node in it. In the HR scheme developed in \citet{Aydodu2019npg} component paring is done by shifting the nodes in each interval to the their nearest interval boundaries and then interpolating new points to any interval boundaries which are empty. In this way the HR method compares ensemble members on the same mesh updating only the physical values of the nodes. Dimension return is then done by deleting interpolated points and shifting the updated physical values back to the previous mesh. In contrast, the HRA method leaves ensemble member points where they are and interpolates new points to empty intervals with the location drawn from a normal  distribution with variance $\delta_1 / 2$ with a check the location resides within that interval. Component paring is then done using like intervals. Next, the state vector is formed with the node locations appended and both physical values and nodes are updated. After the update the remeshing scheme is applied to enforce a valid mesh, and points in previously empty intervals are deleted. 

We find that when updating the node locations ensemble collapse becomes a problem and some additive or multiplicative inflation may be necessary. This is less of an issue for the HR method due to some inherent stochasticity arising from the mapping procedures, although jitter and multiplicative inflation can improve RMSE values there as well. Given an initial mesh size, ensemble size, and observation error, the jitter and inflation can be optimised with twin model experiments. When this is done we find that the HRA method typically provides better performance in terms of the time average RMSE for both the function and the first derivative. When using additive inflation, such as the jitter as we have defined it, there is potential to distort ensemble members while still obtaining a good analysis mean. This could be problematic in some frameworks such as the HMM frame work discussed in section~\ref{results}. To quantify the severity of the distortions we calculate several metrics defined in Eqs.~\ref{sigens},~\ref{kurtens} and~\ref{rmseens}. From this analysis we can see that the addition of jitter is primarily responsible for distorting the ensemble members while multiplicative inflation has less of an effect. One would want to weigh what is more important, low RMSE of the analysis mean or preserving ensemble member fidelity and should be application specific. 

We would also like to address the question of computational efficiency between these two approaches. When using the augmented state vector of the HRA scheme the size of the error covariance matrix is doubled compared to that of the HR scheme. For very high dimensional models this may be problematic, yet as can be seen in Fig.~\ref{optimalKSM} when updating the node locations the first spatial derivative RMSE is much improved and if that information is needed the extra computational cost may be worth it. It remains to be seen how a scheme like this plays out in 2-d or 3-d AMM models however, we speculate that the utility in the inclusion of the cross covariances between physical values and node locations may be far more significant in these higher dimensional cases. The complexity and range of types of patterns that can form in 2-d or 3-d is far grater than is possible in 1-d. This implies that far more information may be carried in the cross-covariances between physical values and node locations. Further, if the motivation for the use of an AMM scheme is to focus computational power in regions of strong gradients, updating those node locations in accordance with where observations of those gradients are large may be very advantageous. If the remeshing rules for the AMM model are based on strict considerations of node distances and mesh geometries a 2-d or 3-d analogue of the HR reference mesh should be attainable enabling the application of the HR or HRA schemes presented here. An example of such a model is the novel lagrangian sea ice model neXtSIM \citet{rampal2016nextsim}. neXtSIM uses a finite element method based on a triangular non-conservative adaptive mesh with strict rules on the distance between nodes and angles between edges and was the motivation behind our exploration in 1-d presented in this work. The authors are currently working on implementing the precursor of the current method \cite[{\it i.e.} without node updates]{Aydodu2019npg} in neXtSIM and shall investigate the joint physics and nodes updates based on this study soon afterwards.

\section{Acknowledgements}\label{acknowledgements}

The research in this work has been funded by the US Office of Naval Research grants Data Assimilation Development and Arctic Sea-Ice Changes (award A18-0960) and DASIM-II (award N00014-18-1-2493), A.C. has been funded by the UK Natural Environment Research Council (award NCEO02004).

\section{Conflict of interest}

{\label{644442}}

The authors attest to no conflicts of interest regarding this work.

\selectlanguage{english}
\bibliography{References.bib}
\clearpage
\graphicalabstract{Figures/COVEX/GRADVSCOVBGMKSM.pdf}{Physically driven adaptive moving mesh solvers offer many advantages over traditional fixed grid solvers. However, they present significant challenges when using ensemble Data Assimilation techniques such as the Ensemble Kalman Filter (EnKF).  We develop an EnKF scheme for non-conservative moving meshes which updates both physical state variables and node locations themselves. This leverages the information carried in the mesh structures of the ensemble members while also updating their locations through the assimilation of the physical variables that drive their locations.}
\end{document}